\documentclass[12pt,preprint]{aastex}

\newcommand{\msun}{\mathrm{M_\sun}}

\slugcomment{\today}

\shorttitle{Star Clusters in Spiral Galaxies}
\shortauthors{Di Nino et al.}

\begin{document}


\title{Star Clusters in Pseudo-bulges of Spiral
Galaxies\altaffilmark{1}}

\author{Daiana Di Nino\altaffilmark{2,3}, Michele
  Trenti\altaffilmark{2,4}, Massimo Stiavelli\altaffilmark{2},
  C.~Marcella Carollo\altaffilmark{5}, Claudia
  Scarlata\altaffilmark{6,5}, Rosemary F.~G.~Wyse\altaffilmark{7,8}}

\altaffiltext{1}{Based on observation with the NASA/ESA \textit{Hubble
    Space Telescope}, obtained at the Space Telescope Science
  Institute, which is operated by the Association of Universities for
  Research in Astronomy, Inc., under NASA contract NAS5-26555.}

\altaffiltext{2}{Space Telescope Science Institute, 3700 San Martin
  Drive, Baltimore, MD 21218, USA}

\altaffiltext{3}{CICLOPS, Space Science Institute, 4750 Walnut St.,
Boulder, CO 80301, USA}

\altaffiltext{4}{Center for Astrophysics and Space Astronomy,
University of Colorado, 389 UCB, Boulder, CO 80309, USA}

\altaffiltext{5}{ETH Zurich, Physics Department, CH-8033 Switzerland}

\altaffiltext{6}{Spitzer Science Center, Caltech, 1200 East California
Boulevard, Pasadena, CA 91125, USA}

\altaffiltext{7}{Department of Physics and Astronomy, Johns Hopkins
  University, 3400 North Charles Street, Baltimore, MD 21218, USA}

\altaffiltext{8}{Institute for Astronomy, University of Edinburgh,
 Blackford Hill, Edinburgh EH9 3HJ, UK}


\begin{abstract}
  We present a study of the properties of the star-cluster systems
  around pseudo-bulges of late-type spiral galaxies using a sample of
  11 galaxies with distances from 17 to 37 Mpc. Star clusters are
  identified from multiband HST ACS and WFPC2 imaging data by
  combining detections in 3 bands (F435W and F814W with ACS and F606W
  with WFPC2). The photometric data are then compared to population
  synthesis models to infer the masses and ages of the star
  clusters. Photometric errors and completeness are estimated by means
  of artificial source Monte Carlo simulations. Dust extinction is
  estimated by considering F160W NICMOS observations of the central
  regions of the galaxies, augmenting our wavelength coverage. In all
  galaxies we identify star clusters with a wide range of ages, from 
  young (age $\lesssim$ 8 Myr) blue clusters, with typical mass of
  $10^3$ ${\mathrm M_{\sun}}$ to older (age $>$ 100-250 Myr), more
  massive, red clusters. Some of the latter might likely evolve into objects
  similar to the Milky Way's globular clusters. We compute the
  specific frequencies for the older clusters with respect to the
  galaxy and bulge luminosities. Specific frequencies relative to the
  galaxy light appear consistent with the globular cluster
  specific frequencies of early-type spirals.  We compare the specific
  frequencies relative to the bulge light with the globular cluster
  specific frequencies of dwarf galaxies, which have a
  surface-brightness profile that is similar to that of the
  pseudo-bulges in our sample. The specific frequencies we derive for
  our sample galaxies are higher than those of the dwarf galaxies,
  supporting an evolutionary scenario in which some of the dwarf
  galaxies might be the remnants of harassed late-type spiral galaxies
  which hosted a pseudo-bulge.

  \end{abstract}


\section{Introduction}\label{sec:intro}

Over the last two decades, observational evidence has been
accumulating that not all spiral galaxies possess a bulge resembling a
small elliptical galaxy with a de Vaucouleurs R$^{1/4}$ light
profile. Bulges with a surface-brightness profile that does not follow
the de Vaucouleurs law also show disk-like, cold stellar kinematics
\citep{Kormendy93} and light profiles that are well-fitted by an
exponential or an intermediate Sersic law \citep{Andredakis}. For
these reasons these bulges are known as pseudo-bulges or exponential
bulges. HST imaging of pseudo-bulges has also revealed that they often
contain central compact sources, most likely star clusters
\citep{wfpc2_II} and that they have shallower nuclear slopes than
traditional bulges with the same magnitude \citep{wfpc2_III}. It is
also intriguing that the nuclear cusp slopes of the exponential bulges
are similar to those of dwarf elliptical galaxies with the same radii
and luminosities. This similarity suggests that there might be an
evolutionary link between pseudo-bulges and dwarf ellipticals, as if
the presence of a disk did not affect the nuclear properties of
pseudo-bulge spirals \citep{Stiavdwarfs}.

In order to study in detail the properties of pseudo-bulges and
investigate possible formation scenarios, we isolated a sample of 11
pseudo-bulges from our previous HST surveys with WFPC2 and NICMOS and
obtained for this sample additional HST/ACS imaging data. The relation
between bulge properties and galaxy properties for these systems has
been discussed already by \citet{acs_I}. These authors found that the
bulge properties correlate very well with those of the host galaxy,
suggesting that evolutionary processes of the disk may be responsible
for the formation of the pseudo-bulges. In this paper we focus rather
on the properties of the population of star clusters within the
pseudo-bulges.

We have two main goals. The first is to characterize the
star-cluster systems of these late-type pseudo-bulge hosts and to
compare them with the growing body of literature focused on spirals
with classical bulges (for a review see
\citealt{brodie_strader_2006}). Under the assumption that globular
  clusters form in association with enhanced bursts of star formation
  (e.g.~\citealt{beasley02}), the abundance of globular clusters is
expected to depend on whether a bulge is formed ``classically'', that
is with a single burst of star formation, or more slowly, through
extended star formation associated with secular processes in the
quiescent galaxy disk
\citep{kormendy_kennicutt2004,brodie_strader_2006}. In this latter
scenario, a lower specific frequency would be expected.
 
The second goal is to continue to investigate the similarities between
pseudo-bulges and dwarf ellipticals, first highlighted by
\citet{Stiavdwarfs}. A spiral galaxy can be stripped of its disk
during close interactions with neighbors, especially effective in a
group environment \citep{Moore96}. This ``galaxy harassment'' scenario
suggests evolution of late-type spirals with a pseudo-bulge into dwarf
ellipticals. The stellar populations of low-mass galaxies in clusters
are indeed consistent with some fraction having evolved in this manner
(e.g.~\citealt{conselice03}). Under this scenario, there should also
be a correlation between the properties of the star-cluster systems of
these two galaxy classes. Here we explore this connection by measuring
the specific frequencies of globular clusters per bulge light in our
sample and comparing it with those of dwarf ellipticals.

This paper is organized as follows. We briefly introduce the sample
and discuss the data reduction in Section \ref{sec:sample}. Source
detection and photometry are described in Section \ref{sec:phot}. Star
clusters are identified in Section \ref{sec:DA} based on a color and
luminosity selection. Systematic uncertainties are discussed in
Section \ref{sec:sys}. Specific frequencies for the star cluster
systems are derived in Section \ref{sec:spec_freq} while Section
\ref{sec:nuclear} briefly illustrates the properties of nuclear star
clusters. Our interpretation and conclusions are discussed in Section
\ref{sec:concl}.

\section{Sample and Data Reduction}\label{sec:sample}

We use images of 11 late-type spiral galaxies obtained during the
HST-ACS pseudo-bulge survey GO-9395 (PI: C.~M.~Carollo). The same
systems were also observed previously with WFPC2 and NICMOS
\citep{wfpc2_I,wfpc2_II,wfpc2_III}. All galaxies were selected on the
basis of: (1) angular diameter larger than 1\arcmin; (2) regular
morphological type; (3) redshift less than 2500 km s$^{-1}$, to
guarantee a high angular resolution in physical size; and (4) an
inclination angle, estimated from the apparent axial ratio, smaller
than 75\degr, to avoid strong obscuration of the nucleus by the
disk. The objects of this study had also be shown to possess an
exponential bulge \citep{wfpc2_III, acs_I}.  The galaxy coordinates,
distance (assuming $H_0=73$ km s$^{-1}$ Mpc$^{-1}$ for the Hubble
constant), absolute magnitude, morphological classification and
galactic extinction are listed in Table \ref{tb:sample}. For each
galaxy, we have ACS/WFC images in the F435W and F814W filters, WFPC2
images in the F606W filter and NICMOS images in the F160W filter.

The exposures in the F606W ($V$) filter were acquired between 1996 and
1997 by WFPC2 with the galaxy nucleus centered on the PC camera (the
only WFPC2 chip considered in this study as its resolution is
comparable with ACS/WFC), which has a scale of 0.046\arcsec $ $
pixel$^{-1}$ and a field of view of about 35\arcsec x 35\arcsec. The
observations were carried out in fine lock with a nominal gain of 15
electrons/DN and a total exposure time per field of 600 s, split in
two to allow removal of cosmic rays. For these images the raw data
were processed with the standard WFPC2 pipeline (CALWP2) in order to
use the most recent reference frames for flat-fielding, bias and dark
current subtractions; the cosmic-rays were removed using the
IRAF\footnote{IRAF is distributed by NOAO, which is operated by AURA
  Inc., under cooperative agreement with the National Science
  Foundation.} STSDAS task CRREJ and the remaining hot pixels were
removed by interpolation. Finally, sky subtraction was performed by
determining the sky values from the WF chips, in areas farthest from
the nucleus.

The images in the F435W ($B$) and F814W ($I$) filters were acquired
between 2002 and 2003. The observations were carried out with the
ACS/WFC channel, whose pixel scale is 0.05\arcsec $ $ pixel$^{-1}$ for
a total field of view of 3.4\arcmin x 3.4\arcmin. Details of these
observations can be found in \citet{acs_I}. The standard ACS pipeline
(CALACS) was used to perform the basic data reduction (flat-fielding,
bias and dark subtractions and removal of the overscan regions), then
the IRAF STSDAS tasks ACSREJ and DRIZZLE were used, respectively, to
achieve cosmic-ray rejection and to correct for geometric
distortions. Finally, any remaining hot pixels were removed by
interpolation and sky subtraction was carried out for all galaxies.

The images in the F160W ($H$) filter were acquired between 1997 and
1998 by NICMOS (Camera 2). The pixel scale is 0.075\arcsec $ $
pixel$^{-1}$ for a total field of view of about 19.2\arcsec x
19.2\arcsec. These observations were carried out in snapshot mode and
split into multiple exposures (to allow better cosmic-ray rejection)
for a total exposure time per field of 384 s for ESO 498G5, ESO
499G37, NGC 1345, NGC 1483, NGC 3259 and NGC2758 and a total exposure
time per field of 256 s for NGC 406, NGC 2082, NGC 3455, NGC4980 and
NGC 6384.  These images were reduced through the standard pipeline
software (CALNICA). The various NICMOS anomalies (e.g. the pedestal
anomaly) were corrected on a case-by-case basis. We direct the reader
to \citet{nicmos_I} for further details.

\section{Photometry}\label{sec:phot}

\subsection{Source Detections \& Aperture Photometry}\label{ssec:detection}

Our goal is to identify star-cluster candidates and to characterize
their integrated photometric properties. Therefore we aim at
constructing a uniform catalog of sources, focusing on detections that
lie in the common area of the F435W, F606W and F814W images. The
resulting search radius for each galaxy is given in the last column of
Table \ref{tb:sample}. The NICMOS F160W images have a more limited
field of view and we use them only to evaluate the possible impact of
dust extinction (see Section~\ref{sec:dust}).

Our study is limited to star clusters close to the nucleus of the host
galaxy, where there are significant inhomogeneities in the image
background, mainly due to dust lanes and gradients in the
integrated-light profile. Therefore, we identify star-cluster
candidates on a `variance-normalized' image, that allows improved
detection of point-like sources in regions of rapidly varying
background \citep{Miller97}. The `variance-normalized' image is
obtained by summing the F435W and F814W images to improve the signal
to noise ratio and then by dividing the result by a smoothed version
of itself, realized with a 11x11 pixel median filter.

We use the IRAF task DAOFIND to search for star clusters in the
images. This task, optimized for point-source detection, is adequate
for our goal, due to the compact size (FWHM $\approx$ 1 or 2 pixels)
of the star clusters in our sample.  Indeed, for our purposes, with
our data, DAOFIND marginally outperforms extended-source detection
packages such as SExtractor \citep{bertin}.

The coordinates of the sources detected on the `variance-normalized'
image are then mapped into the local pixel coordinates of each image
using the IRAF tasks XY2RD, RD2XY and GEOMAP. We visually inspected
all sources, with particular attention paid to those with FWHM
$>0.25\arcsec$ in the F435W band, corresponding to a half-light radius
$R=0.15\arcsec$ after deconvolution with the PSF, if the source has a
Plummer surface brightness profile (see Section~\ref{sec:ap_corr}).
The angular scale $R = 0.15\arcsec$ corresponds to R $\approx$ 13 pc
at 17 Mpc --- the distance of our closest galaxy --- and R $\approx$
27 pc at 38 Mpc --- the distance of the farthest galaxy. These more
extended sources have in general very low surface brightness and often
appear to be artifacts and are thus excluded from the subsequent
analysis. This cut is primarily motivated by the size distribution of
galactic GCs, which have radii smaller than about 20 pc. In addition,
all the sources in our final catalog have been visually inspected to
remove bright foreground stars identified from their diffraction
spikes.

We perform circular aperture photometry of all the detected sources
using an aperture radius of 0.15\arcsec $ $ for each image and
estimating the background level in an annulus between 0.5\arcsec $ $
and 1\arcsec $ $ in radius. We verify that this method yields a
correct estimate of the background including the galaxy luminosity. We
checked this by choosing the central part of the galaxies, where their
luminosity gradient is steepest and comparing our photometry of the
nuclear star clusters against the published photometry of
\citet{wfpc2_I,wfpc2_II}, which took into account a detailed model of
the galaxy light profile. In addition, the aperture correction and
error estimates have been performed via Monte Carlo simulations on the
actual images (see Section~\ref{sec:ap_corr}), therefore any residual
bias is corrected to first order.

We consider only sources that in all bands have a magnitude greater
than a chosen threshold magnitude, corresponding to a signal-to-noise
ratio (SNR) greater than 5 (see Table \ref{tb:basic_phot}). The
photometric calibration is done by converting instrumental magnitudes
to the VEGAMAG magnitude system by applying the photometric zeropoints
listed in Table \ref{tb:basic_phot}. All magnitudes are then corrected
for Galactic extinction following \citet{red}. Since the HST filters
differ from the Johnson-Cousins band-passes, reddening corrections are
computed using the effective transmission curve for each HST filter,
created with the IRAF task CALCBAND in the SYNPHOT package (see Table
\ref{tb:basic_phot}).

In this paper we indicate with uppercase letters (e.g. $V$) the
apparent magnitude in the HST filters, while we use symbols like
$M_{V}$ for the corresponding absolute magnitude. Absolute magnitudes
in Johnson-Cousins bands are instead indicated with symbols like
$M_{V_{JC}}$.

\subsection{Aperture Correction \& Uncertainty Estimates}\label{sec:ap_corr}

Total magnitudes are determined by correcting for flux outside the
measurement aperture. Aperture corrections, photometric uncertainties
and the completeness of our observations are estimated by means of
Monte Carlo simulations. We add artificial star clusters of known
luminosity and size to each image in random positions within the
search area defined in Section \ref{ssec:detection} and we measure the
retrieved flux with the same aperture photometry procedure adopted for
the real sources. The luminosity profile of the simulated star
clusters follows a projected Plummer law:
\begin{equation}
L(R)=\frac{L_{tot}}{\pi b^2}\left(1+\frac{R^2}{b^2}\right)^{-2}
\end{equation}
where $b$ is the assigned scale length and $L$ is the assigned total
luminosity. This is then convolved with the Point Spread Function
(PSF) in the appropriate filter obtained with TinyTim \citep{TinyTim}
before being added to the image.

\begin{figure}
\epsscale{1.0}
\plotone{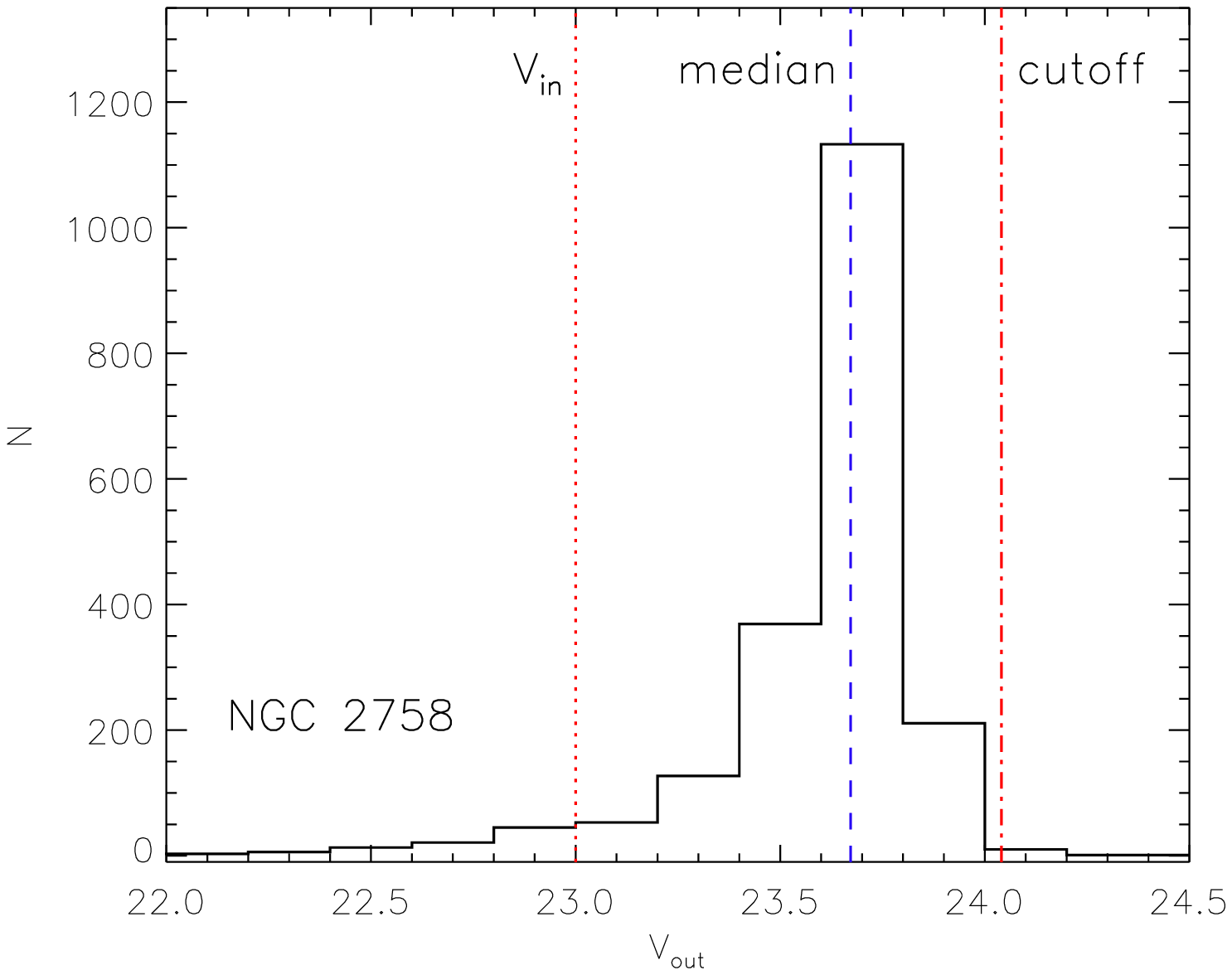} 
\caption{Distribution of retrieved magnitudes ($V_{out}$) obtained
  through a Monte Carlo simulation with $V_{in} = 23$ mag (red dotted
  line) for the NGC 2758 galaxy in the $V$-band. The resulting median
  value of the distribution is also shown (blue dashed line). Sources with 
 retrieved magnitude brighter than the $2 \sigma$ cutoff (red dot-dashed line) 
 and fainter than $V_{in}$ (red dotted line) are defined as successfully recovered.  
  \label{fig:m_out}}
\end{figure}

The result for a representative photometric Monte Carlo simulation is
shown in Fig.~\ref{fig:m_out}. Note that the distribution of the
output magnitude ($m_{out}$) at a fixed input magnitude ($m_{in}$) has
a width much smaller than the difference $m_{out}-m_{in}$. We define
the aperture correction as the difference between $m_{in}$ and the
median of the distribution of the output magnitude $m_{out}$. We find
that the aperture correction is almost independent of $m_{in}$, and
depends mildly on the size of the artificial source, with variations
of about 0.2 mag for clusters with observed half-light radii between
$0.05\arcsec$ and $0.15\arcsec$. We therefore apply the aperture
correction to the observed sources according to their apparent size,
using linear interpolation across our grid of models.

\begin{figure}
\epsscale{.50}
\plotone{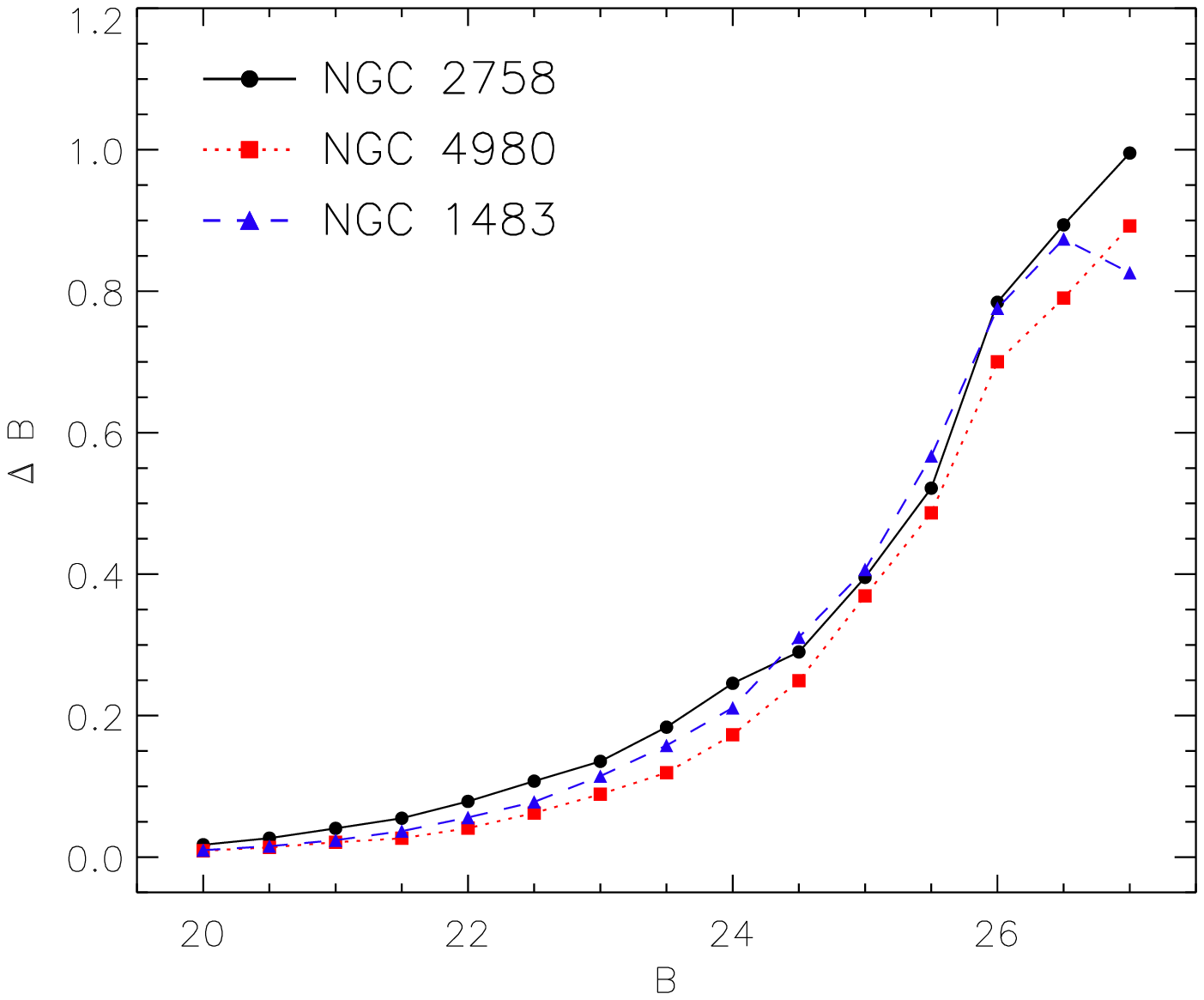} 
\plotone{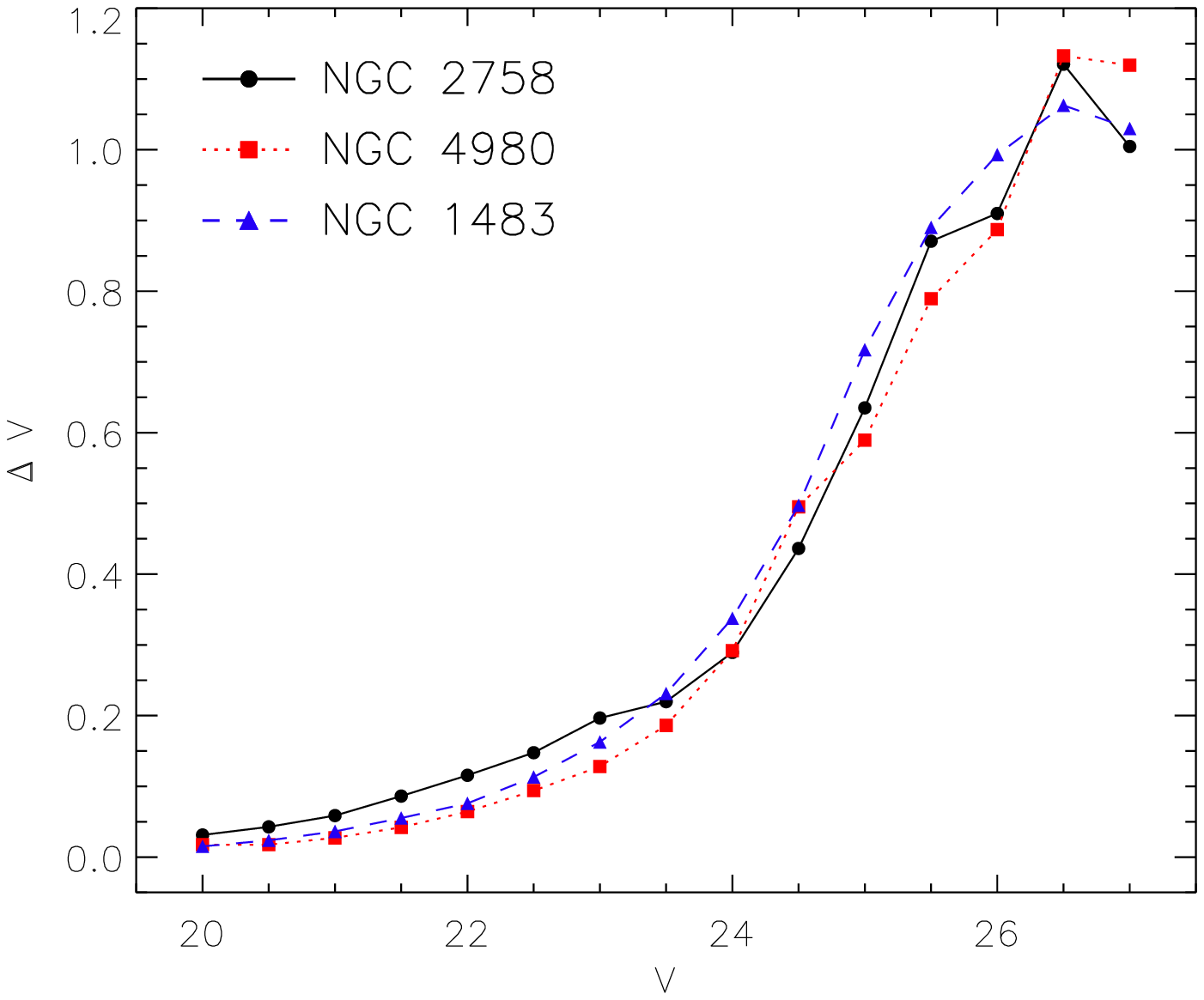} 
\plotone{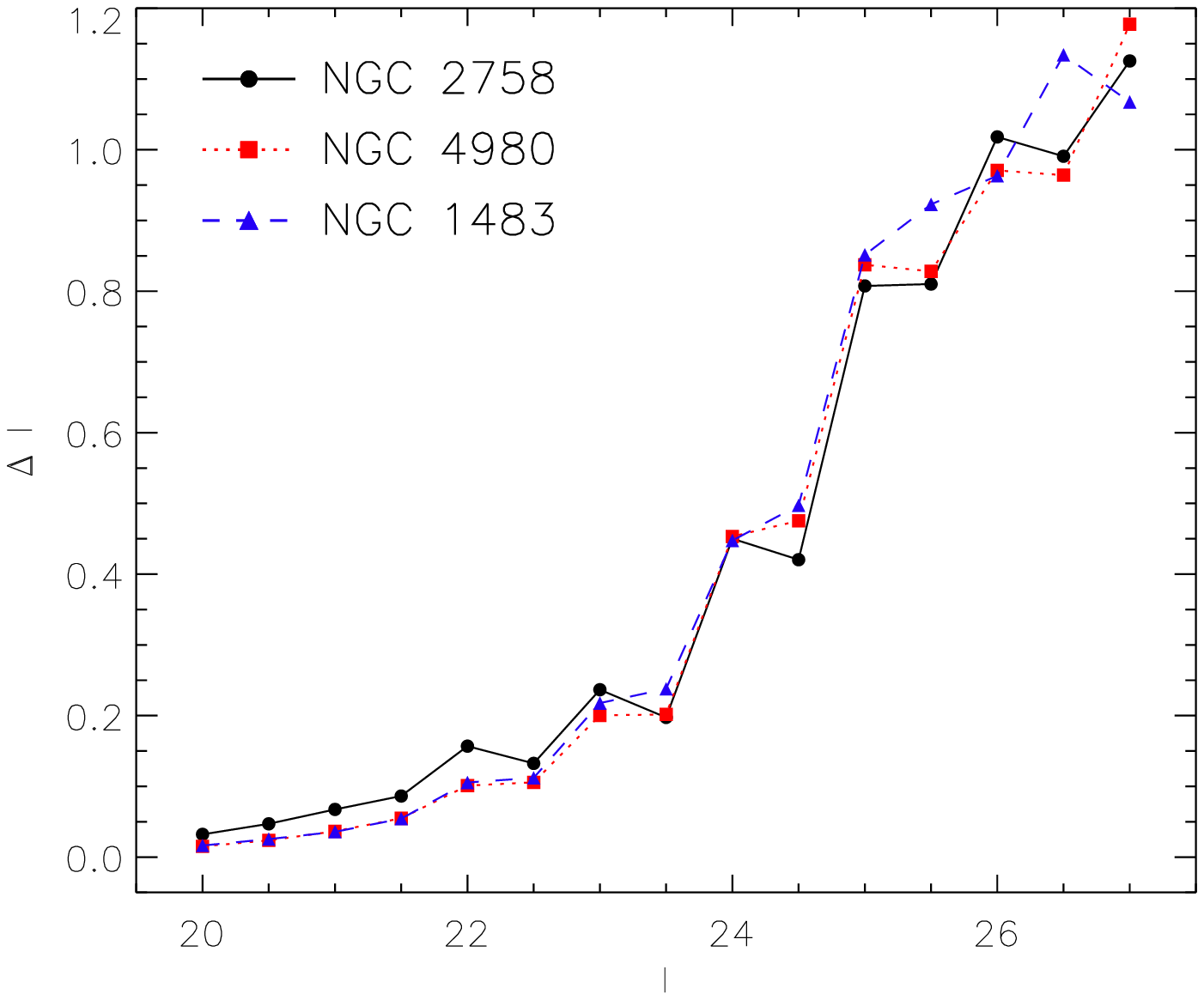} 
\caption{Photometric uncertainties for star cluster-like sources in
  three galaxies representative of our sample (NGC 2758 solid line,
  NGC 4980 dotted line and NGC 1483 dashed line). These uncertainties
  have been estimated via Monte Carlo simulations by placing and
  recovering artificial sources in the original images.
  \label{fig:error_trend}}
\end{figure}

The error on the photometry is similarly estimated from the Monte
Carlo simulations as the variance of the output magnitudes of the
artificial sources. The typical uncertainties are plotted in
Fig.~\ref{fig:error_trend} as a function of the magnitude for three
galaxies representative of our sample, NGC 1483 ($D = 17.1$ Mpc), NGC
4980 ($D = 23.9$ Mpc) and NGC 2758 ($D = 31.3$ Mpc). Note that the
photometric error so obtained is larger than the formal statistical
photometric error reported by the IRAF aperture photometry task.

We neglect the effect of possible varying charge transfer efficiency
(CTE) because the observations were carried out when the instruments
were still young, thus the effect is not severe. Further, the
background levels in the images are provided by the underlying
galaxies, rather than just the sky, and are high enough to make the
effect minimal \citep{ZP_wfpc2}.

\begin{figure}
\epsscale{.50}
\plotone{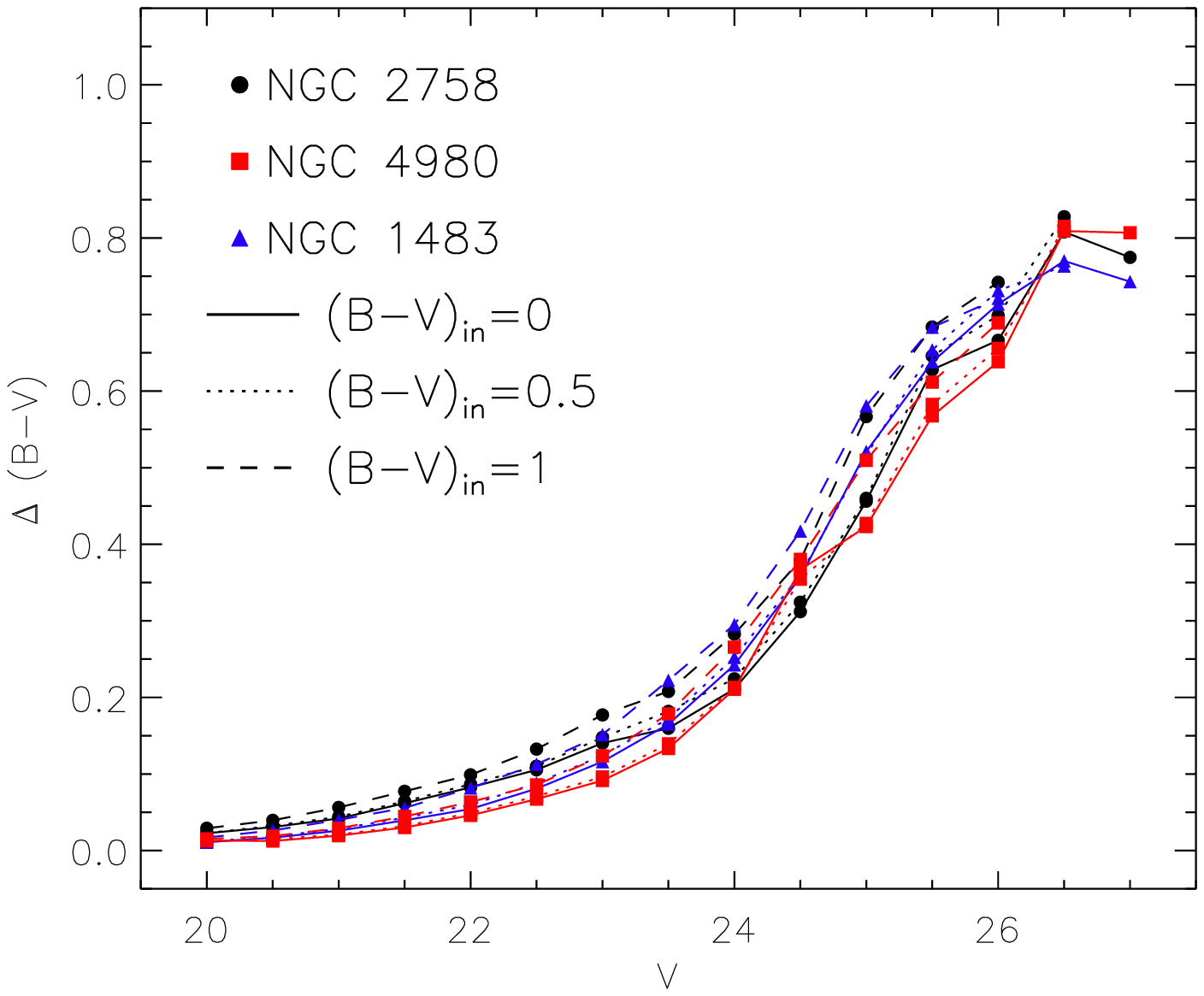} 
\plotone{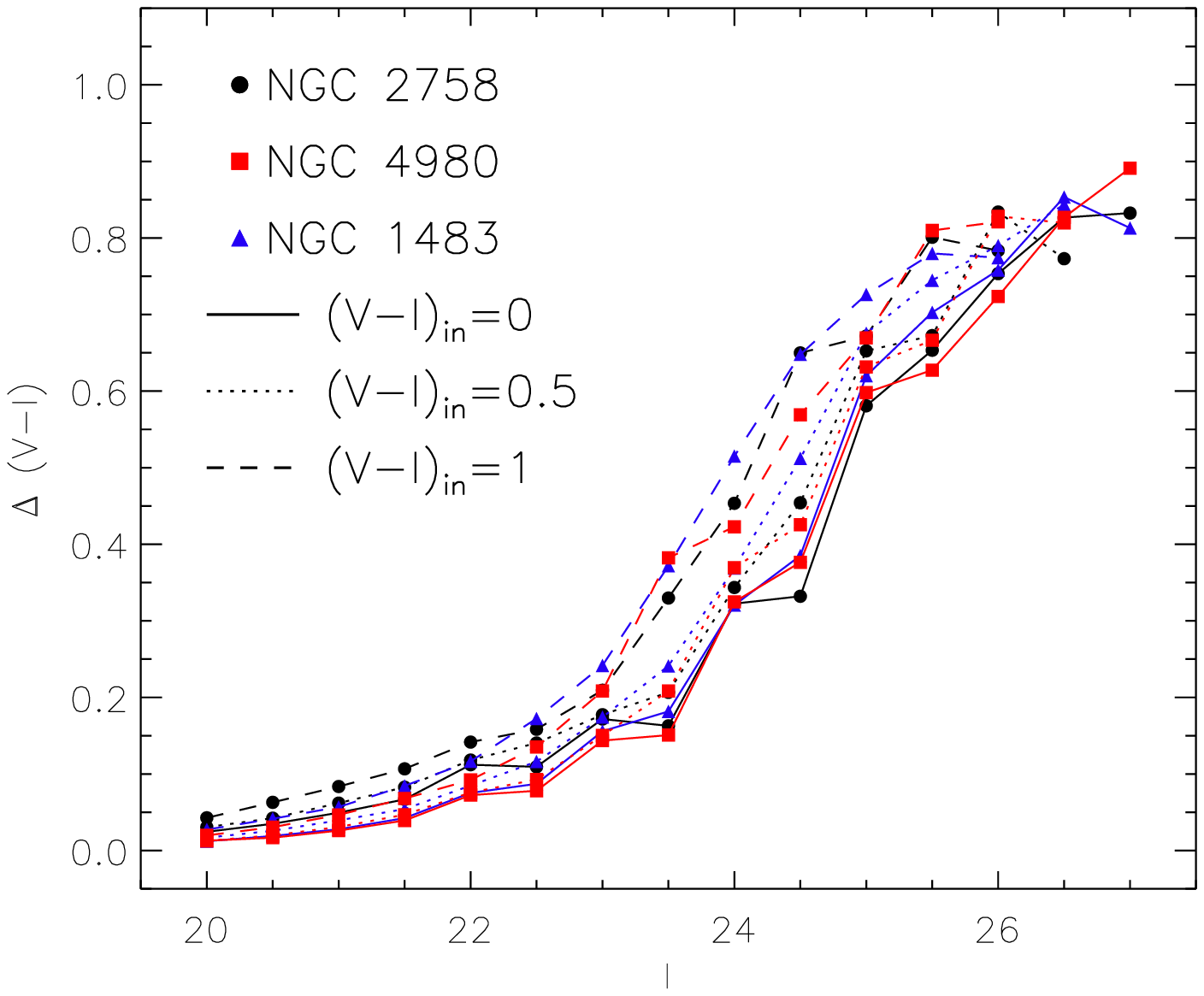} 
\plotone{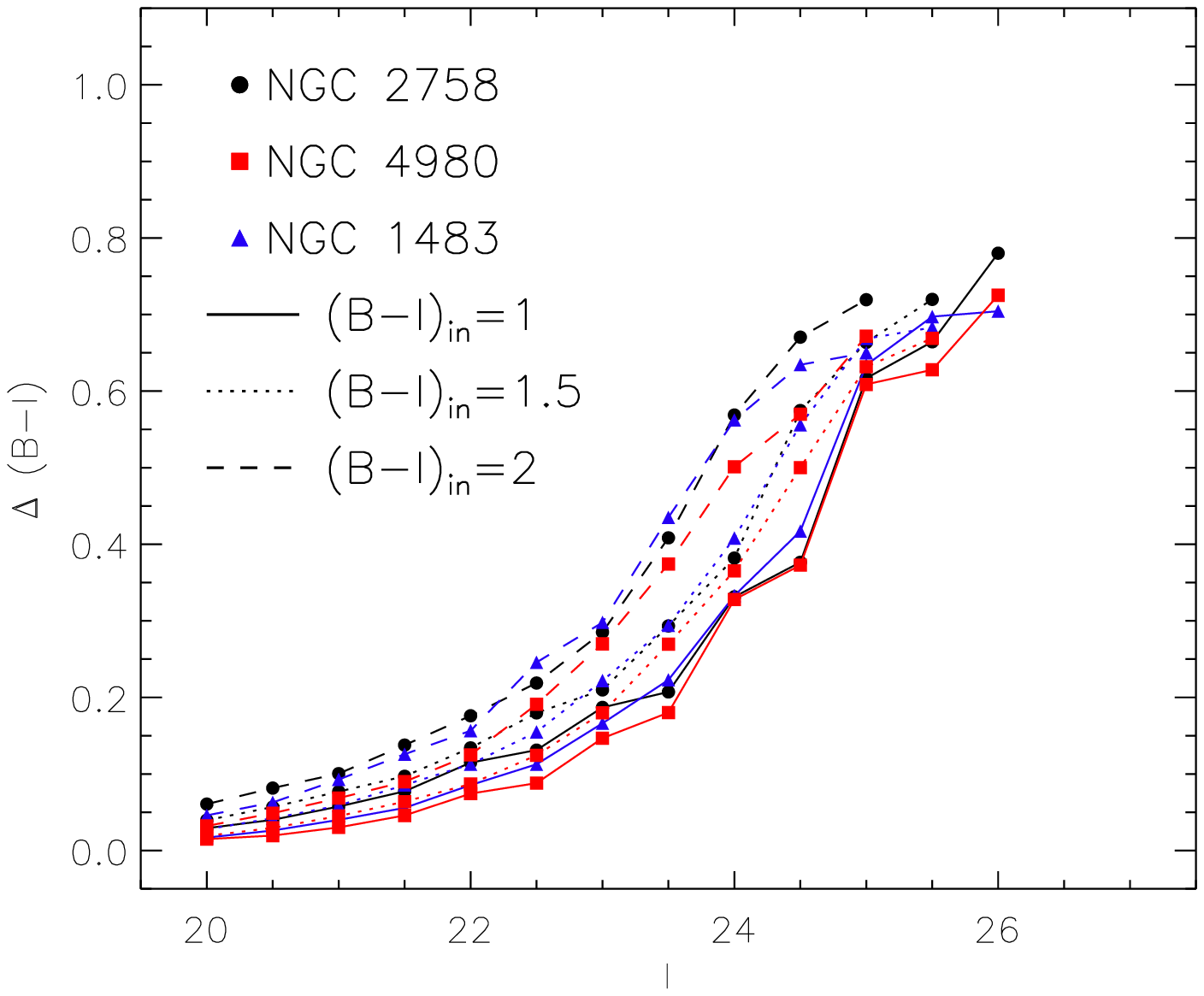} 
\caption{Photometric uncertainties in the color measurements relative
  to input magnitudes for representative input colors and galaxies in
  the sample, estimated from Monte Carlo simulations as in
  Fig.~\ref{fig:error_trend}.
\label{fig:color_error_trend}}
\end{figure}

The errors on the colors are also quantified using our Monte Carlo
simulations. We place and recover artificial sources with a given
input color ($M_B-M_I\equiv B-I=1,1.5,2$; \hbox{$M_V-M_I$} $\equiv
V-I=0,0.5,1;M_B-M_V\equiv B-V=0,0.5,1$) on the corresponding
images. From the recovered photometry we measure the error on the
color and correlate it with the errors in each band, under the
assumption of a linear relation. For example, in case of the $B-I$
color the error-relation reads:
\begin{equation}
  \sigma_{B-I}^2=\sigma_{B}^2+\sigma_{I}^2-2\cdot r_{BI}\cdot \sigma_{B}
  \cdot \sigma_{I}
\end{equation}
where $r_{BI}$ is the linear correlation term. The typical
uncertainties on the colors are plotted in
Fig.~\ref{fig:color_error_trend} as a function of the magnitude and
the color itself for the three reference galaxies introduced
above. Our analysis yields that the typical value of the correlation
$r$ is only mildly dependent on the color itself, with variations from
$r\approx 0.5$ to $r\approx 0.9$. We assume a reference value of $r
\approx 0.7$ (the average value of the sample), which turns out to
describe the color errors satisfactorily.

Finally, through the same Monte Carlo simulations we also determine
the completeness of our observations. For this purpose we define as
`successfully recovered' a synthetic source whose output magnitude
satisfies:
\begin{equation}\label{eq:comp}
m_{in}< m_{out} <m_{in}+m_{corr}+2 \sigma
\end{equation}
where $m_{in}$ is the assigned input magnitude and $m_{corr}$ and
$\sigma$ are respectively the aperture correction and the one-sigma
photometric error previously estimated. Note that (i)
$m_{corr}>>\sigma$ (for example, see Fig.~\ref{fig:m_out}), thus a
source which is recovered as brighter than its input luminosity has an
error at several standard deviations (essentially this happens only
when a synthetic source falls on top of another existing source),
which implies failure in the recovery; (ii) the error distribution is
highly skewed (see Fig.~\ref{fig:m_out}). Thus the completeness is
similar if we were to change the cutoff in Eq.~\ref{eq:comp} to either
$1 \sigma$ or $3\sigma$.

\begin{figure}
\epsscale{1.0}
\plotone{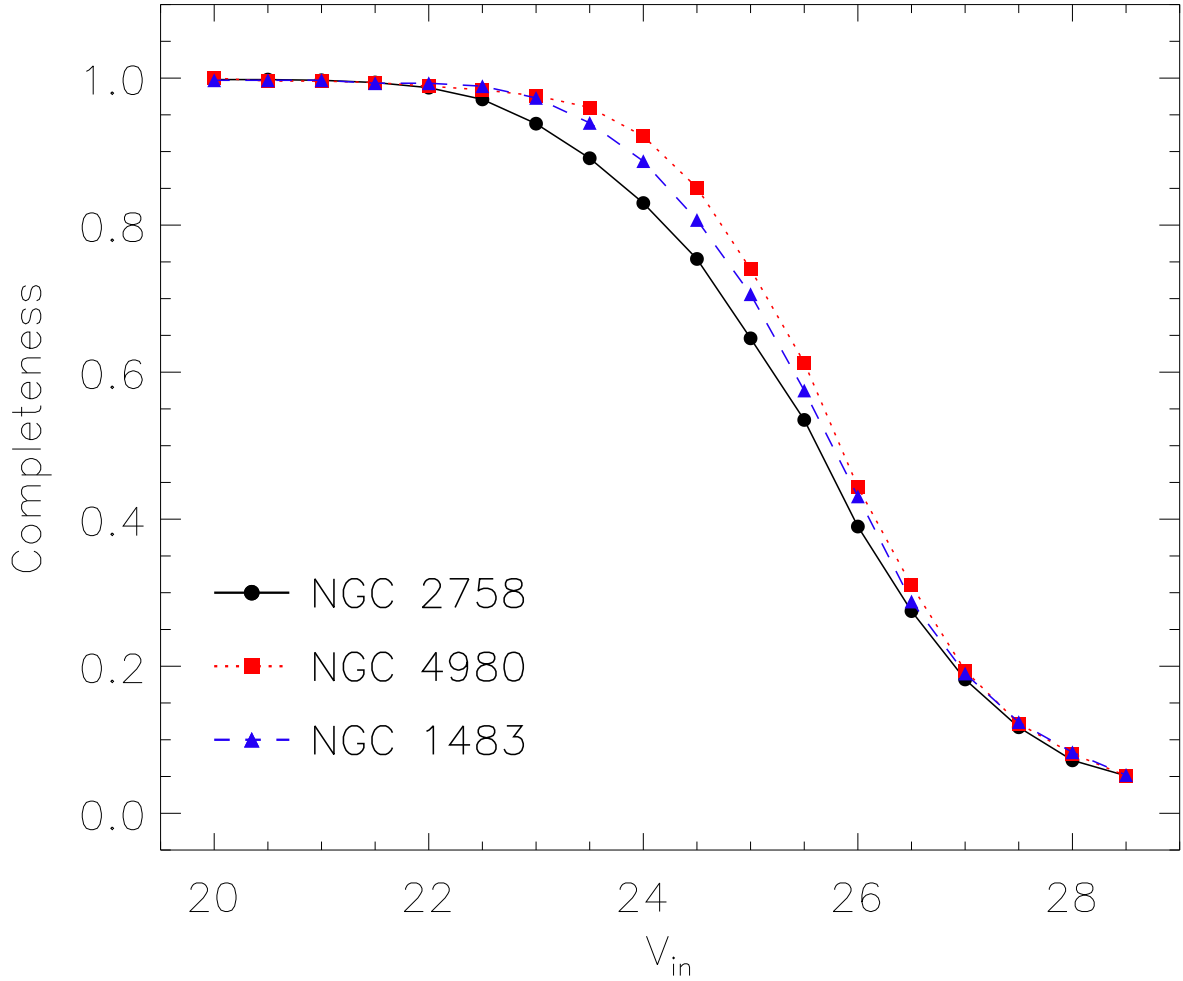} 
\caption{Completeness curves for our survey, determined via Monte
  Carlo simulations by placing and recovering artificial star-cluster
  sources with a Plummer surface brightness profile and half-light
  radius of $0.075\arcsec$, which corresponds to the mean size of the
  stellar clusters in our sample.
\label{fig:completeness}}
\end{figure}

Since we found that the completeness does not significantly depend on
the radius of the objects, we show in Fig.~\ref{fig:completeness} only
the completeness curves derived for the mean size of the stellar
clusters ($0.075\arcsec$) for three galaxies representative of our
sample and we note that in all cases the incompleteness becomes severe
only at $V > 25.5$.

\section{Identification of star clusters}\label{sec:DA}

To classify star clusters, we compare the colors of the sources in the
final catalog with those expected from synthetic stellar populations.
For this we use the spectral energy distribution of \citet{BC03}
models based on the Padova (1994) tracks, a \citet{Salpeter} initial
mass function (IMF) with masses between 0.1 and 100
$\mathrm{M_{\sun}}$, a range of metallicities from 0.02 to 1
$\mathrm{Z_{\sun}}$ and ages from 1 Myr to 15 Gyr. In addition we have
also included self-consistently Hydrogen and Helium recombination
lines as well as metal lines, as described in \citet{Oesch07}. The
integrated model colors for the HST filters of the observations have
been obtained by processing the synthetic spectra through the IRAF
task CALCBAND within the SYNPHOT package. With the CALCBAND task we
also include optional dust extinction ($E(B-V)\leqslant 1$ mag). For
comparison we also created \citet{BC03} models using a
\citet{chabrier} IMF with the same mass cutoffs. As expected, we
verified that the resulting optical colors are identical in both cases
(in fact the two IMFs are different only for stars below
$1~\mathrm{M_{\sun}}$). The mass-to-light ratio is a factor
$\approx 1.8$ smaller (with very low dependence on age and
metallicity) when using a Chabrier IMF instead of a Salpeter IMF, due
to the lower fraction of low-mass stars.

\begin{figure}
\resizebox{130pt}{!}{\includegraphics{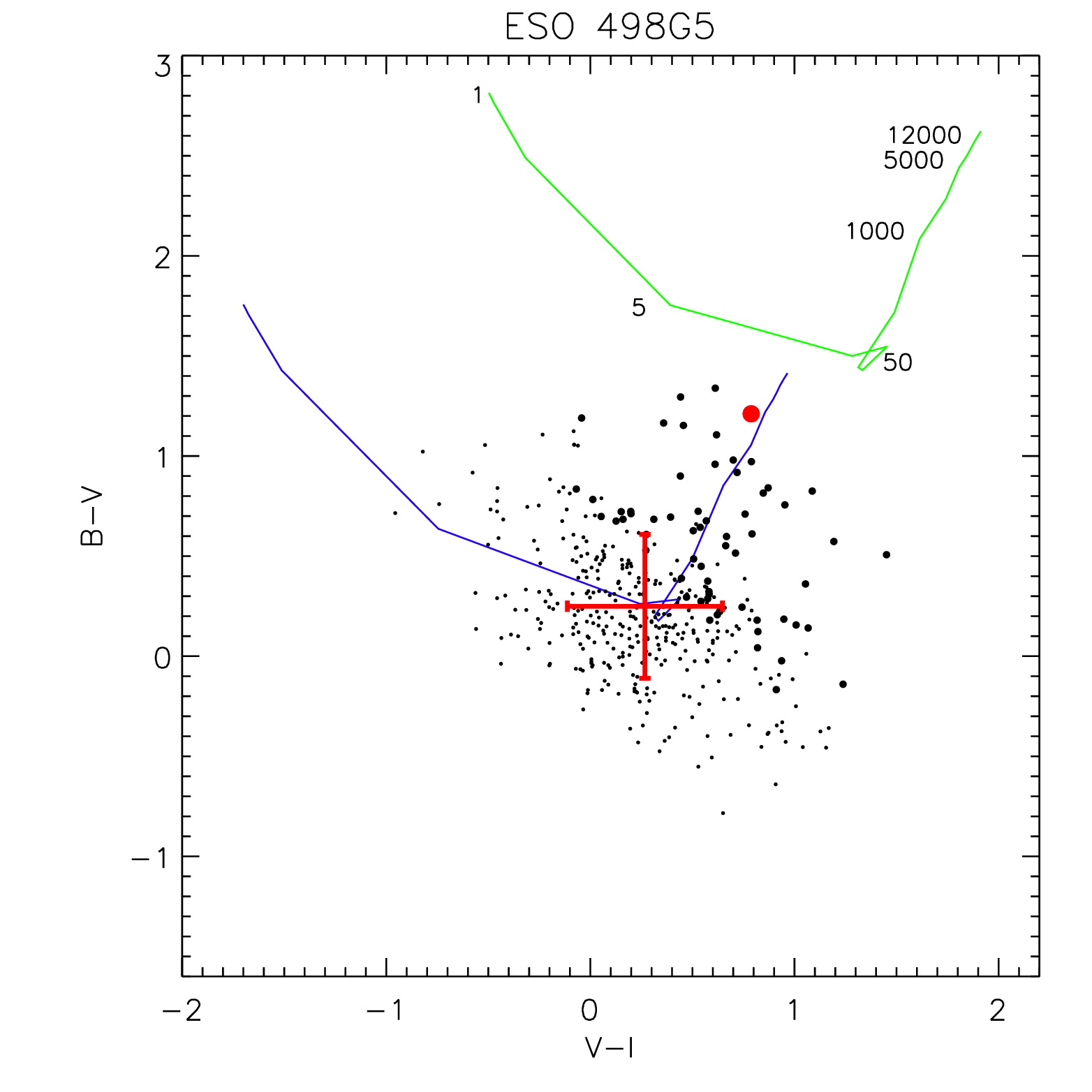}} 
\resizebox{130pt}{!}{\includegraphics{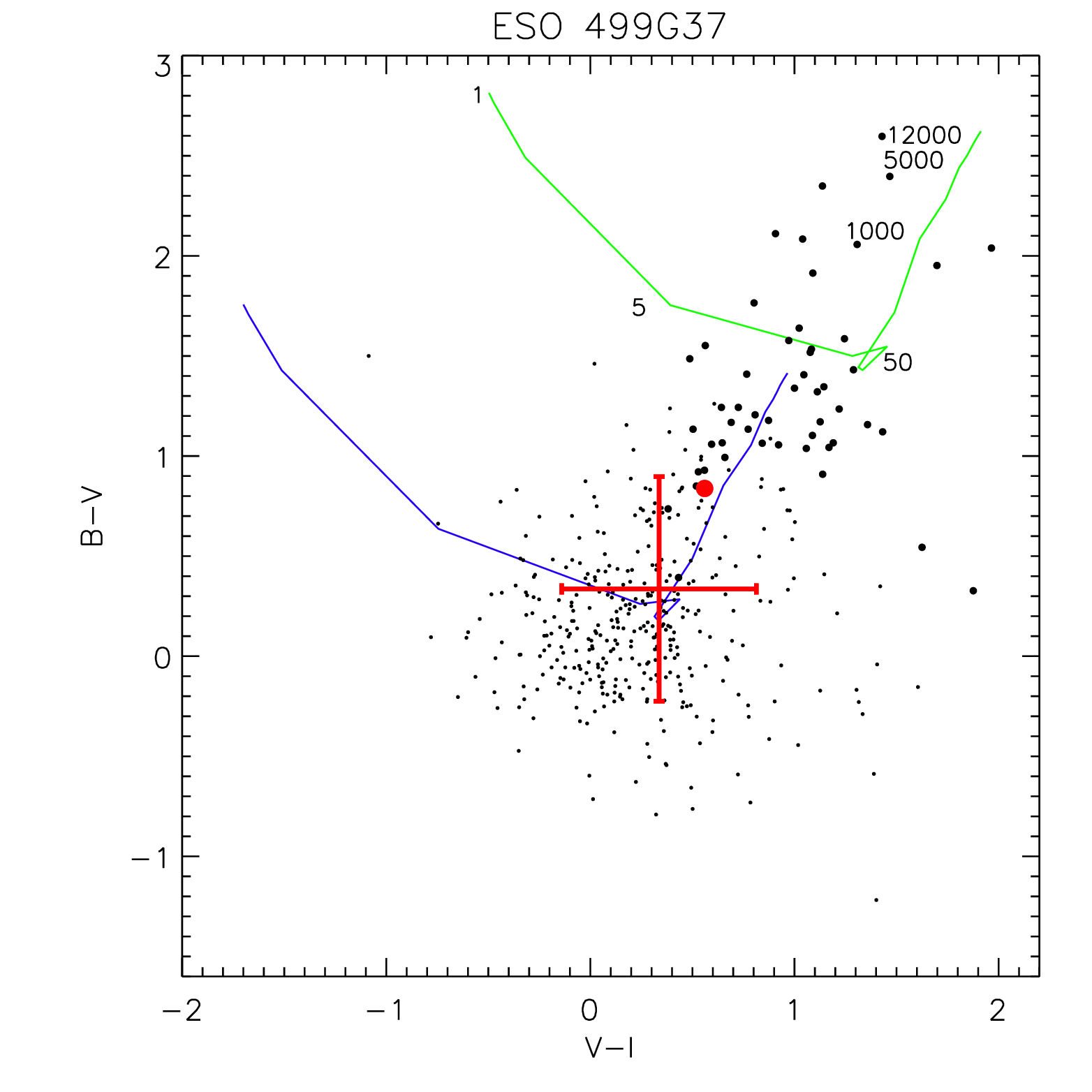}} 
\resizebox{130pt}{!}{\includegraphics{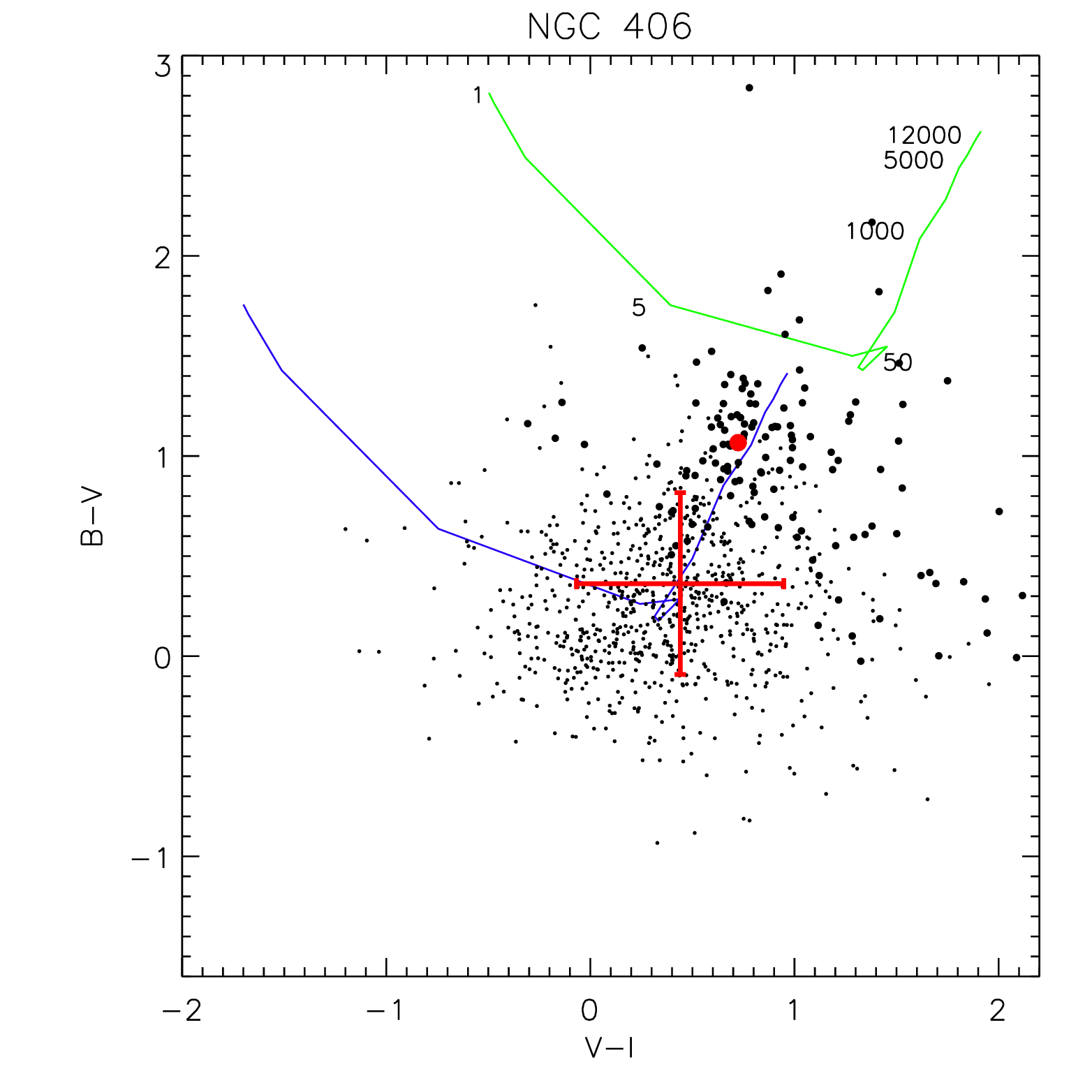}} 
\resizebox{130pt}{!}{\includegraphics{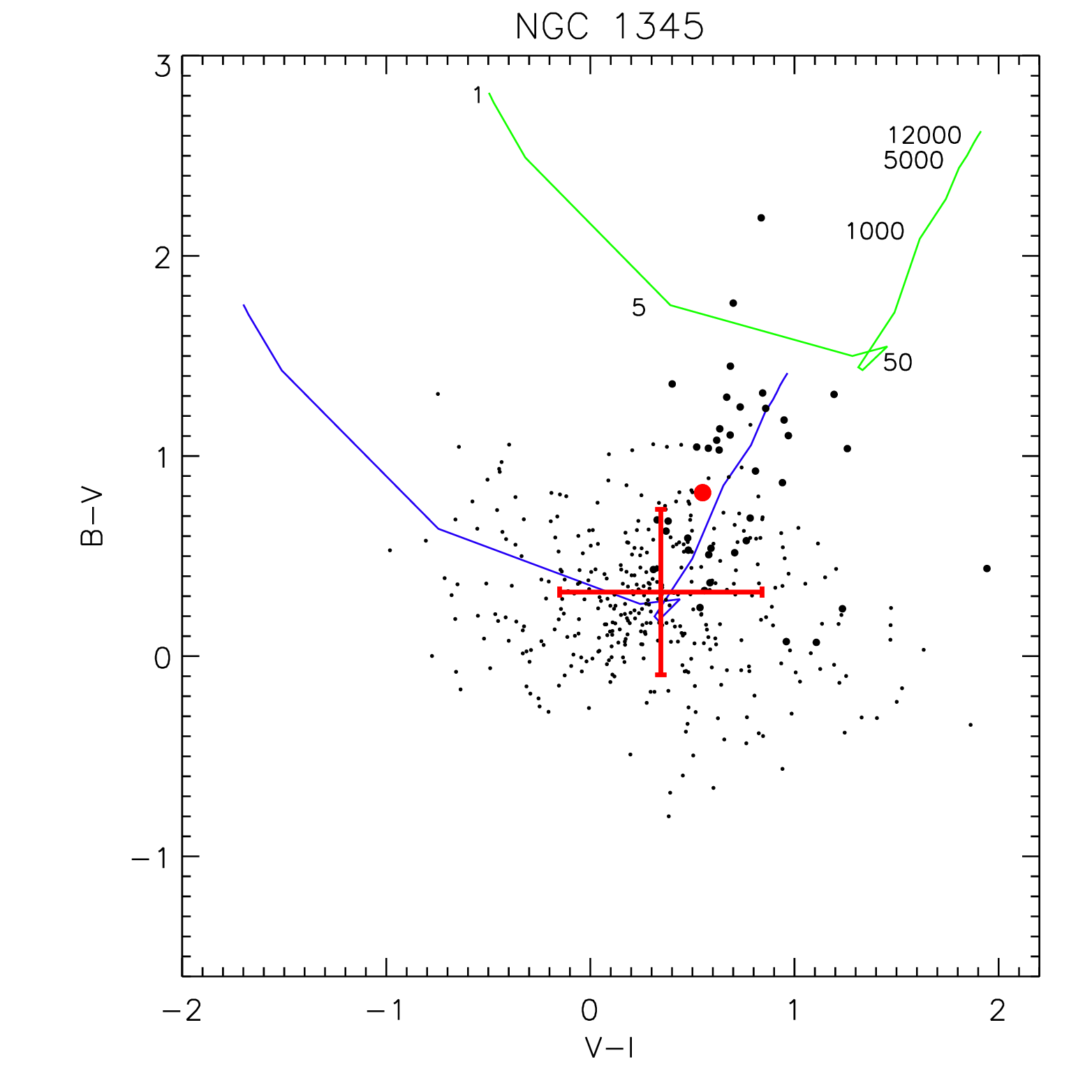}} 
\resizebox{130pt}{!}{\includegraphics{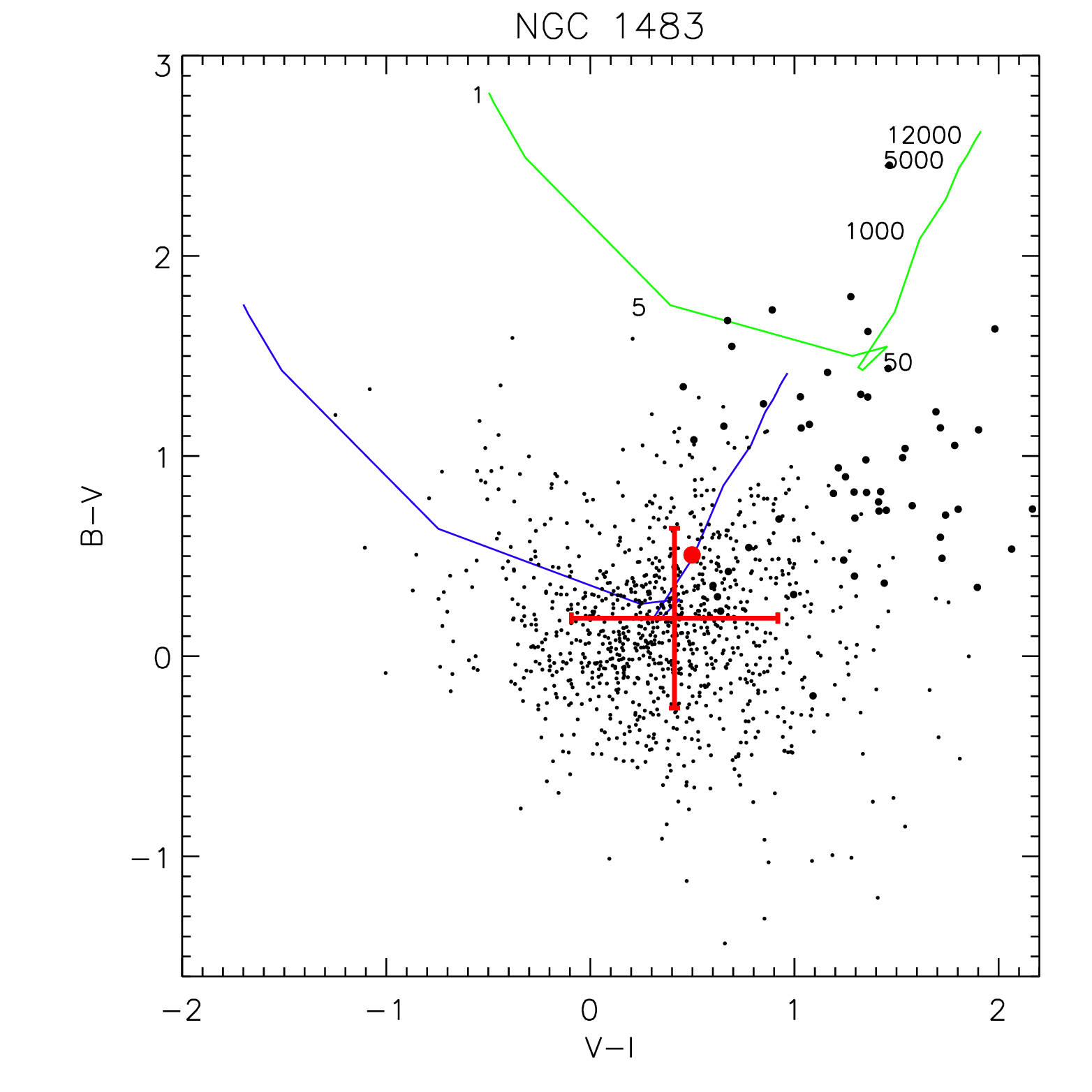}} 
\resizebox{130pt}{!}{\includegraphics{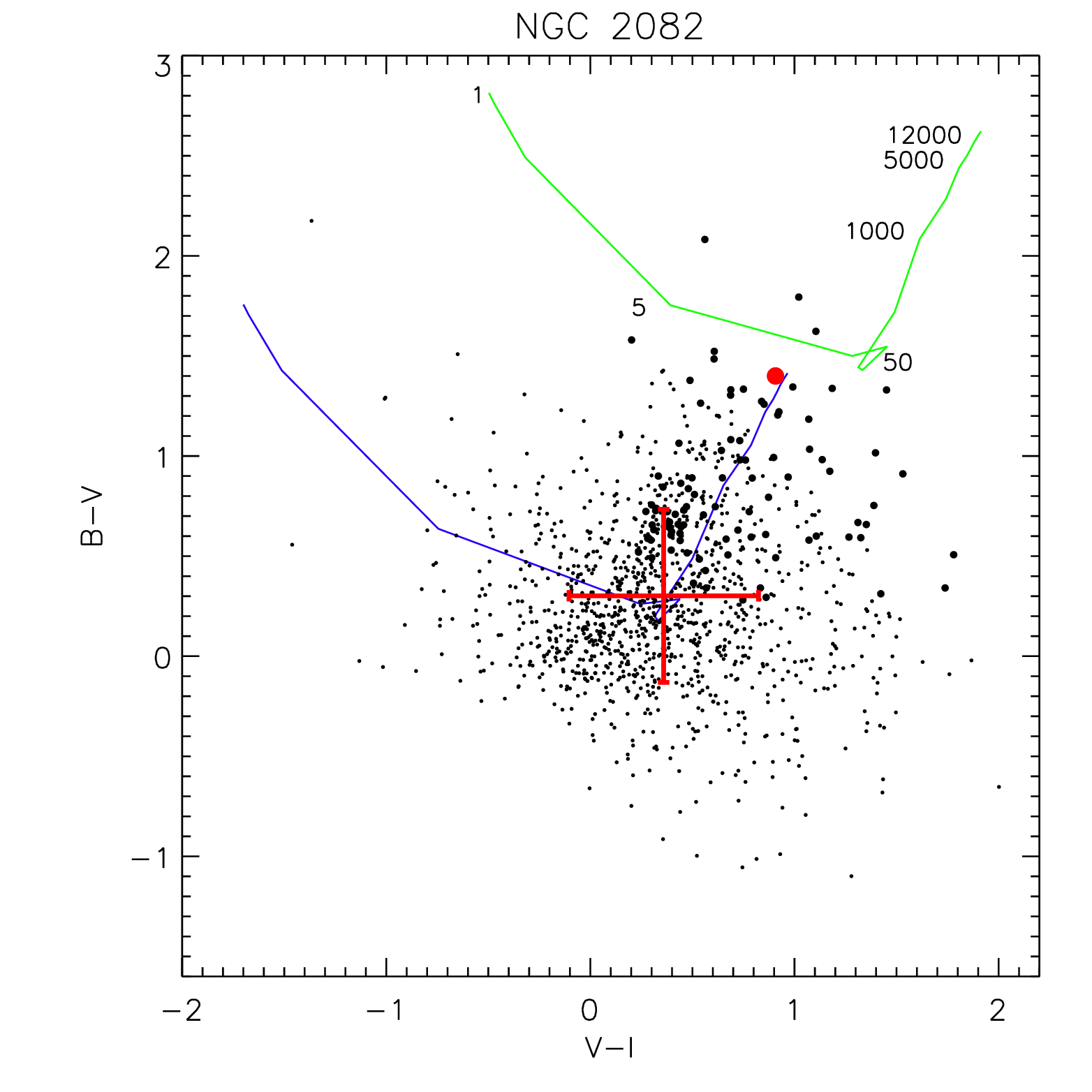}} 
\resizebox{130pt}{!}{\includegraphics{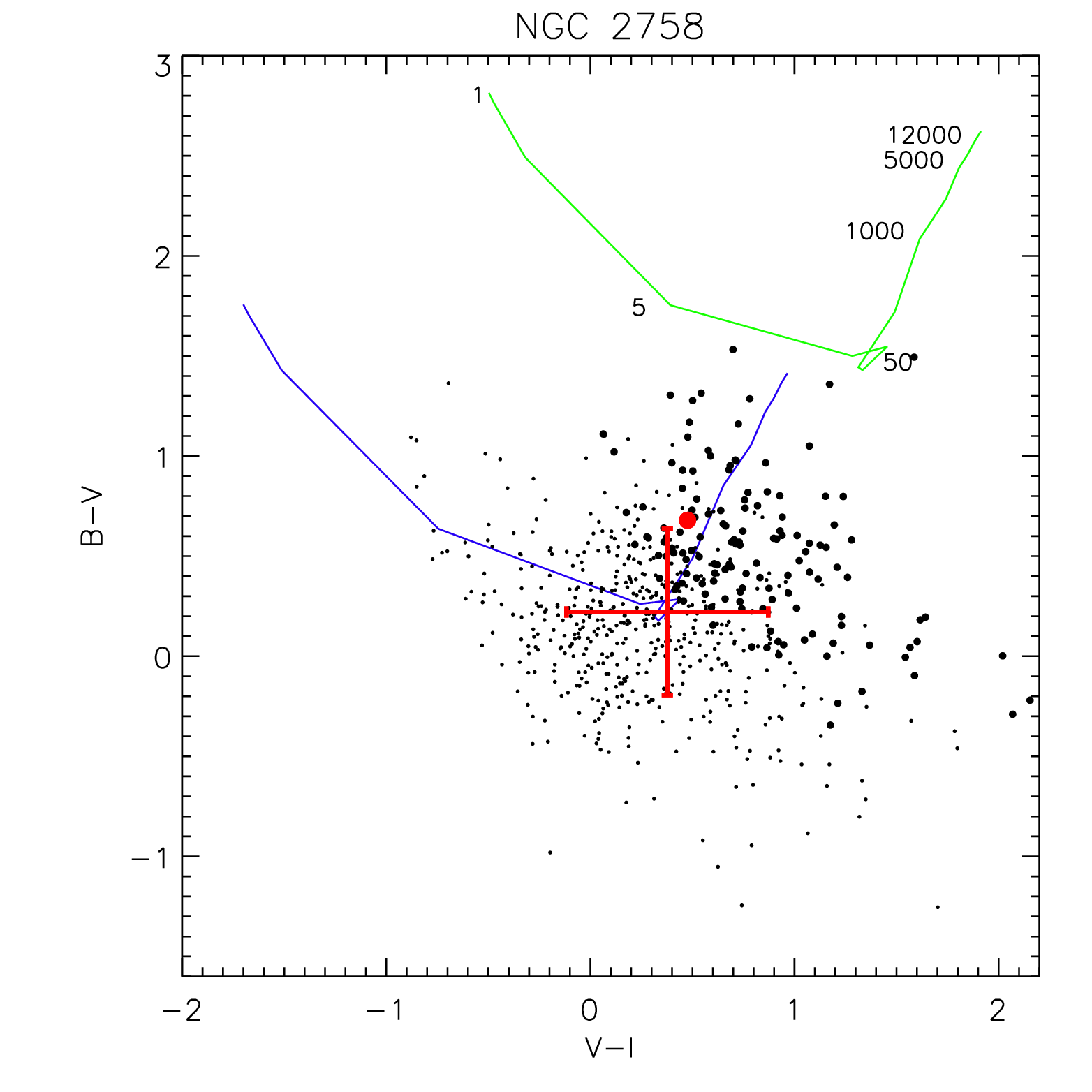}} 
\resizebox{130pt}{!}{\includegraphics{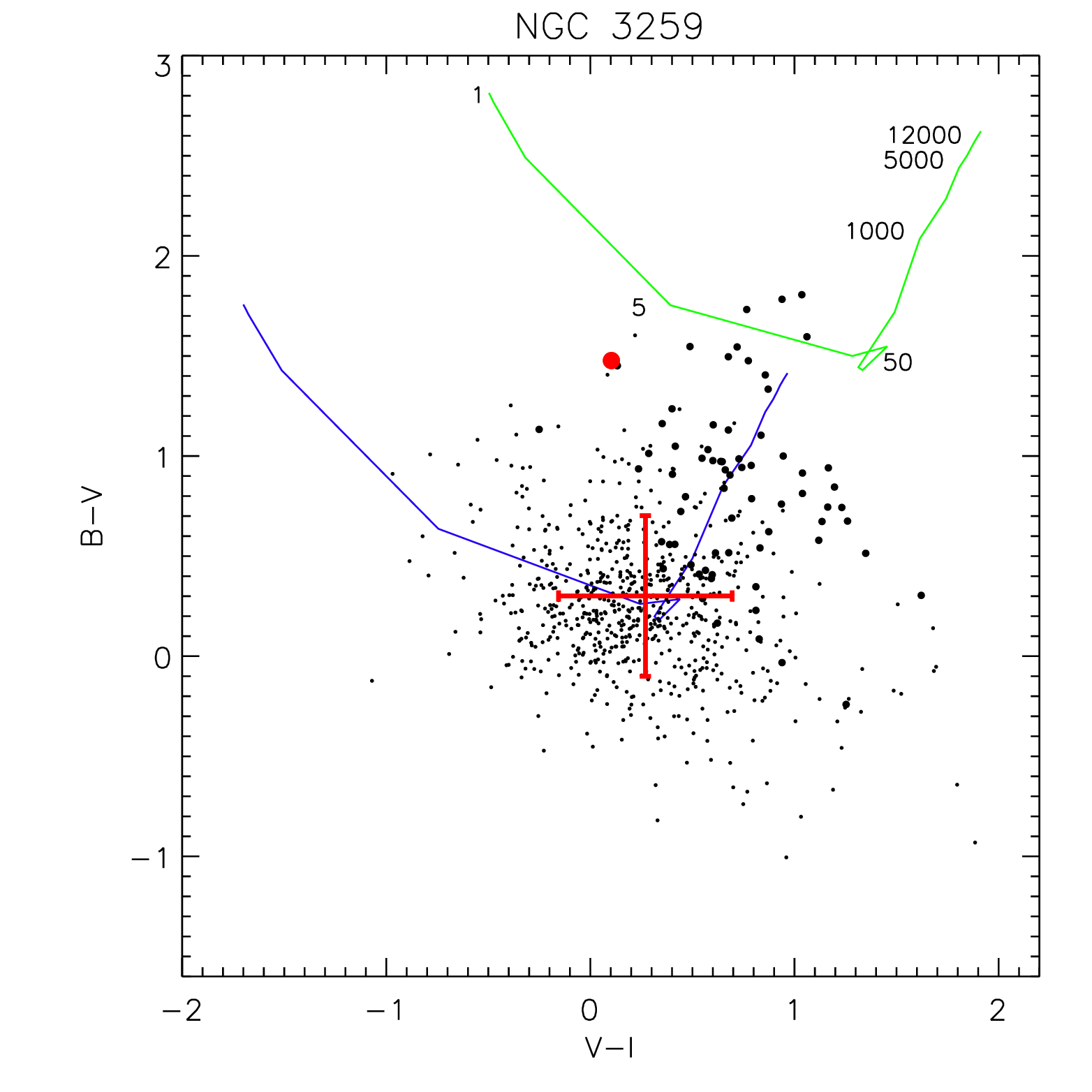}} 
\resizebox{130pt}{!}{\includegraphics{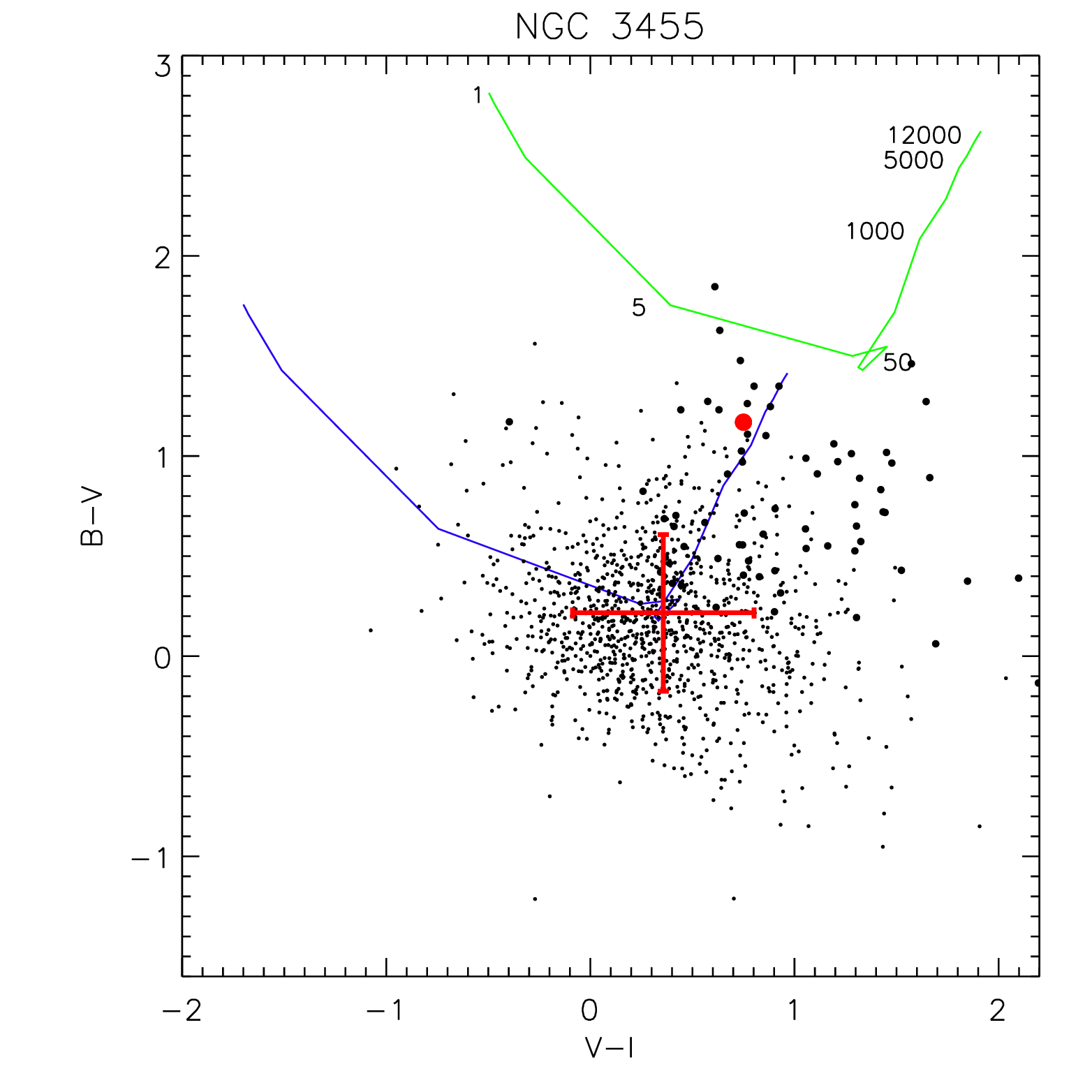}} 
\resizebox{130pt}{!}{\includegraphics{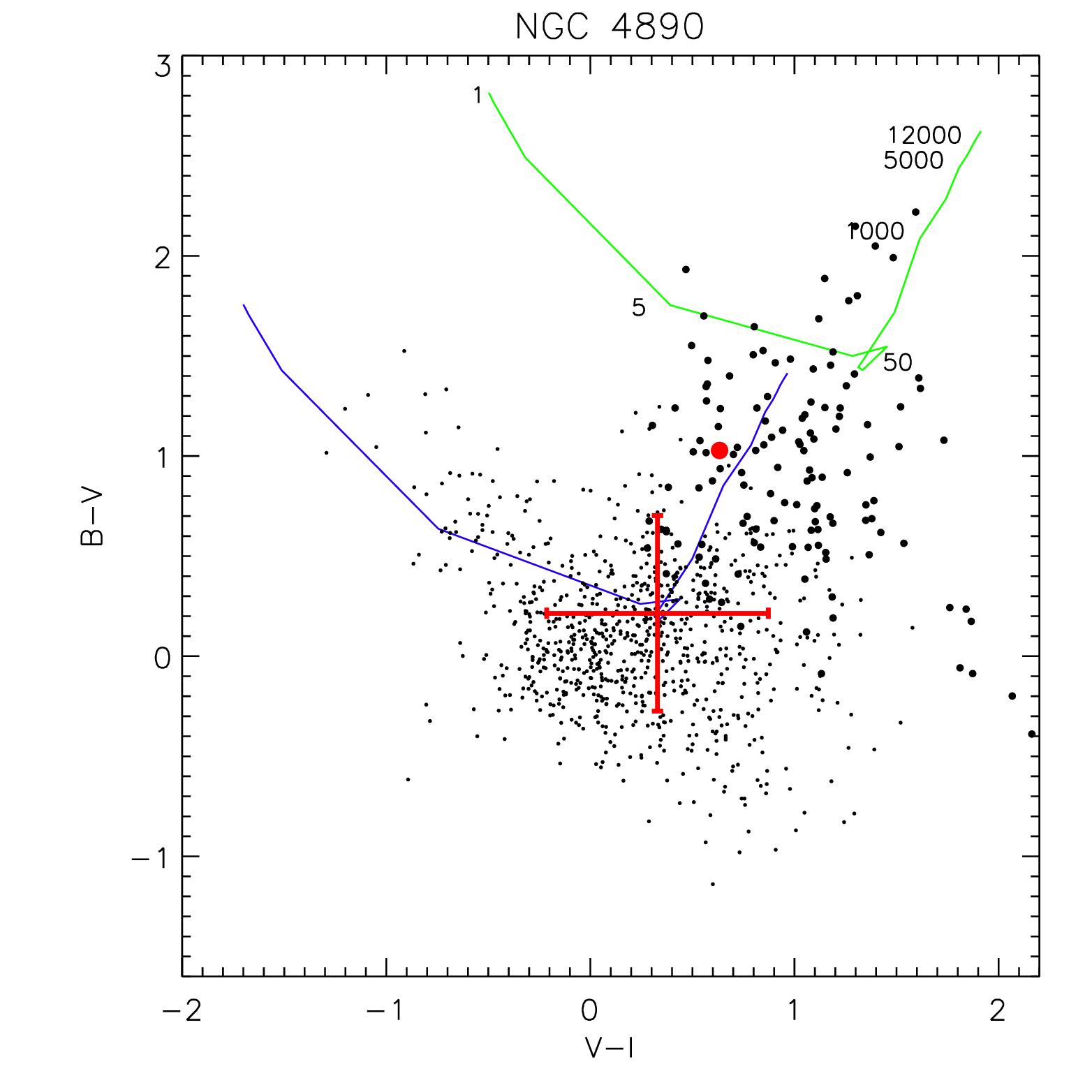}} 
\resizebox{130pt}{!}{\includegraphics{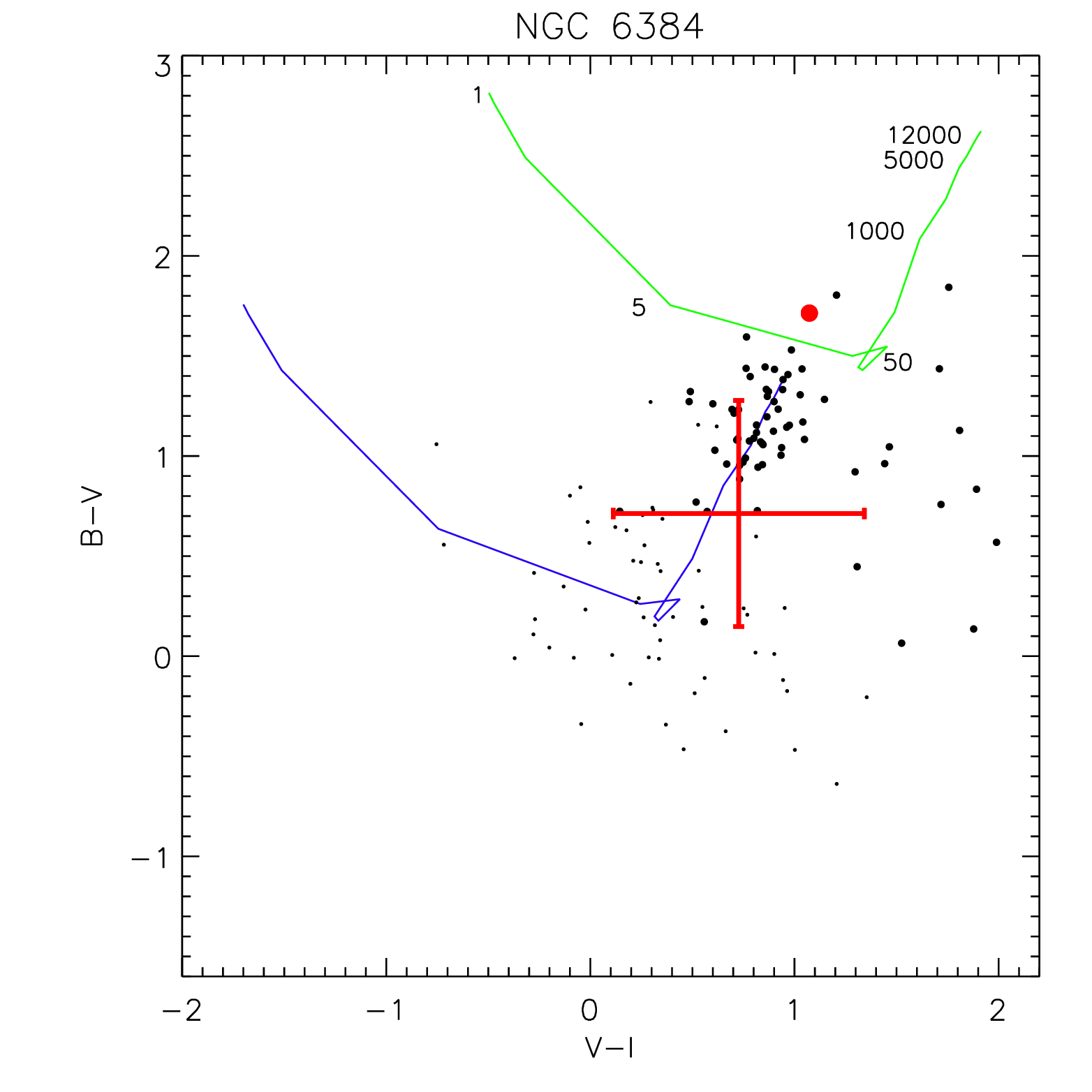}} 
\caption{{\small{Color-color diagrams ($B-V$ vs. $V-I$) for star
      clusters in each galaxy, showing a comparison of our data to
      \citet{BC03} models. Black dots are the data with the ``old''
      (age $\geqslant 250$ Myr) and massive (mass $\geqslant 10^5$
      $\mathrm{M_{\sun}}$) clusters highlighted in bold. The large red
      dot corresponds to the nuclear star cluster of each galaxy. The
      models include emission lines and have solar metallicity, single
      stellar population, a \citet{Salpeter} IMF and no reddening
      [$E(B-V)=0$, blue line] or a 1 mag of reddening [$E(B-V)=1$,
      green line]. Along the theoretical tracks we mark the age of the
      synthetic population in Myr. The star cluster average colors
      with their $1 \sigma$ dispersions are shown in red.}}
  \label{fig:bc_vs_ALLdata}}
\end{figure}

The Salpeter-IMF tracks in the $V-I$ vs. $B-V$ color-color plane are
shown in Fig.~\ref{fig:bc_vs_ALLdata} and compared with the observed
sources. Based on these colors, most of the detected sources are
consistent with being relatively young star clusters (age $\approx$ 50
Myr). The data-model comparison for all galaxies, except NGC 6384,
suggests that only a modest extinction is present (but see
section~\ref{sec:dust} for a more detailed discussion). The spread of
the observed sources in the color-color plane (standard deviation and
central point of the distribution reported in
Fig.~\ref{fig:bc_vs_ALLdata}) is comparable with the photometric
errors on the colors (see Fig.~\ref{fig:color_error_trend}).  Hence it
is not possible to infer the star-cluster formation history
from this plot.

\begin{figure}
\resizebox{130pt}{!}{\includegraphics{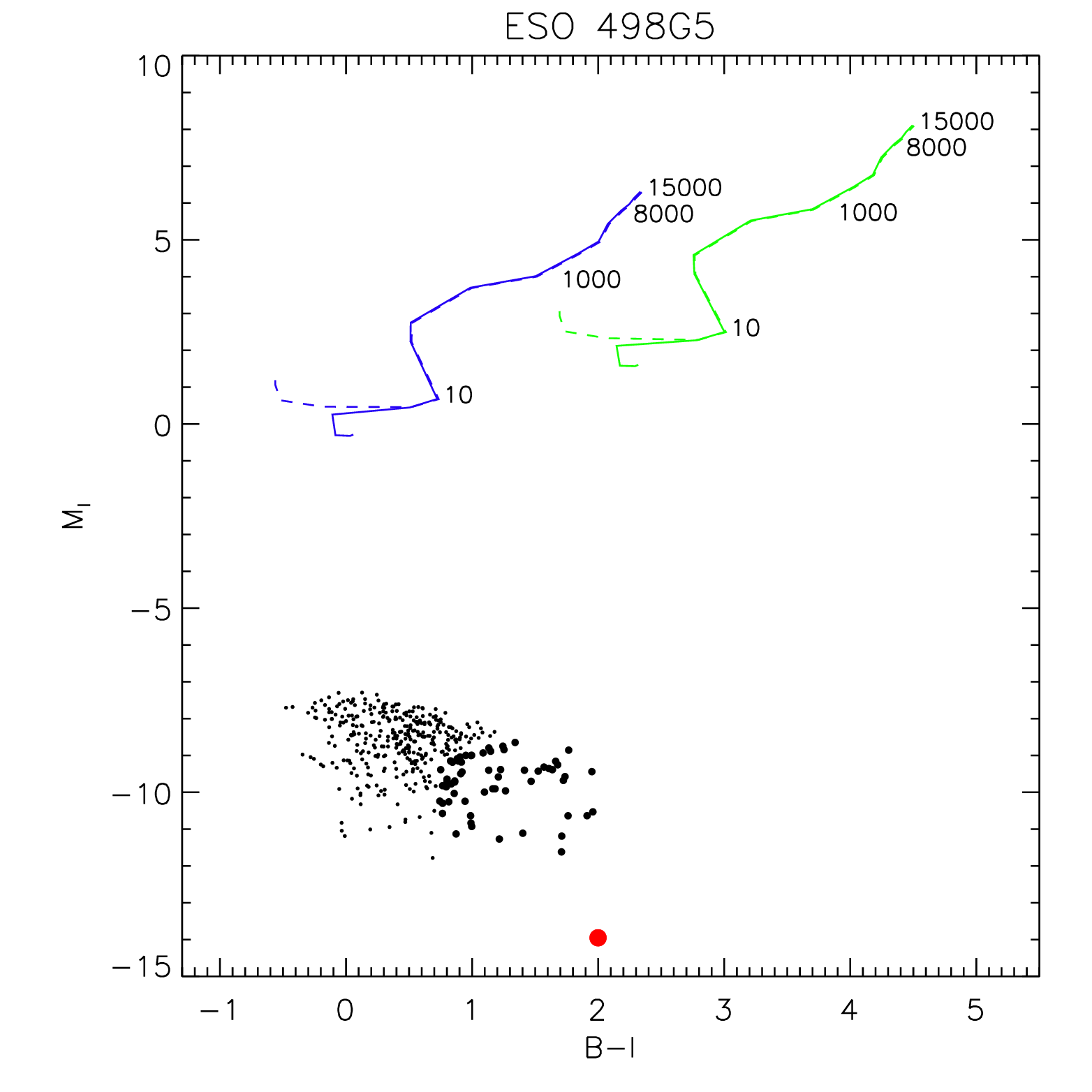}} 
\resizebox{130pt}{!}{\includegraphics{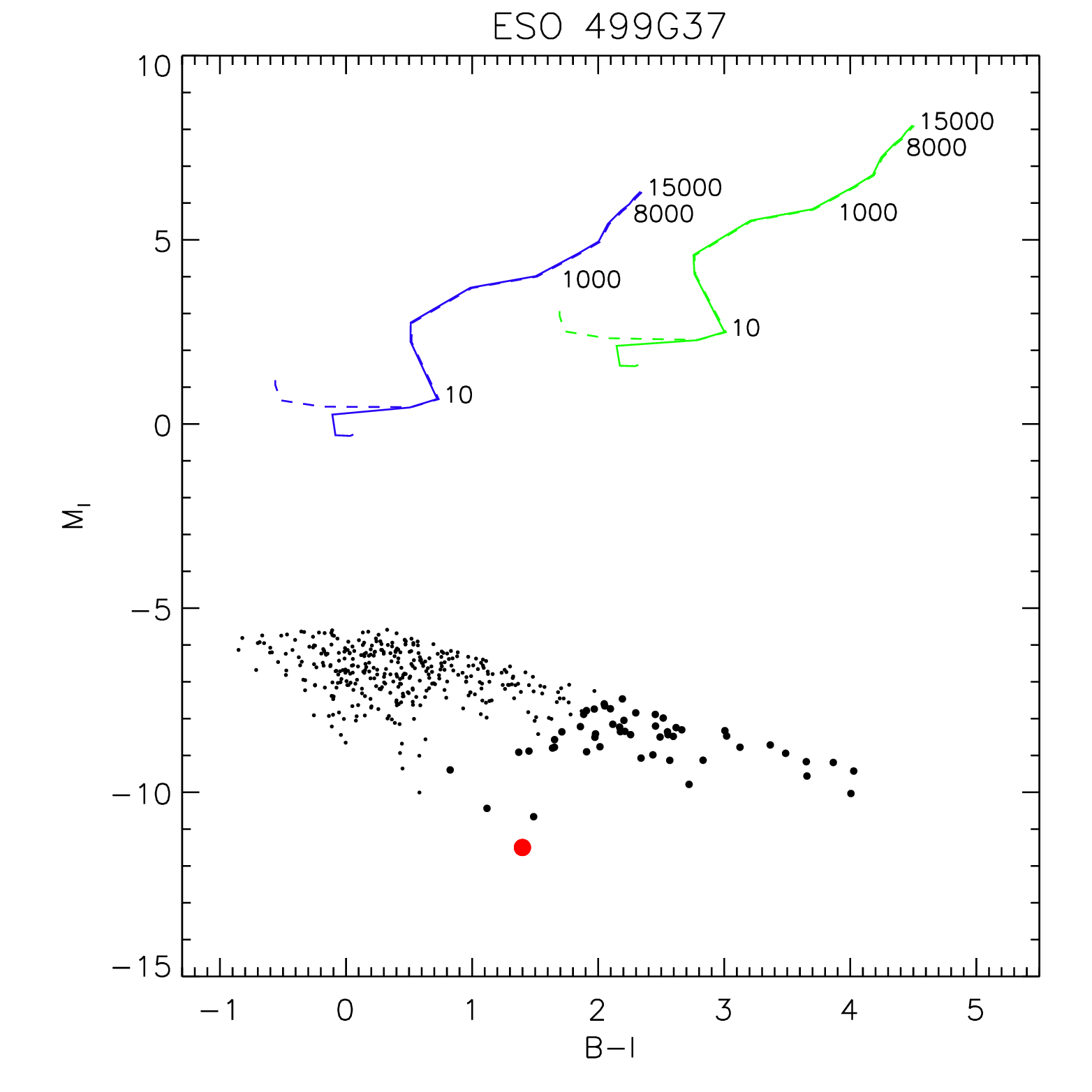}} 
\resizebox{130pt}{!}{\includegraphics{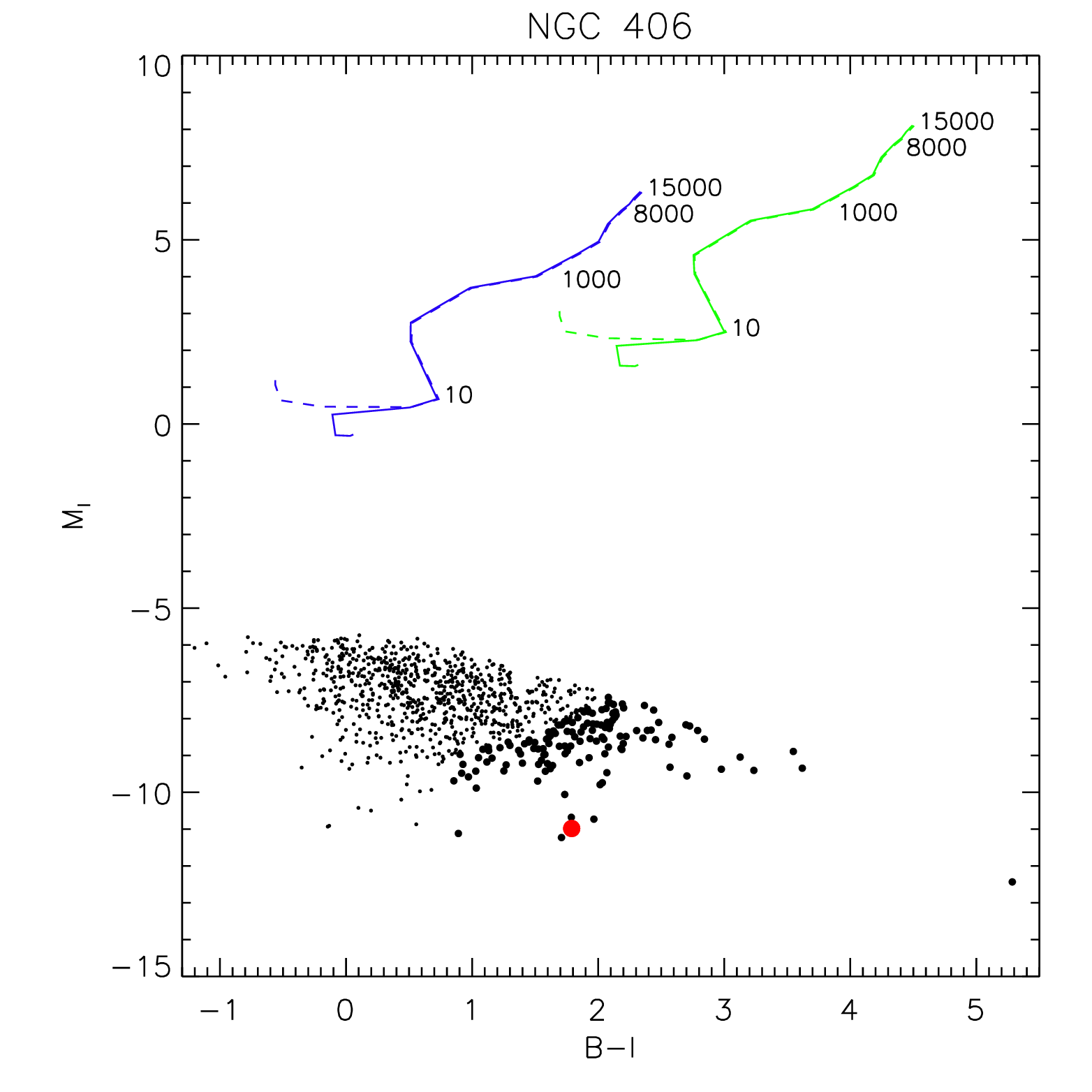}} 
\resizebox{130pt}{!}{\includegraphics{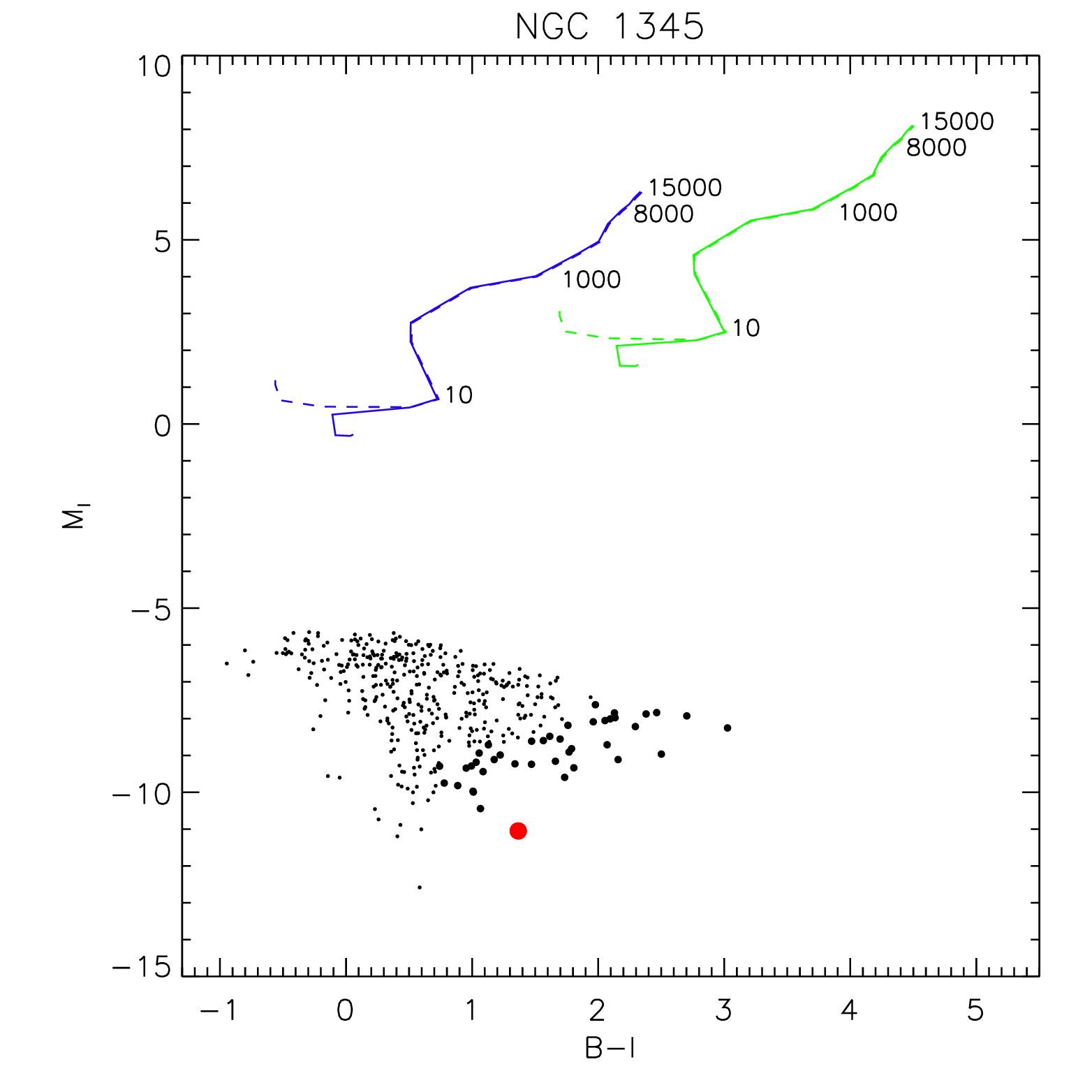}} 
\resizebox{130pt}{!}{\includegraphics{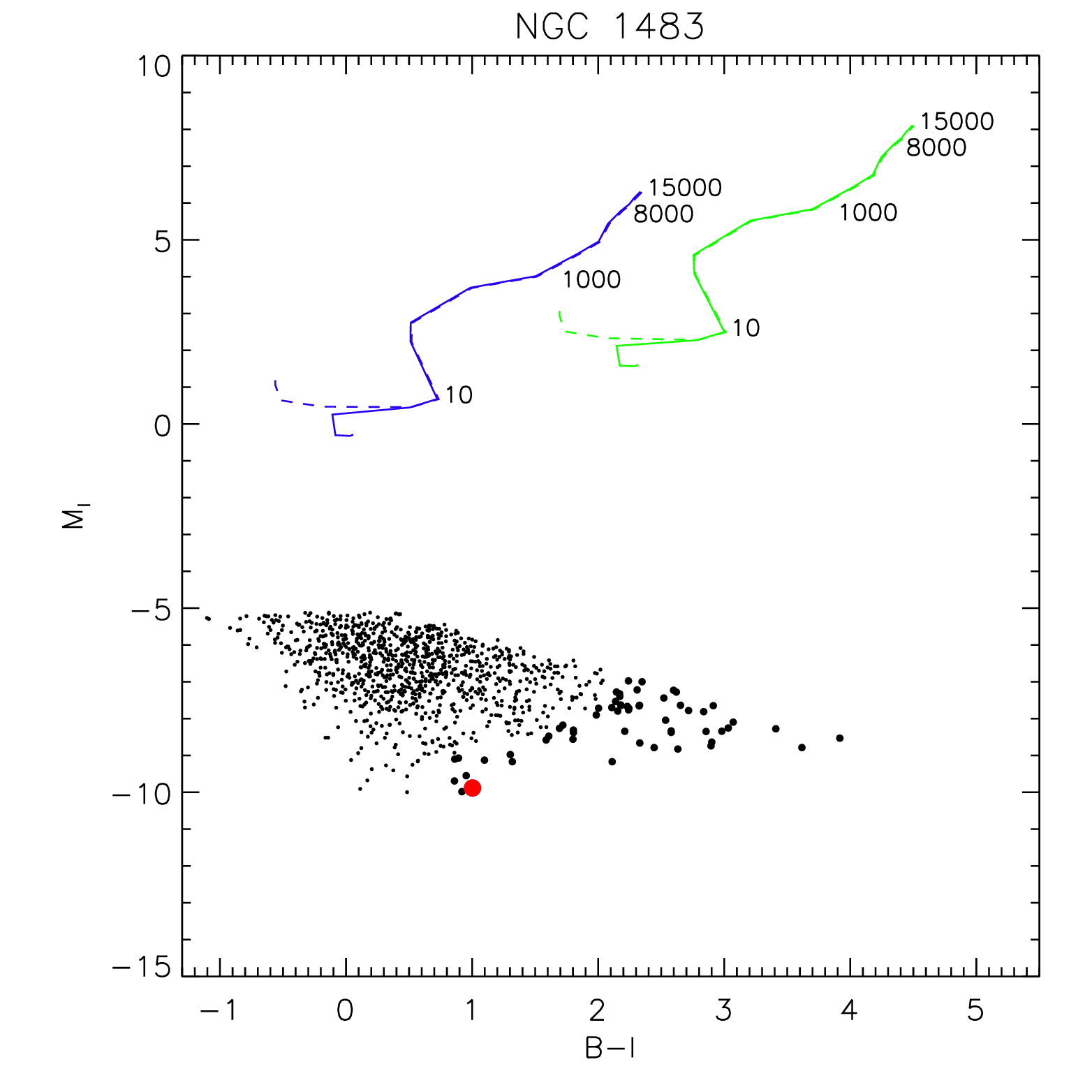}} 
\resizebox{130pt}{!}{\includegraphics{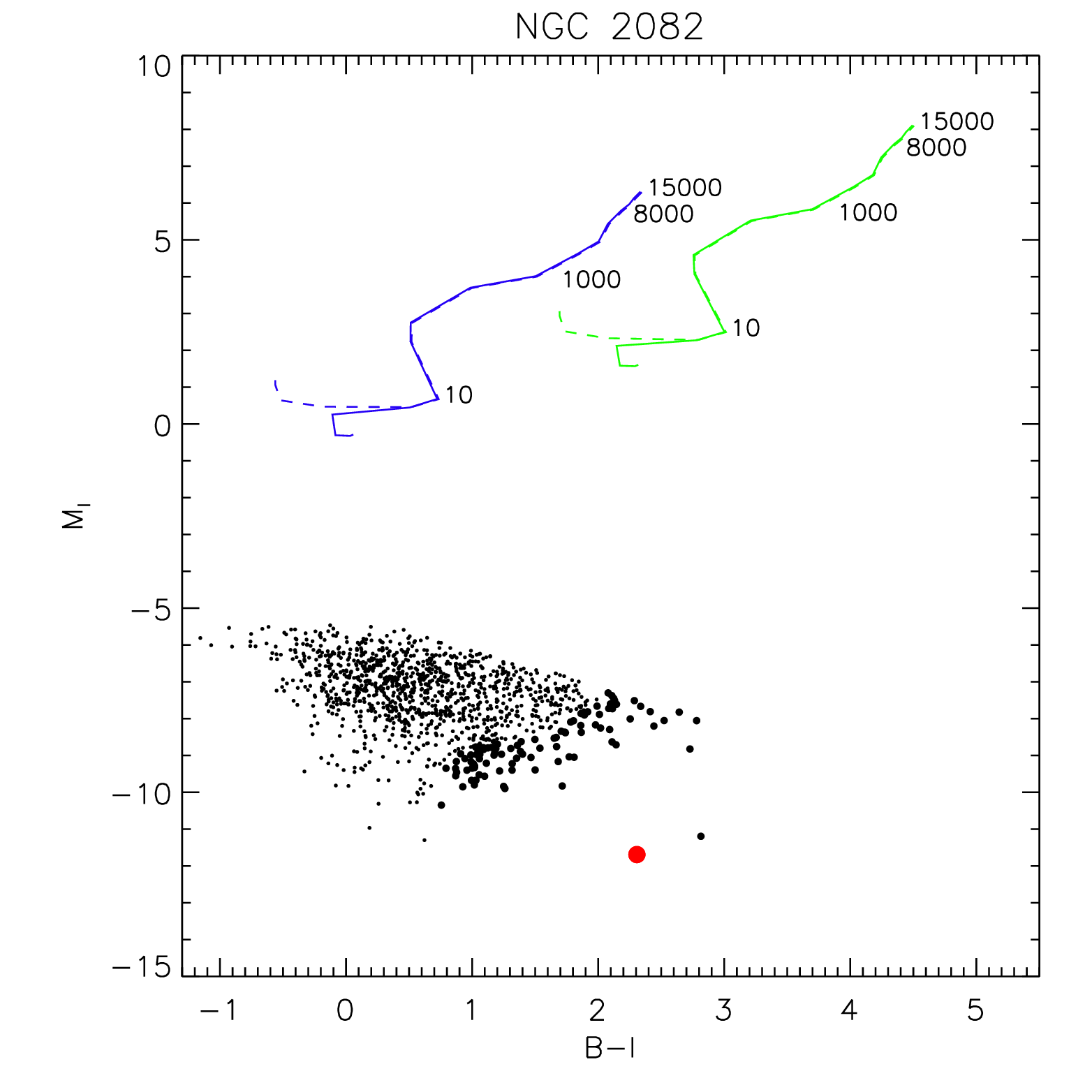}} 
\resizebox{130pt}{!}{\includegraphics{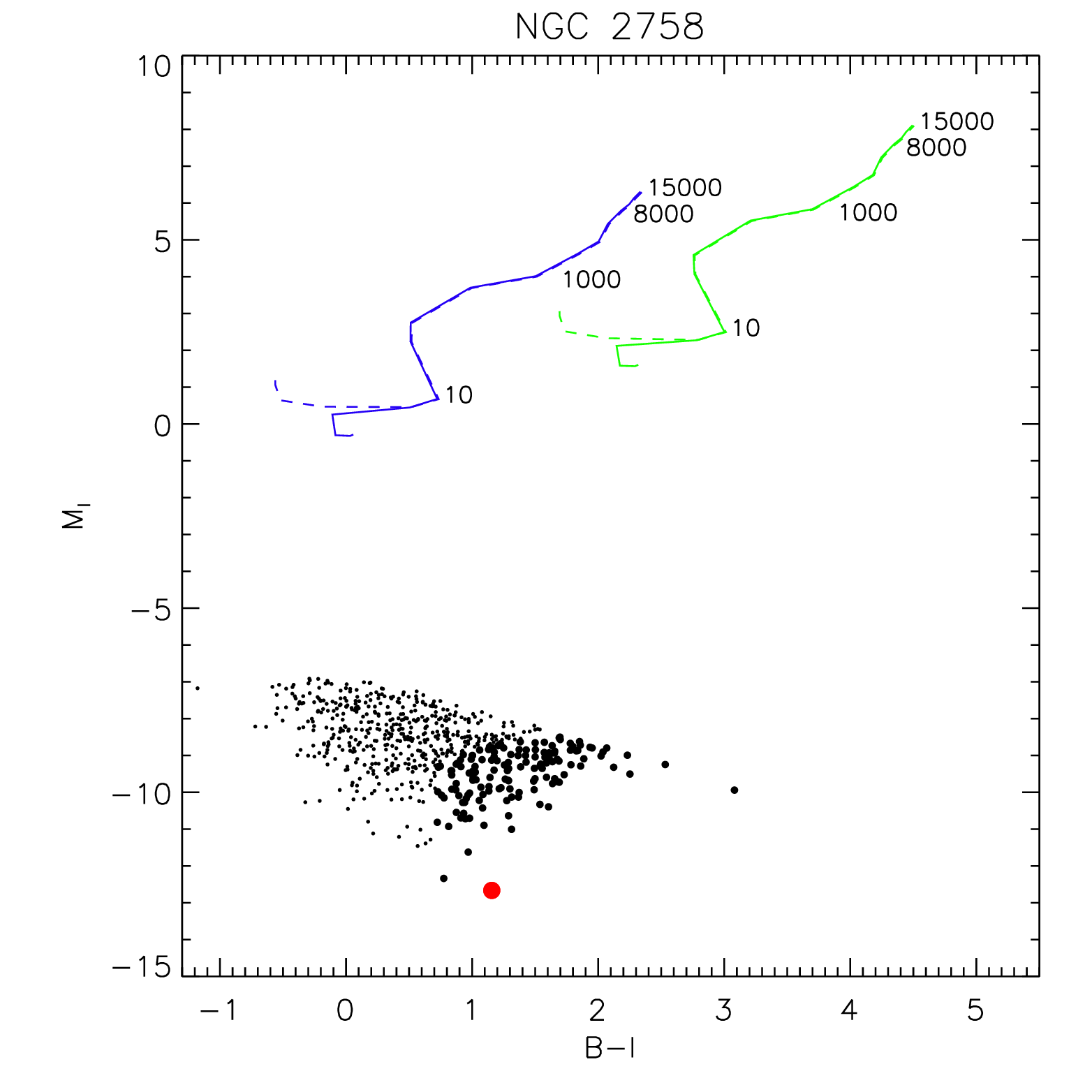}} 
\resizebox{130pt}{!}{\includegraphics{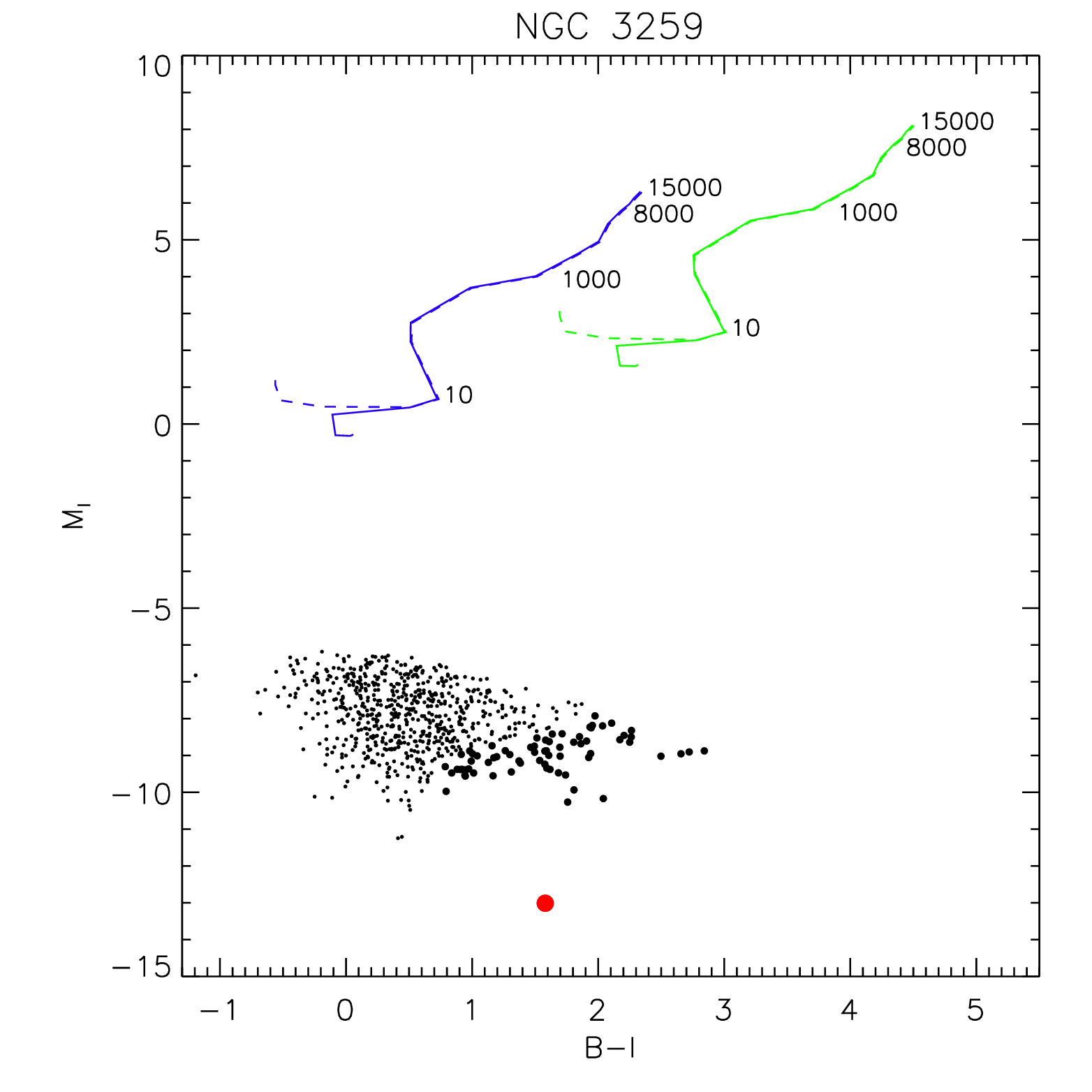}} 
\resizebox{130pt}{!}{\includegraphics{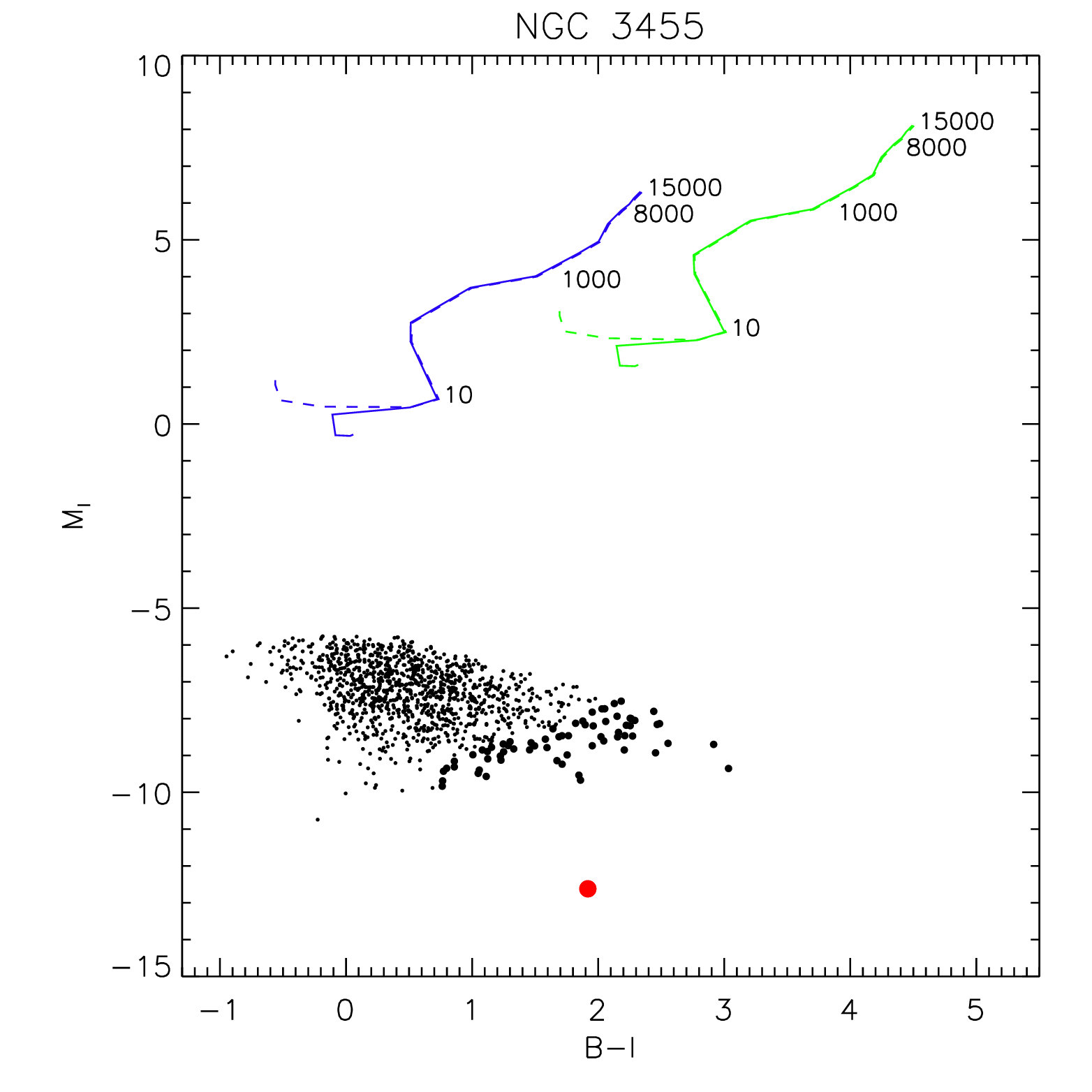}} 
\resizebox{130pt}{!}{\includegraphics{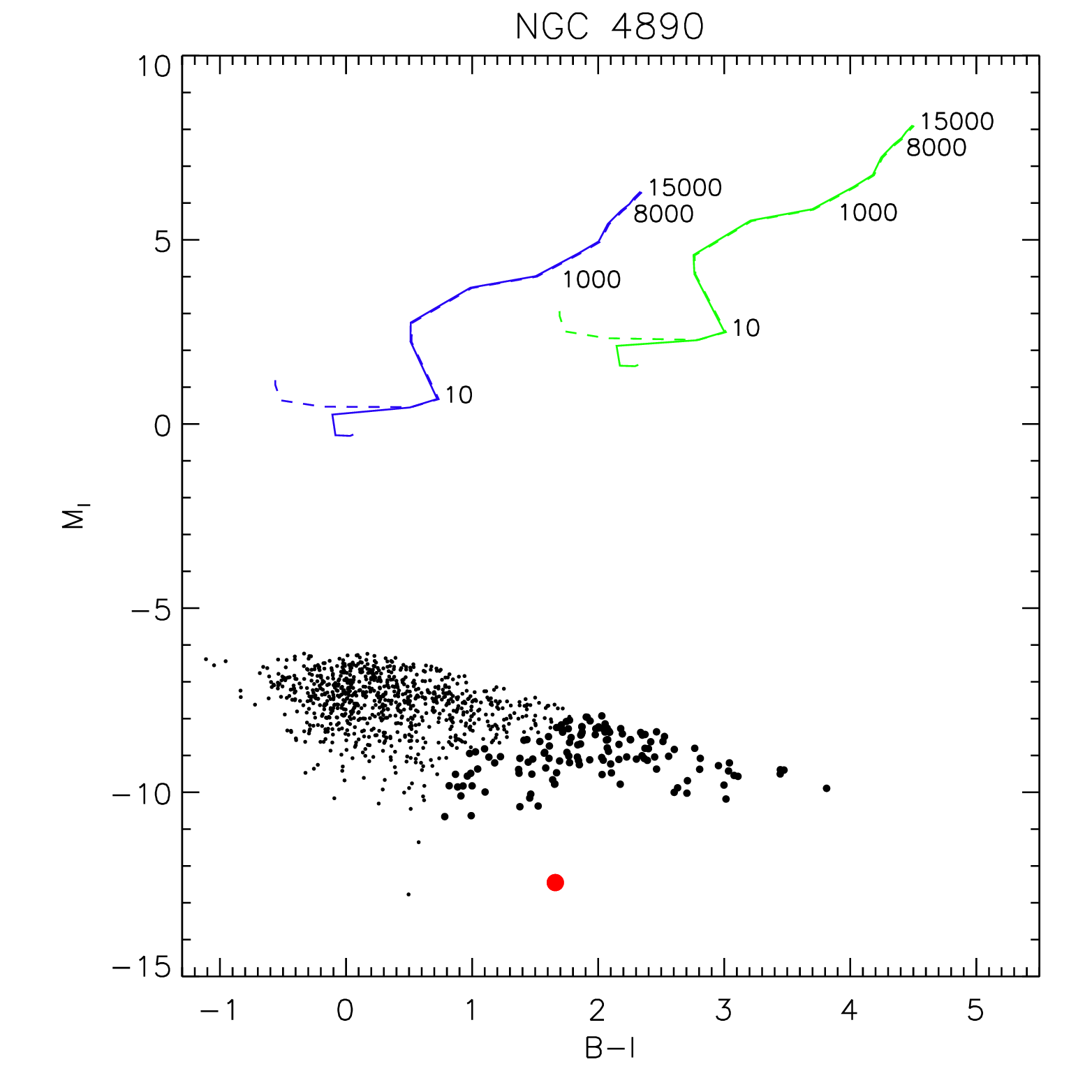}} 
\resizebox{130pt}{!}{\includegraphics{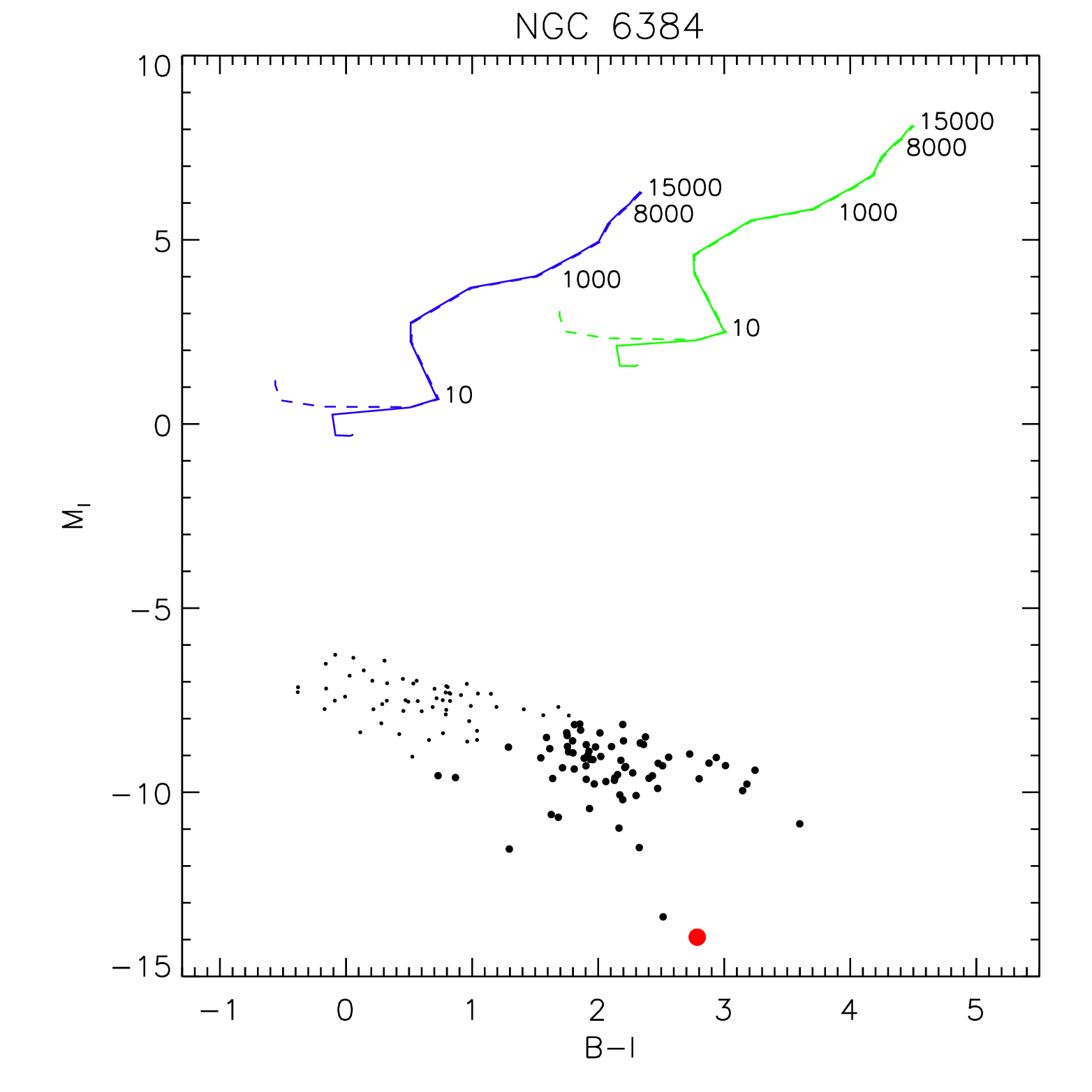}} 
\caption{As in Fig.~\ref{fig:bc_vs_ALLdata} but for the
  color-magnitude diagram ($B-I$ vs. $M_I$) for each galaxy compared to
  the \citet{BC03} models with a Salpeter IMF. Here the tracks for the
  models are shown for a total model mass of 1
  $\mathrm{M_{\sun}}$. Therefore the difference in $M_I$ magnitude
  between an observed source and the model with a corresponding $B-I$
  color gives the total mass of the observed star cluster. In
  addition, we show, as dashed lines, tracks without emission
  lines. The upper envelope of the data points is a selection effect
  given by the magnitude threshold of the survey.\label{fig:BI_vs_I}}
\end{figure}

A better diagnostic is provided by a color ($B-I$) vs. magnitude ($M_I$)
diagram (see fig.~\ref{fig:BI_vs_I} where the $M_I$ magnitude of the
model tracks is computed for a reference 1 $\mathrm{M_{\sun}}$
mass). We can estimate the total mass of each observed source based on
the difference in $M_I$ magnitude compared to the model with the same
$B-I$ color. The $B-I$ color also highlights that up to $50\%$ of the
total detected sources are younger than 8 Myr (see Table
\ref{tb:Ntot}) under the assumption of no dust (again, see
section~\ref{sec:dust} for a discussion of the possible impact of dust
extinction). In all galaxies, the observed sources are clearly not
distributed as an evolutionary sequence at constant mass. Instead the
older the stellar population, the greater is its luminosity (and hence
mass, because the mass-to-light ratio increases with age). This is
consistent with the infant-mortality scenario for star clusters
\citep{Fall05}, in which only a small fraction of star clusters
(usually the most massive) survives over time.

Every galaxy in our sample has a number of young star clusters, with
ages $\lesssim$ 8 ${\rm Myr}$ and masses of the order of $10^3$
${\mathrm M_{\sun}}$. These can be used to reconstruct the recent star
formation history of the parent galaxies.  Taken at face value, our
results indicate an average recent star formation rate, in such
clusters, of $0.1-0.01$ $\mathrm{M_{\sun}/yr}$. Linking this SFR to the
actual SFR is however challenging, both because we need to extrapolate
our results to young clusters below the detection threshold and
because the photometric uncertainties might introduce a Malquist bias
difficult to quantify. Conservatively we can consider our results a
lower limit to the overall galaxy SFR.

\begin{figure}
\epsscale{1.0}
\plotone{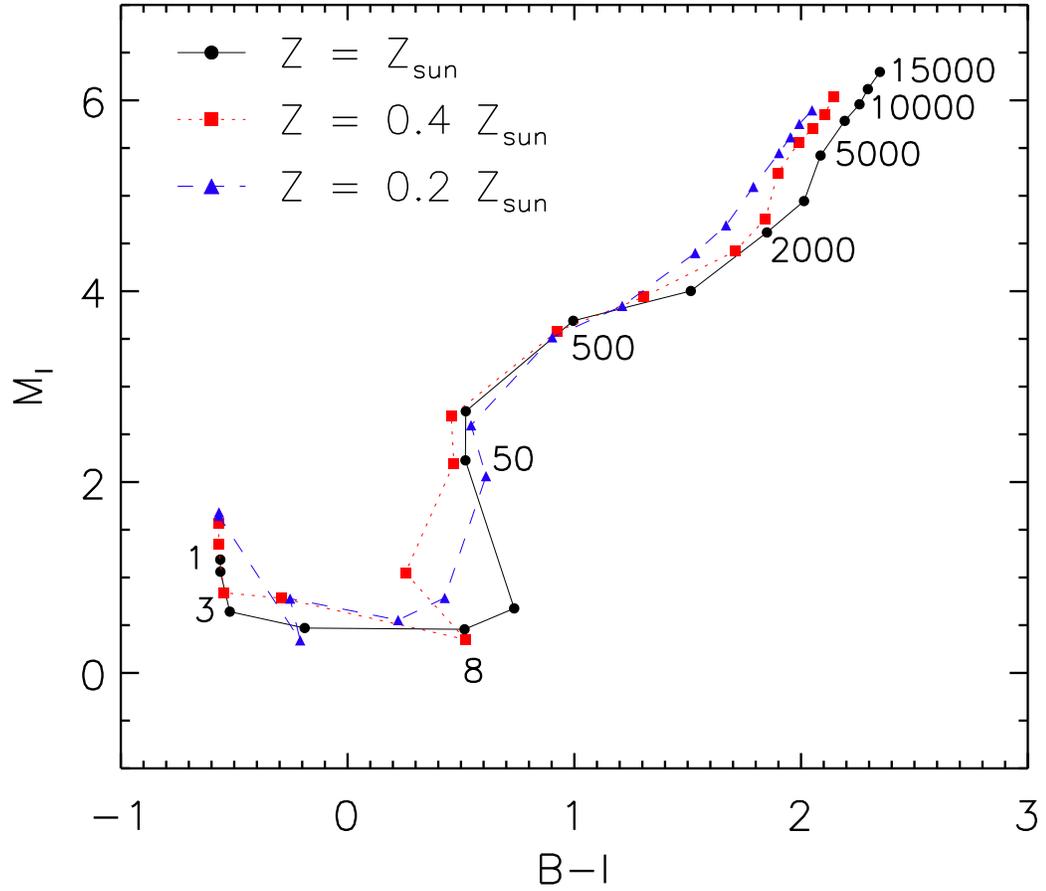} 
\caption{Color-magnitude diagram ($B-I$ vs. $M_I$) tracks for
  \citet{BC03} models for different metallicities (solid black line
  and dots: ${\mathrm Z_{\sun}}$; red dotted line and squares: 0.4
  ${\mathrm Z_{\sun}}$; blue dashed line and triangles: 0.2 ${\mathrm
    Z_{\sun}}$). The plot highlights that there is relatively little
  metallicity dependence. Ages in Myr are superimposed on the
  tracks.
\label{fig:tracks}}
\end{figure}

In this paper we focus on the properties of the oldest (age
$\geqslant$ 250 Myr) and most massive ($M \geqslant 10^5$ ${\mathrm
M_{\sun}}$) stellar clusters; for reasons discussed further below, we
identify these as young candidate globular clusters. Our choice in age
is an operational choice to avoid (i) the youngest systems, where
stellar evolution of the most massive stars can influence the
dynamical evolution of the system by their high mass-loss rates and
(ii) a knee in the luminosity color relation (see
Fig.~\ref{fig:BI_vs_I}) that would lead to a degeneracy in the mass
estimate for the sources. Once a metallicity for the stellar
population is assumed, the age and mass selection translates into a
single color and luminosity selection respectively. If we assume a low
metallicity ($Z=0.02$ ${\mathrm Z_{\sun}}$), the nuclear star cluster
age turns out to be longer than the Hubble time, suggesting that these
galaxies are more metal enriched (at least in their central regions
studied here). For a metallicity range from 0.2 to 1 ${\mathrm
Z_{\sun}}$ the total number of sources selected as globular cluster
candidates varies at the 15\% level at most. In fact, the tracks at
different metallicities are close to each other for objects older than
250 Myr (see Fig.~\ref{fig:tracks}).

Therefore, in the following we adopt solar metallicity for the
clusters. The age selection $\geqslant 250$ Myr then translates into a
color selection $(B-I) \geqslant 0.73$ mag, assuming no dust. The same
color cut corresponds instead to a younger age selection (age $\gtrsim
100$ Myr) if dust is present (see section \ref{sec:dust}). Our mass
selection ($M \geqslant 10^5$ ${\mathrm M_{\sun}}$) translates into a
cut-off luminosity that depends on the color of the source: all the
sources whose magnitude is brighter than the corresponding 1 $M_\odot$
BC03 track, shifted by 12.5 mag, are accepted. Our selection has been
carried out with the assumption of a Salpeter IMF. The luminosity
cutoff we use refers instead to a mass $M \geqslant 5.6\cdot10^4$
${\mathrm M_{\sun}}$ if we were to assume a Chabrier IMF. We identify
these older clusters as young candidate globular clusters since we
calculate \citep{Trenti07} that they have more than a 50\% probability
of surviving tidal dissolution over a Hubble time, assuming they orbit
within a point-like potential well. In addition, the age selection we
apply (age $>$ 100-250 Myr) is sufficiently long to rule out early
dissolution by supernova feedback.

\begin{figure}
\resizebox{130pt}{!}{\includegraphics{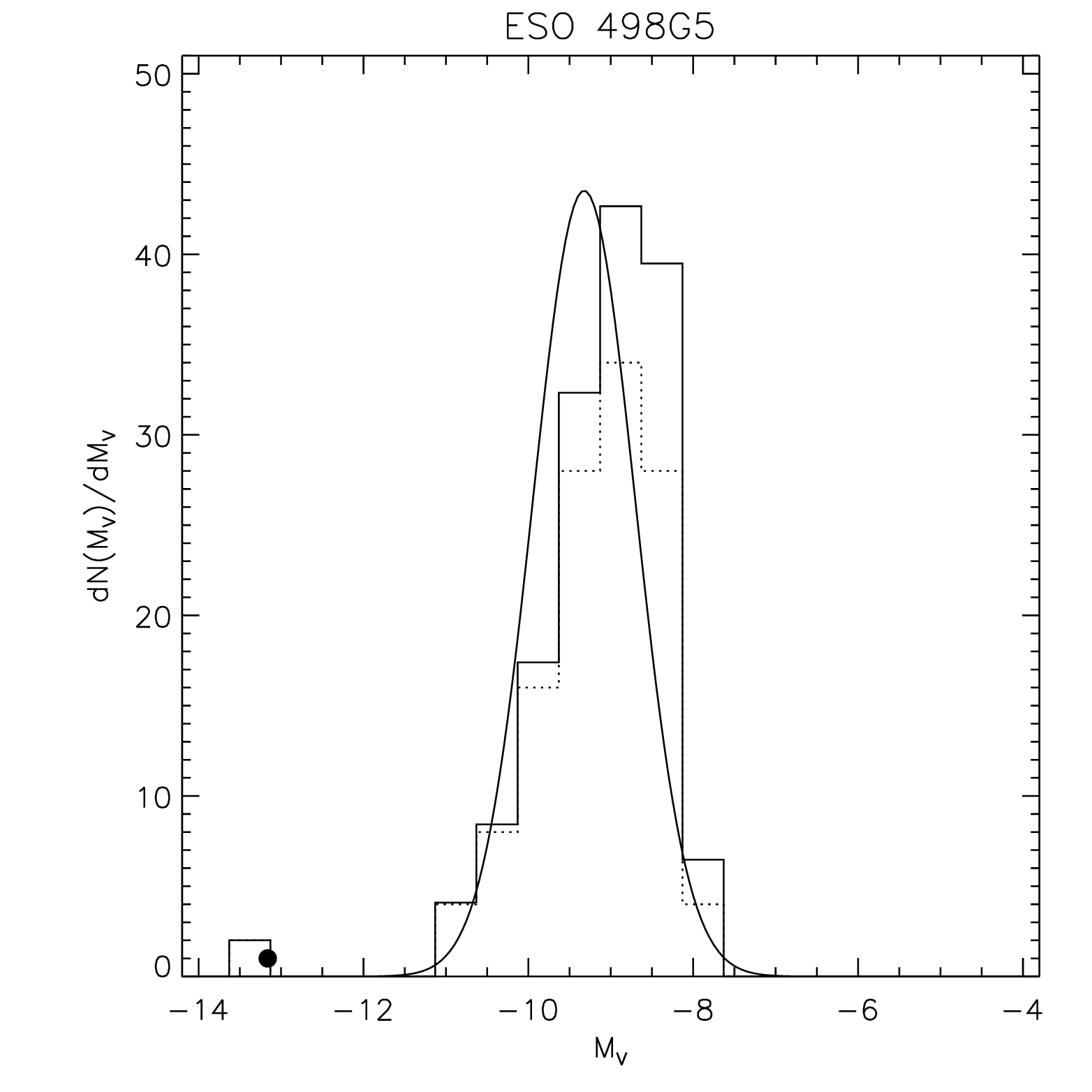}} 
\resizebox{130pt}{!}{\includegraphics{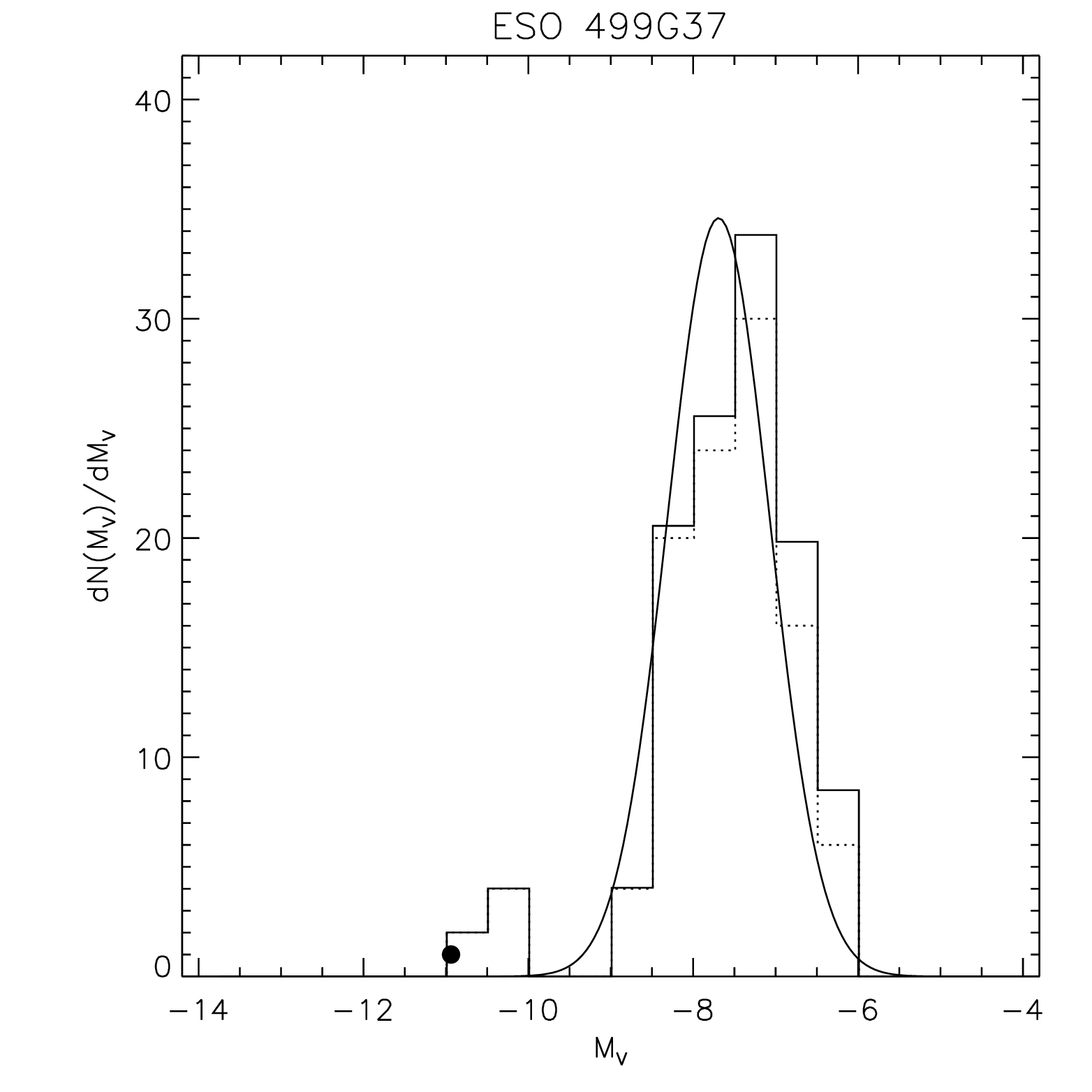}} 
\resizebox{130pt}{!}{\includegraphics{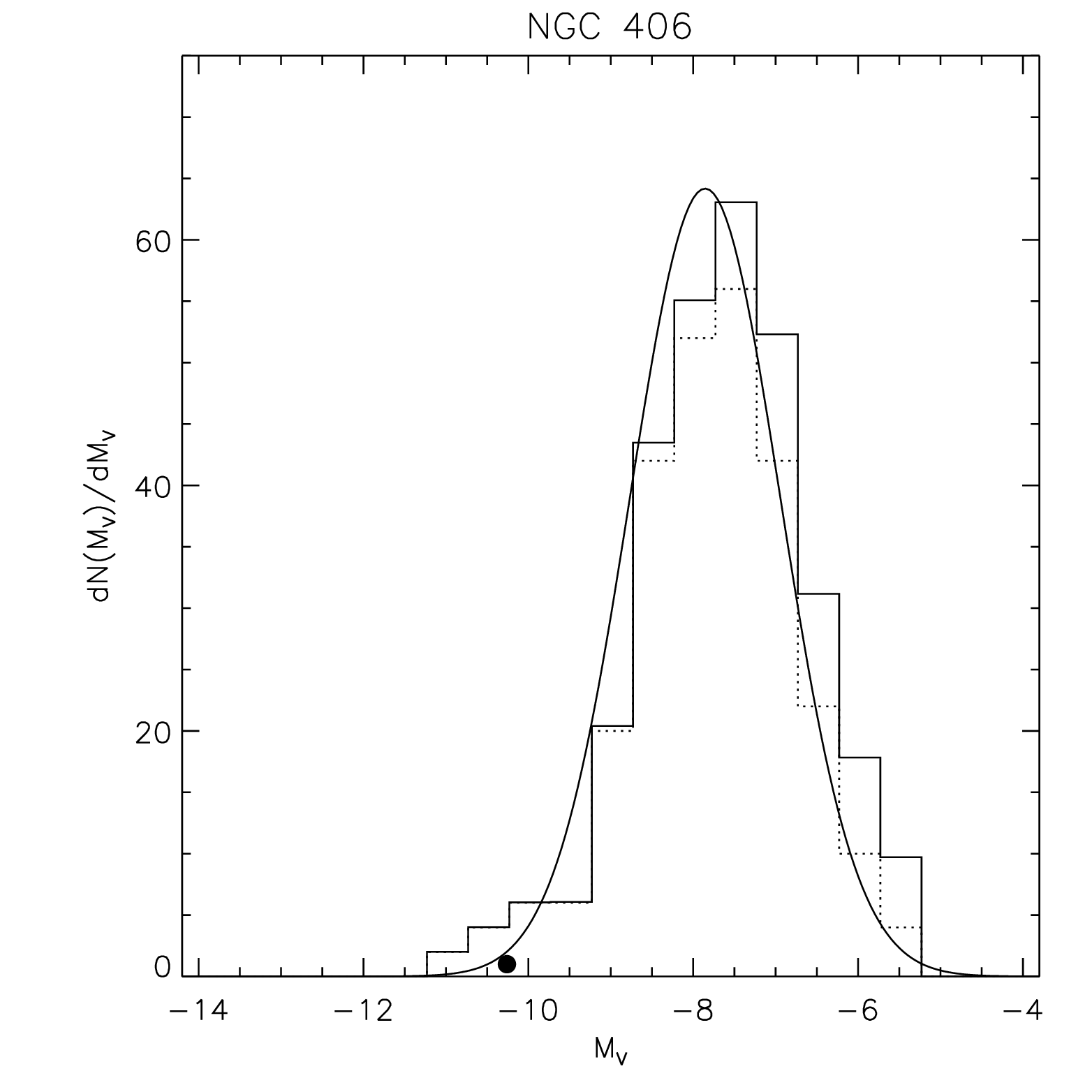}} 
\resizebox{130pt}{!}{\includegraphics{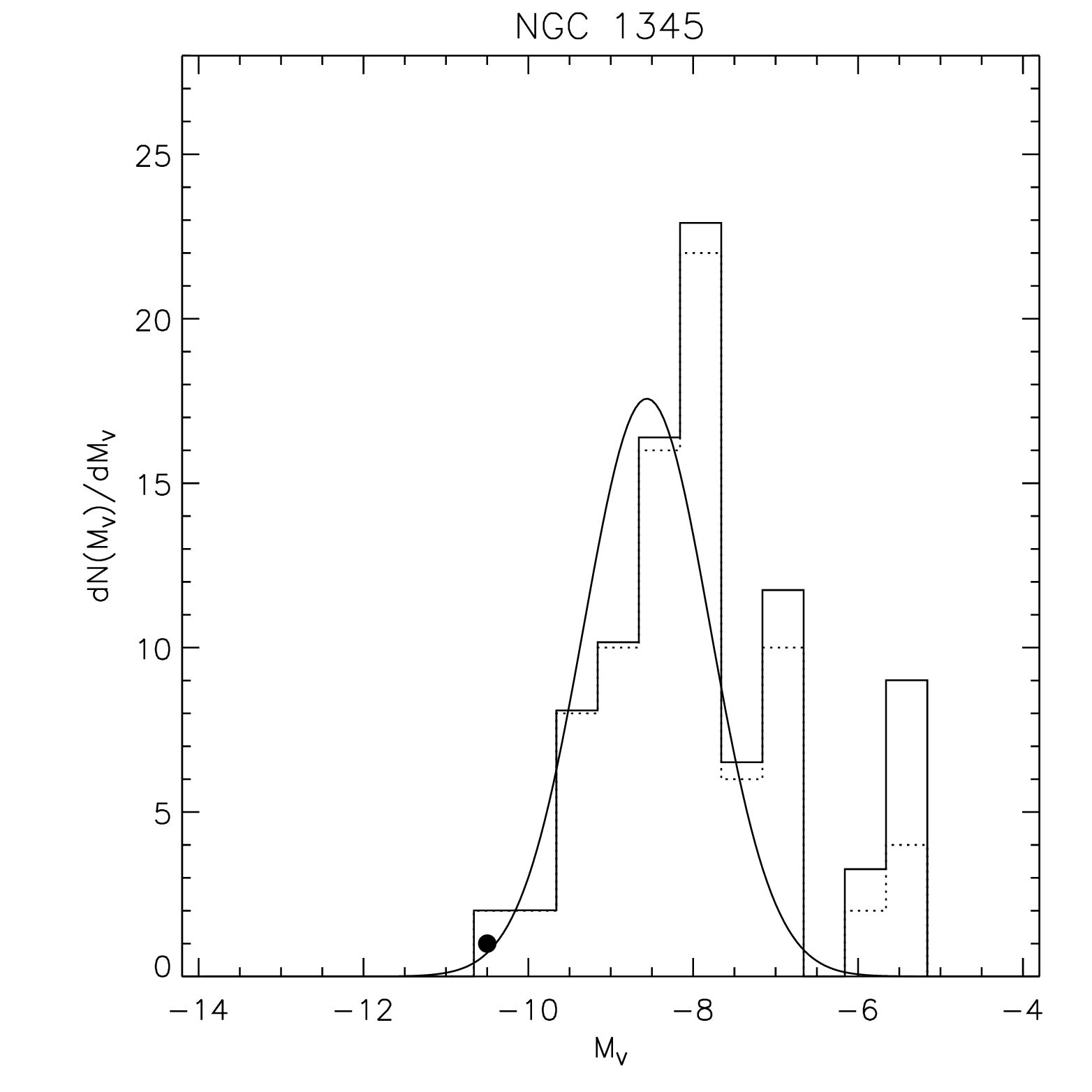}} 
\resizebox{130pt}{!}{\includegraphics{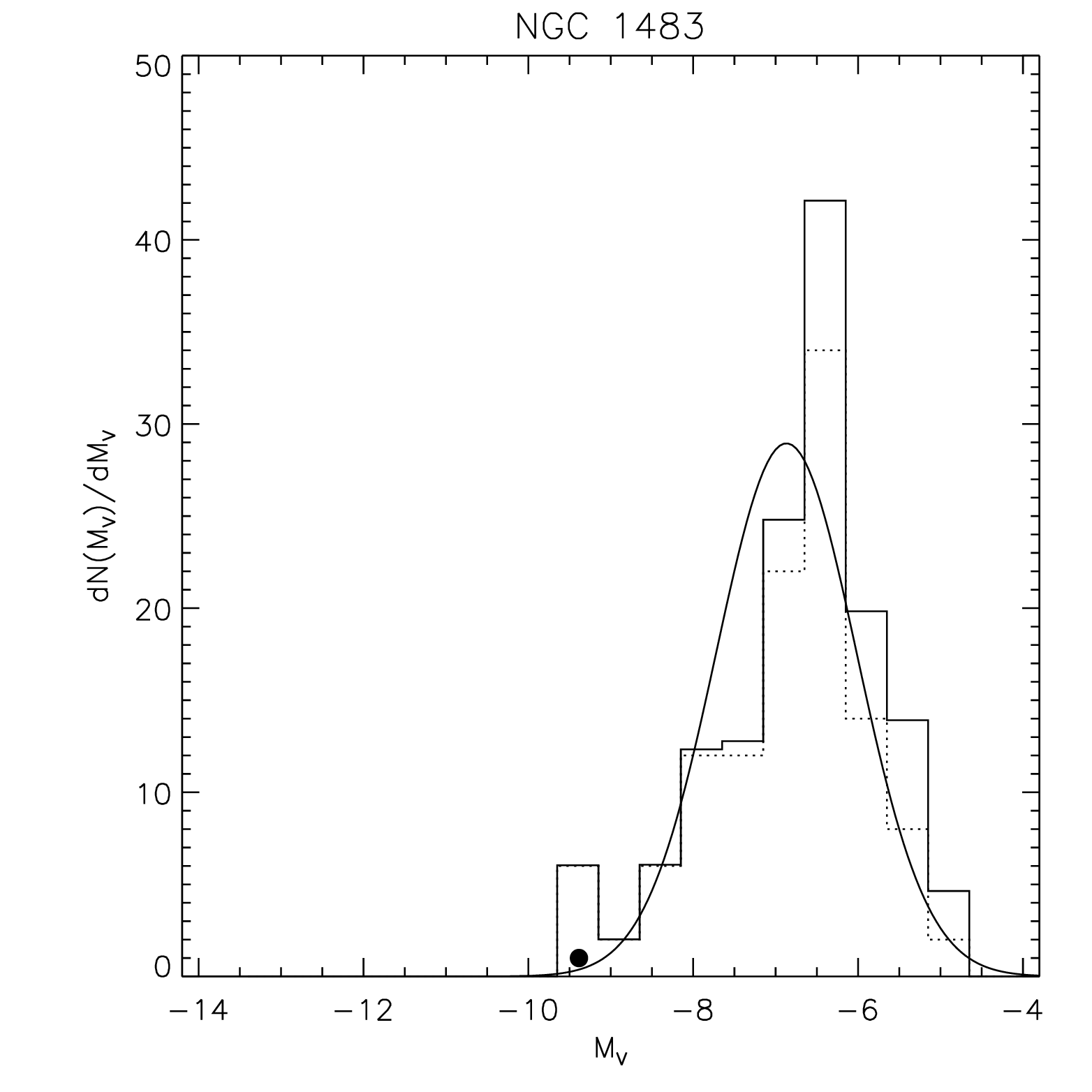}} 
\resizebox{130pt}{!}{\includegraphics{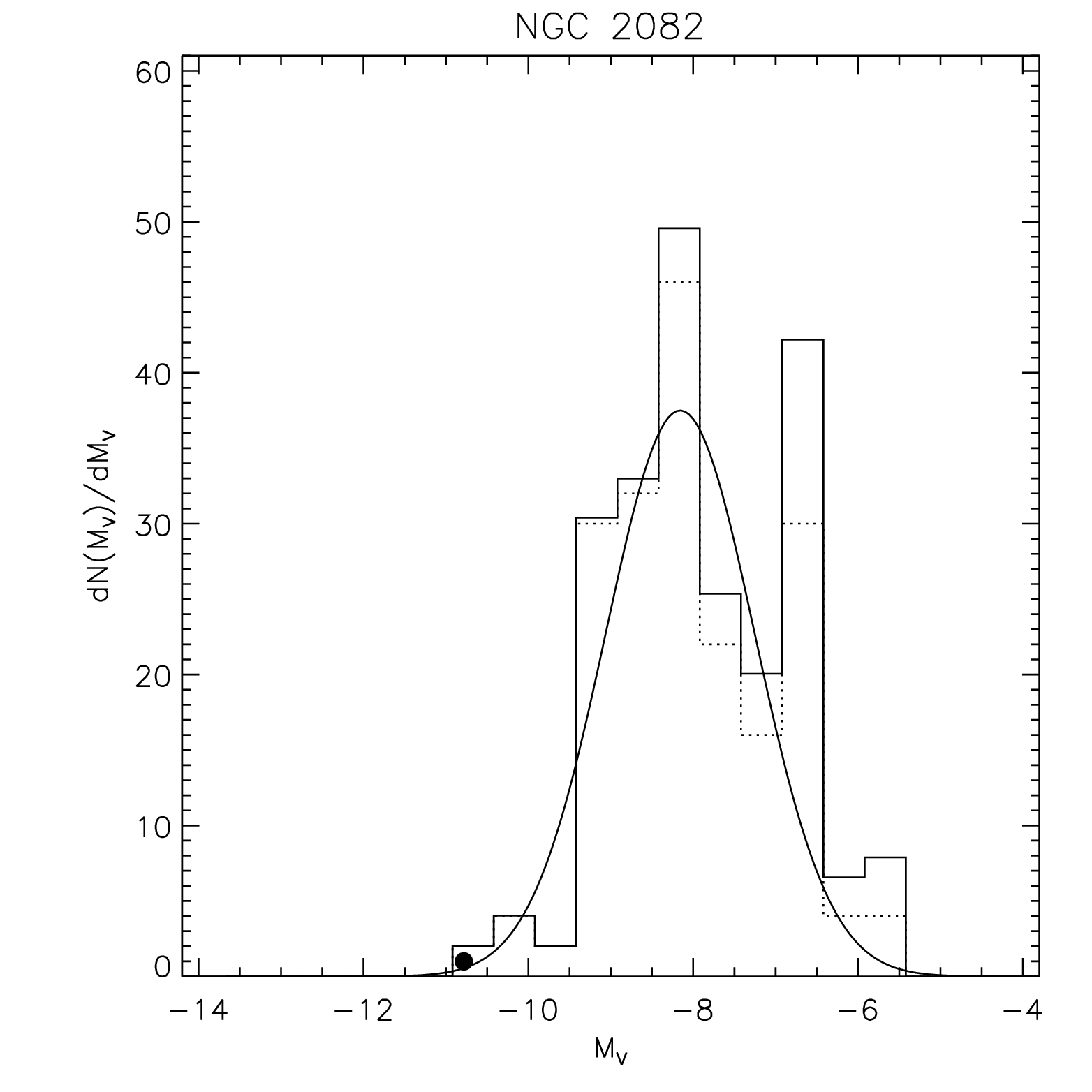}} 
\resizebox{130pt}{!}{\includegraphics{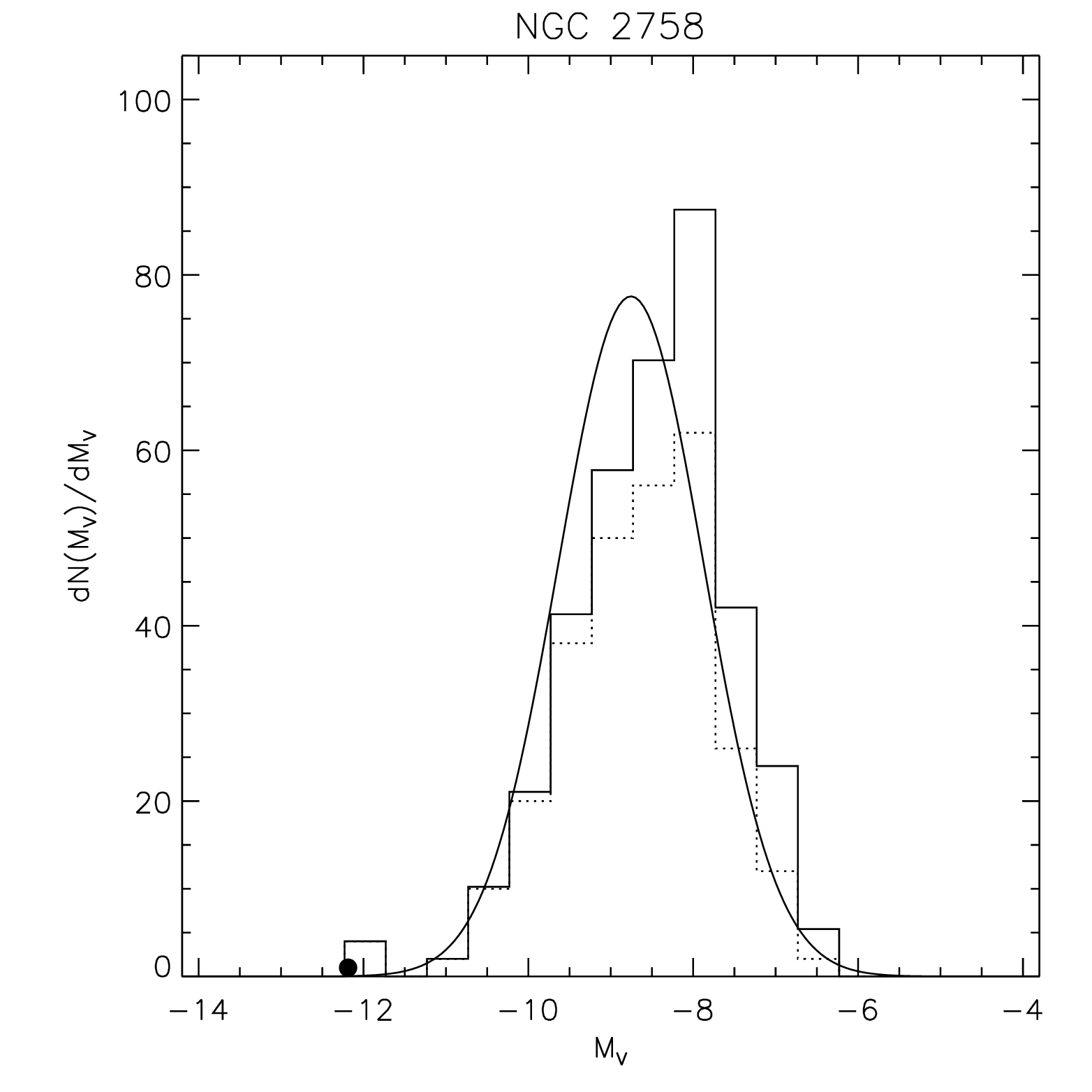}} 
\resizebox{130pt}{!}{\includegraphics{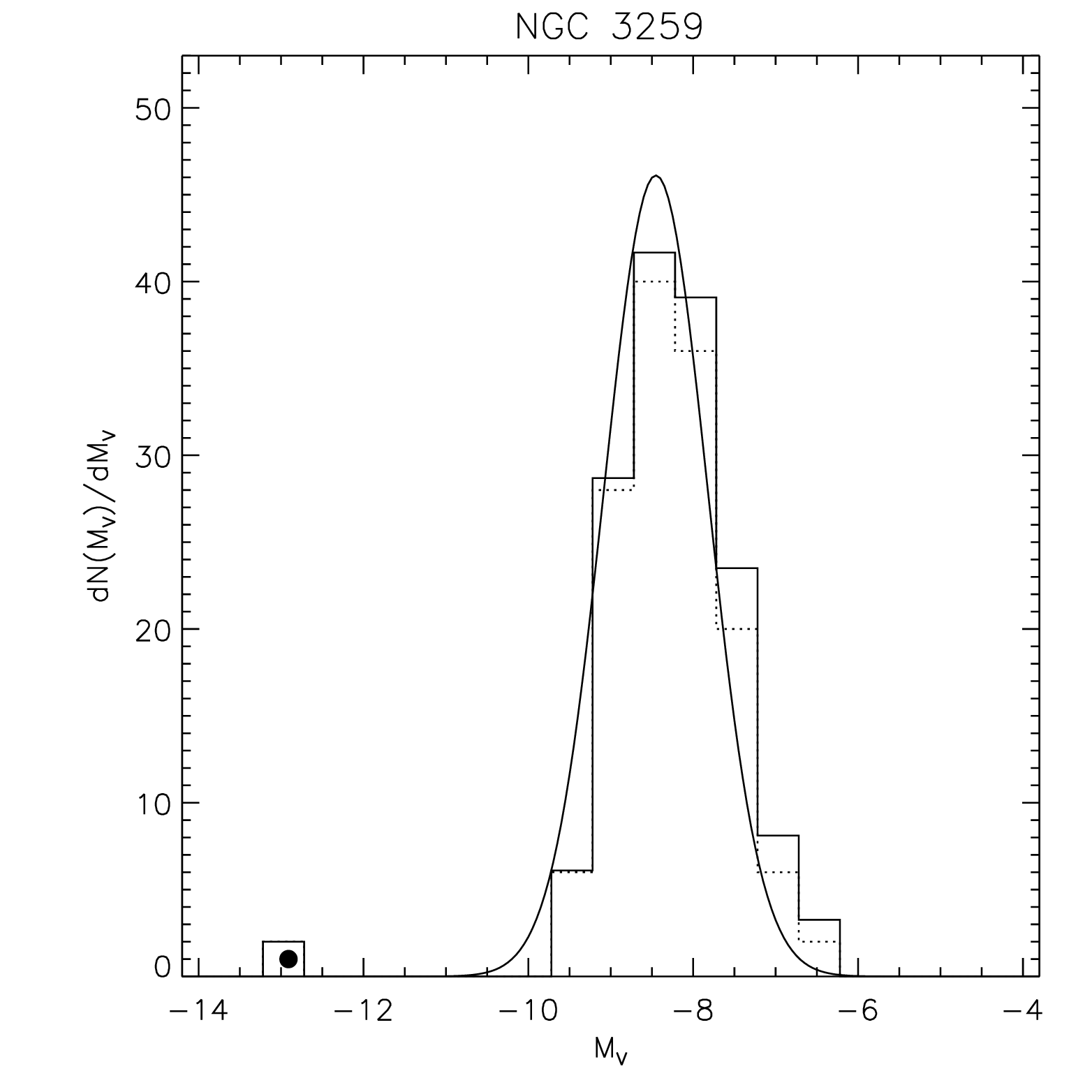}} 
\resizebox{130pt}{!}{\includegraphics{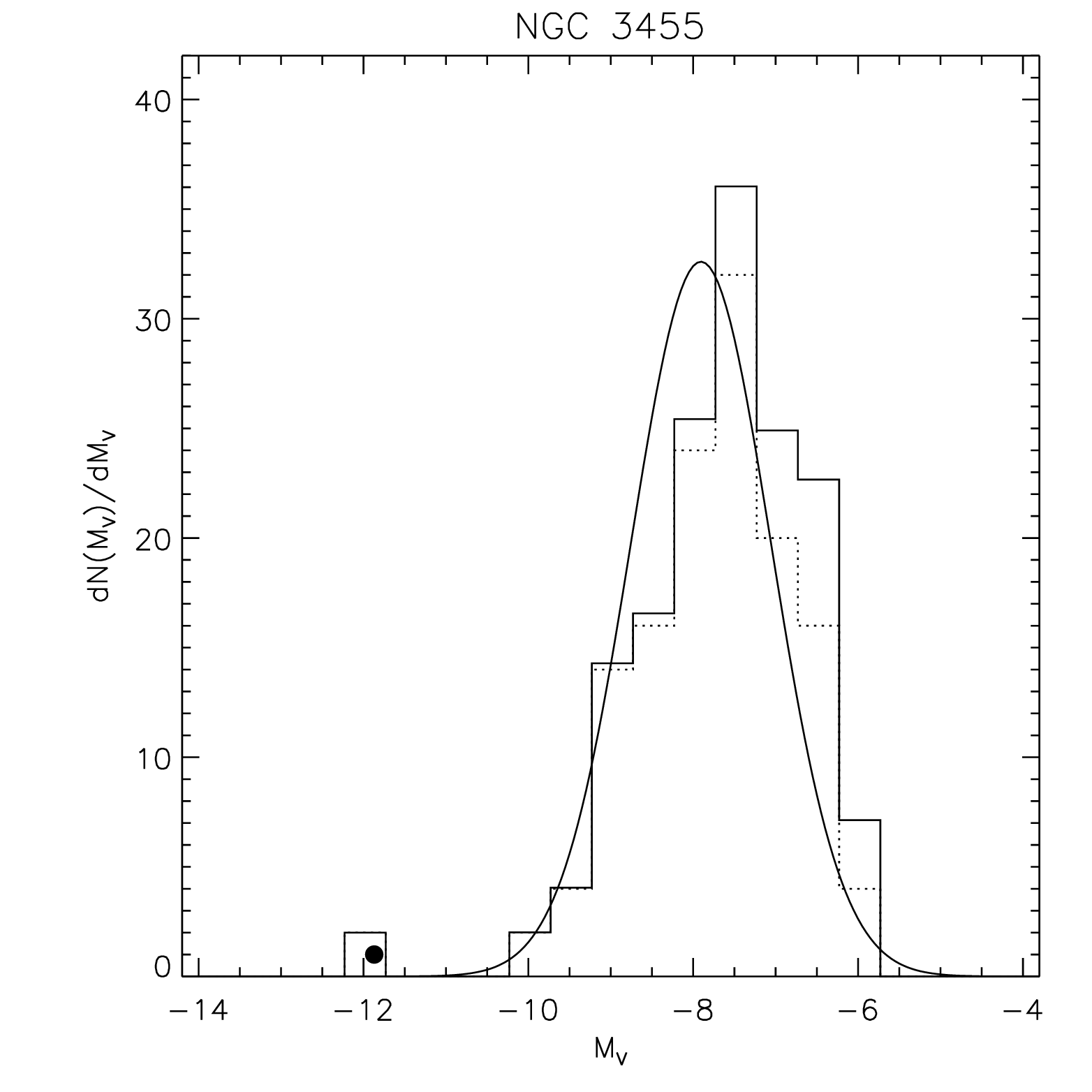}} 
\resizebox{130pt}{!}{\includegraphics{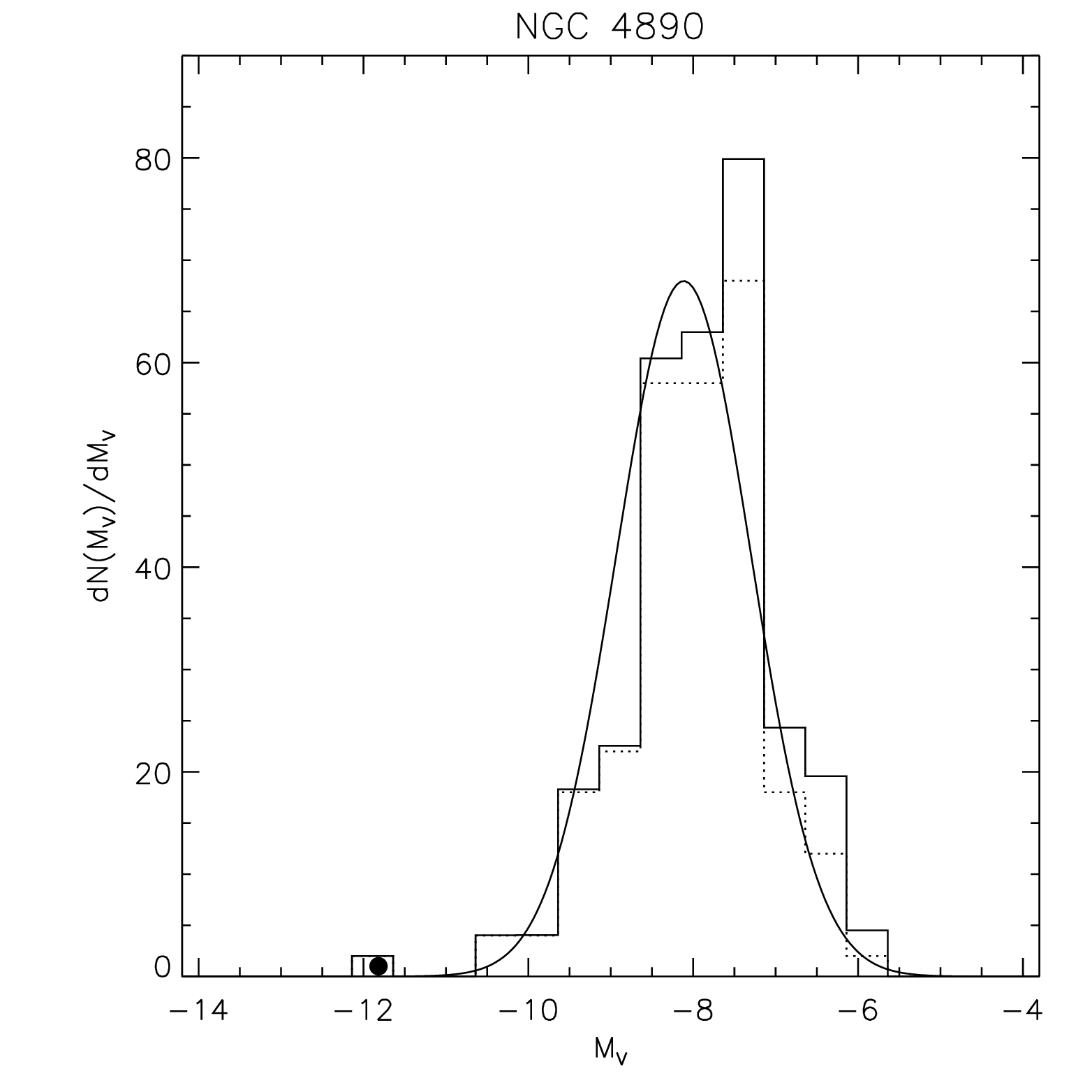}} 
\resizebox{130pt}{!}{\includegraphics{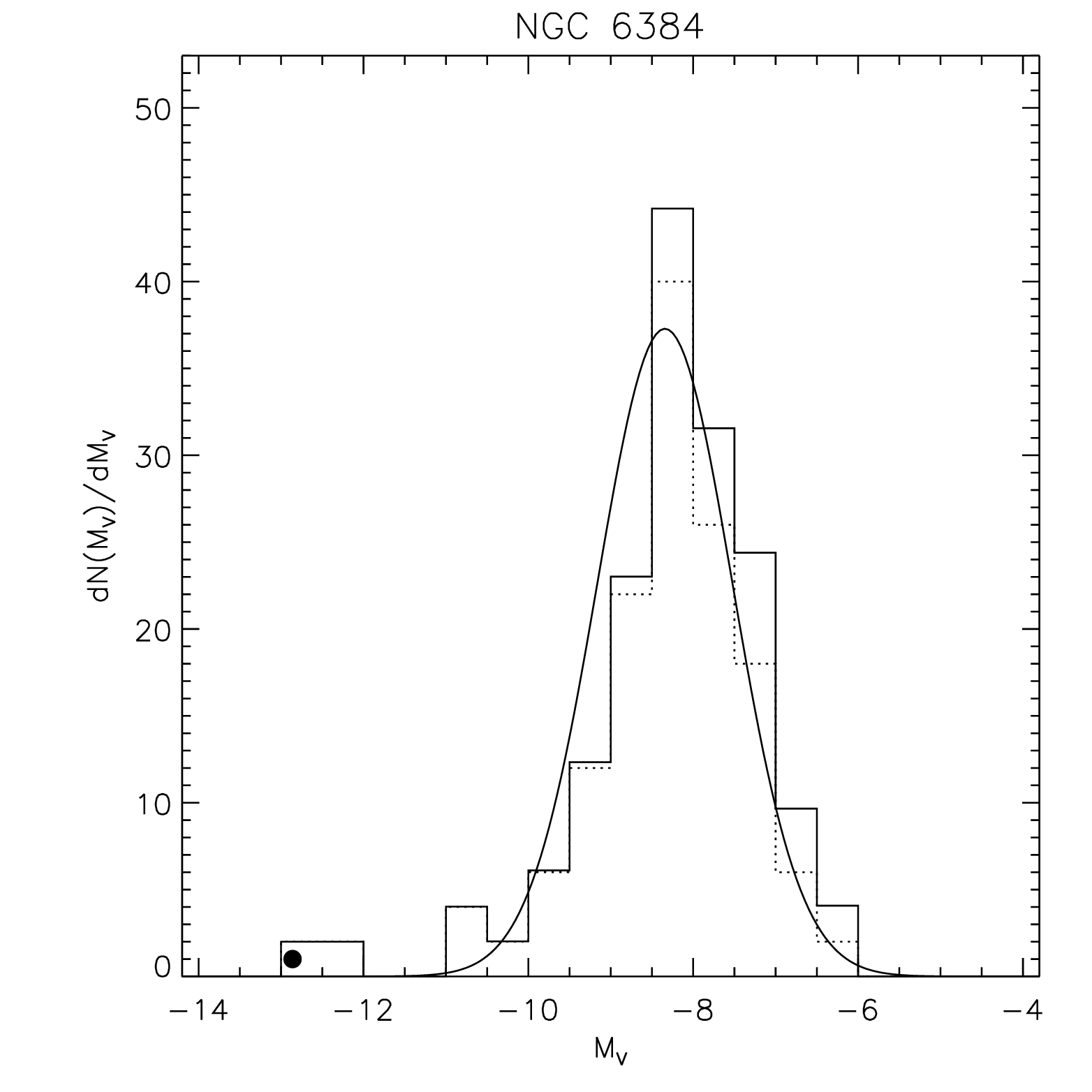}} 
\caption{Solid lines: Candidate Globular Cluster Luminosity Functions
  in the V-band, completeness-corrected, with the best-fitting
  Gaussian superposed. Dotted lines show the GCLF without completeness
  correction. No extinction or contamination correction has been
  applied. The black dot corresponds to the nuclear star cluster of
  each galaxy. The histograms ($ \frac{dN(M_V)}{dM_V}$) are normalized
  so that the total number of observed sources $N= \int_{M_V}
  \frac{dN(M_V)}{dM_V} dM_V$.  \label{fig:gaussian}}
\end{figure}

For each galaxy, the candidate globular cluster luminosity function
(GCLF), i.e.~the number of candidate globular clusters per unit of
magnitude, is plotted in Fig.~\ref{fig:gaussian}. For all the
galaxies, the GCLF (completeness corrected) is reasonably fitted by a
Gaussian distribution, whose FWHM and peak magnitude are reported in
Table \ref{tb:gaussian}. The residuals from the fit are somewhat
asymmetric with an excess of faint sources. This might reflect an
intrinsic skewness of the GCLF, which is however difficult to assert
based on our data. Alternatively, it might be due to the presence of a
small fraction of stellar contaminants with a luminosity distribution
peaked at the faint end of the candidate globular cluster luminosity
distribution (see Section~\ref{sec:contamination}). Note that dust
extinction does not introduce skewness in the GCLF, but only changes
its peak value unless the amount of dust present correlates with the
luminosity of the star clusters.

\section{Systematic Uncertainties}\label{sec:sys}

The results in section~\ref{sec:DA} are derived under the assumption
of no dust extinction present outside the Milky Way and of absence of
contamination in our sample. Here we discuss these two issues.

\subsection{Dust Extinction}\label{sec:dust}

All the galaxies in our sample have NICMOS F160W coverage in the
central region, which we use to quantify the impact of dust
extinction. The F160W band is ideal, since in combination with the
other bands we consider, it significantly increases the wavelength
baseline. However the more limited field of view of NICMOS does not
allow us to apply this diagnostic to the complete sample.

For the subsample of star cluster sources within the NICMOS field of
view we perform a minimum chi-square fit on the data in four
photometric bands comparing them to single stellar population models
with a variable amount of dust extinction.
We have three free parameters: the total stellar mass of the source,
its age and its dust content. The fit is performed assuming different
extinction laws within SYNPHOT: Galactic, LMC and SMC. As expected,
the best fitting models with dust tend to be younger than their
counterparts with no extinction. The estimation of the total mass of
the sources is instead not strongly modified by allowing this
additional degree of freedom.

We summarize the results of these fits in Table~\ref{tb:extinction},
where we give the number of sources older than $100$ Myr and more
massive than $10^5 \mathrm{M_{\sun}}$, for different assumed
extinction laws. These results are compared, in the same table, to
those obtained assuming an age $ \geqslant 250$ Myr and no dust, and
it may be seen that the estimates are within a factor two of each
other. The difference between an age of $100$ and $250$ Myr is not
very significant within the context of infant star-cluster mortality:
in fact, most mortality happens within the first $50$ to $100$ Myr and
is connected to mass loss induced by the rapid stellar evolution of
the most massive stars in the cluster \citep{parmentier07}.

To take into account the effects of dust on the general sample, where
NICMOS coverage is absent, we introduce a galaxy-dependent statistical
correction $f_{ext-bias}$ to the total number of sources we define
``massive'' and ``old'' star clusters. $f_{ext-bias}$ (reported in the
last column of Table~\ref{tb:gaussian}) is computed within the NICMOS
field of view as the fractional difference between the average number
of sources identified in the presence of dust vs. the number identified
under the no-dust assumption. This correction for dust extinction is
then applied when we use the number of old star clusters obtained from
the no-dust scenario to derive their specific frequencies (see Section
\ref{sec:spec_freq}). 

The correction $f_{ext-bias}$ is conservative. In fact, based on the
likelihood ratio test, the statistical significance of dust-extinction
is above the 90\% confidence level only for about one third of the
sources with NICMOS coverage. This means that for some of the sources
the apparent detection of non-zero extinction might simply be a
consequence of allowing an additional degree of freedom in the
fit. Finally, the NICMOS field of view is located at the center of the
galaxy, where the amount of extinction is maximal (e.g. see
\citealt{holwerda05}). The star clusters outside the NICMOS field of
view are thus expected to be less influenced by dust.


\subsection{Sample Contamination}\label{sec:contamination}

A second systematic source of uncertainty that needs to be evaluated
is the presence of stellar contaminants in our sample. These could
either be stars in the Milky Way too faint to be identified by the
presence of diffraction spikes or super-giants in the host galaxy. 

To quantify the impact of Milky Way interlopers, in each galaxy of our
sample we selected a region of the ACS field as far away from the
galaxy as possible, with an area comparable to that used to search for
star clusters. On these external regions we selected sources using the
B and I bands according to the selection criteria we apply to old star
clusters. The surface density of sources identified in the outskirts
of the field of view is an upper limit to the number of Galactic
contaminants. For all the 11 galaxies we obtained surface densities
from 1 to 5 \% (average 2.5\%) of that of old star clusters selected
in our main search region, therefore we can rest assured that Galactic
stars are a negligible source of contamination compared to the larger
uncertainties related to the treatment of dust extinction.

To quantify the impact of bright stars in the host galaxy, we consider
the models of \citet{marigo08}, that provide luminosities for
giant-branch stars. Even for very low metallicities ($Z=10^{-4}
Z_{\sun}$), the maximum luminosity of giant stars is $M_V>-7$; for
higher metallicities the peak luminosity decreases \citep{marigo08}.
From our Fig.~\ref{fig:gaussian} it is immediate to see that at most a
few percent of the sources we select as star clusters are fainter than
$M_V=-7$. Brighter contaminants are possible if they are young,
extremely massive stars. These rare hypergiants can reach $M_V
\sim~-10$ and exhibit a wide range of colors (e.g. see
\citealt{dejager98} for a review). However, they are very short lived
(a few million years) and thus not likely to be found outside the HII
regions where they were born. Ideally we would need images in a narrow
filter centered around the H-$\alpha$ emission line to identify these
sources. As this is not available for our galaxies, we estimate their
impact on the sample based on the measured star formation
rate. Assuming a Salpeter IMF, we expect from a few to $\sim 10$ stars
with mass $M>60~\mathrm{M_{\sun}}$ (bright enough to reach $M_V<-7$,
e.g. see \citealt{stothers99} ) to enter in our selection for each
galaxy.  Combining these two sources of stellar contaminants, that is
Galactic and extragalactic stars, we assume a fraction 10\% of
sample-contamination (indicated as $f_{cont}$ in
Table~\ref{tb:gaussian}). While this might carry some uncertainty, its
impact is certainly secondary compared to the one induced by
extinction, except for ESO 498G5 and NGC 6384.

\section{Specific Frequency}\label{sec:spec_freq}

For each galaxy, we calculate the specific frequency of the candidate
globular clusters in our field of view, i.e.~the older, more massive
cluster population normalized to the host galaxy luminosity (see
Table~\ref{tb:spec_freq}). This is defined as follows:
\begin{equation}
  {S_N}_{gal}=N_{tot}\cdot10^{0.4({M_{V_{JC}}}_{gal}+15)},
\end{equation}
where ${M_{V_{JC}}}_{gal}$ is the absolute magnitude of the galaxy
within the search area in the Johnson-Cousins V-band and $N_{tot}$
(reported in Table \ref{tb:gaussian}) is the total number of candidate
GCs, obtained by integrating the completeness-corrected GCLF and
applying a correction $f_{cont}=0.1$ for the estimated fraction of
stellar contaminants and a galaxy-dependent correction $f_{ext-bias}$
to take into account dust extinction (see
section~\ref{sec:contamination}). The luminosity of the host galaxy
within the star-cluster-search area in the HST-F606W filter is
calculated through aperture photometry. The background level is
estimated using the WF chips of WFPC2, which cover a larger region of
the sky compared to the star-cluster-search area, limited within the
PC chip.  The F606W magnitude ($V$) is then converted to the
Johnson-Cousins $M_{V_{JC}}$ absolute magnitude using:
\begin{equation}\label{eq:conv}
  M_{V_{JC}}=V+0.287\cdot(V-I)+dm,
\end{equation}
where $dm$ is the distance modulus, listed in Table \ref{tb:sample}
and the color $V-I$ is from \citet{acs_I}, except for NGC 1345 (for
this galaxy the color is calculated from the photometry of the galaxy
search area in both images). The coefficient 0.287 is obtained by
fitting with a linear law the relation between $M_{V_{JC}}$ and $V-I$ for
the \citet{BC03} models of old star clusters (magnitudes computed with
the IRAF task CALCBAND). Eq.~\ref{eq:conv} agrees very well with the
calibration given by \citet{ZP_wfpc2} in his Table 10 (maximum
difference of 0.05 mag. in the range $-1<V-I<2$), but has two
advantages. First it is calibrated exactly on the color range we are
interested in (see the caveats in \citealt{ZP_wfpc2} on the limited
range of validity for his conversion) and second our formula is
calibrated on the class of sources in which we are interested (star
clusters, i.e. adopting an IMF) rather than individual stars as in
\citet{ZP_wfpc2}. In computing the specific frequencies we have
excluded NGC 1483 (the closest galaxy in our sample) because its bulge
is very extended spatially and it gives the dominant contribution to
the light within the star-clusters search area.

The resulting values of ${S_N}_{gal}$ are listed in Table
\ref{tb:spec_freq}, where uncertainties on $S_N$ take into account the
errors on the candidate GC count ($N_{tot}$), including those derived
from dust extinction (see section~\ref{sec:dust}), but not those in
the distance modulus as they are negligible compared to the other
errors.

The definition of specific frequency adopted here is a generalization
of the standard \citet{spec_freq} definition, which is based on the
total luminosity of the galaxy and on the total number of GCs. We note
though that we do not expect a large difference in the two
definitions, because the PC camera of WFPC2 contains a large fraction
of the total light of the galaxies. Indeed, the absolute visual
magnitude ${M_{V_{JC}}}_{gal}$ we measure in our sample differs by less
than about 1 magnitude compared to the total $M_{V_{JC}}$ luminosity of
the galaxy (estimated from the absolute blue magnitude reported in
Table \ref{tb:sample}, assuming $B-V = 0.5$).

We also compute the specific frequency relative to the bulge
luminosity (${S_N}_{bul}$), defined as:
\begin{equation}
  {S_N}_{bul}=N_{tot}\cdot10^{0.4({M_{V_{JC}}}_{bul}+15)},
\end{equation}
where again ${M_{V_{JC}}}_{bul}$ is expressed in the Johnson-Cousins
system, obtained as above from the bulge $V$-magnitude published by
\citet{acs_I}. The resulting frequencies of candidate globular
clusters per bulge light are also shown in Table \ref{tb:spec_freq}.

These frequencies for our sample of galaxies can be compared with
published values for nucleated dwarf galaxies \citep{Miller98}, as
well as for early and late type galaxies with classical bulges (see
\citealt{brodie_strader_2006}). The ${S_N}_{gal}$ we measure in our
late type galaxies is qualitatively consistent with that of other
spiral galaxies, but there is a moderate tendency toward higher
$S_N$. This finding seems to be at odd with predictions from
semi-analytics models of globular clusters formation which link the
specific frequency with the overall star formation rate of the galaxy
\citep{beasley02}: based on their scenario we would have expected a
lower $S_N$ for our sample compared to that of spiral galaxies with
classical bulges. In fact, we expect a lower star formation rate in
pseudo-bulges if they form from the inner disk. However the
\citet{beasley02} study was aimed at reproducing the properties of GCs
in elliptical galaxies, therefore the comparison is only indirect,
based on the assumption that the bulge formation process is similar to
that of small ellipticals. In addition, it did not include evolution
of the GCs system of the simulated galaxies, but rather its properties
were fixed and fine tuned to match the observed data at the time of
birth.

Another possibility to reconcile with these theoretical expectations
is that of disruption of our candidate globular clusters as they age.
Most of the `old' star clusters in our sample appear in fact to be
relatively young compared to Galactic globular clusters (ages from a
few hundred Myr to one Gyr), so this suggests that tidal interactions
with the parent galaxies will reduce the number of star clusters as
they age. For example, \citet{Gnedin97} estimate that for the Milky
Way, the Galactic globular cluster system has an half life of the
order of the Hubble time. This is consistent with detailed N-body
simulations of the dynamics of star clusters in the presence of a
tidal field (e.g. see \citealt{Trenti07} and references therein).
Under a scenario of significant adult mortality for star clusters, the
specific frequency for our sample will become lower than that of
spirals with a classical bulge, as expected on the basis of
theoretical modeling of star formation (SF): in the case of
pseudo-bulges, with a likely extended but low-efficiency star
formation history, fewer globular clusters are formed than in an
equivalent burst of SF, such as that considered to create classical
bulges \citep{kormendy_kennicutt2004}.

\begin{figure}
\epsscale{1.0}
\plotone{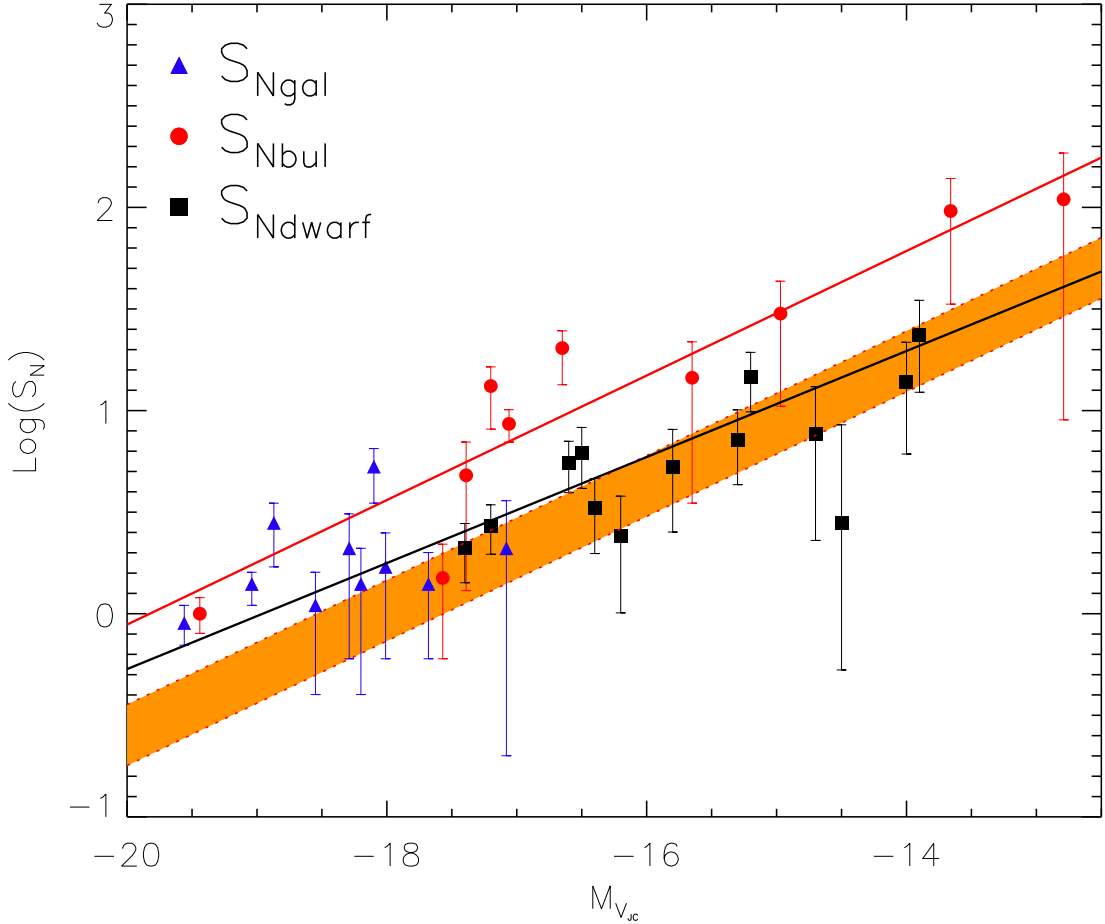} 
\caption{Candidate globular cluster specific frequencies in our sample
  (normalized to the galaxy luminosity: blue triangles; normalized to
  the bulge luminosity: red circles). For comparison the globular
  cluster specific frequencies for the sample of dwarf ellipticals of
  \citet{Miller98} are plotted as black squares. The solid red
  (specific frequency normalized to bulge luminosity) and black
  (specific frequency for dwarf ellipticals) lines represent a best
  fitting linear relation in the ($Log(S_N);M_{V_{JC}}$) space. The orange
  region delimited by the red dotted lines is a passive evolution of
  the solid red line, assuming that the bulge population ages from 1
  to 10 Gyr (corresponding to a dimming of $\approx 2.2$~mag in $M_V$)
  and a reduction in the number of star clusters due to tidal
  dissolution (50\%) and stripping due to galaxy harassment (50\% to
  75\%).
  \label{fig:logSN}}
\end{figure}

Mean specific frequencies for nucleated dwarf galaxies ($\langle
{S_N}_{dwarf} \rangle = 7.5$) --- whose GCs are similar to candidate GCs in
our sample --- lie instead between our $\langle {S_N}_{gal} \rangle =
2.0$ and $\langle {S_N}_{bul} \rangle = 30$. In addition, the trend of
frequencies of candidate globular clusters per bulge light versus
luminosity is very similar to that of dwarf galaxies (see
Fig.~\ref{fig:logSN}). This is suggestive of a stripping scenario in
which dwarf galaxies originate from the bulges of late-type spirals.
One possibility is that galaxy-galaxy fly-by encounters (galaxy
harassment: \citealt{Moore96}) may strip away stars in the disk,
leaving preferentially behind bulge members and the central star
clusters, both more protected because they sit deeper in the potential
well of the system. In fact, passive evolution of the bulges (aging
them from 1 to 10 Gyr, to model dwarf galaxies that have older stellar
populations than pseudo-bulges) and accounting for tidal dissolution
and stripping of star clusters (50 \% tidal dissolution and 50\% to
75\% stripping) shift the candidate GC specific frequency vs. luminosity
relation to match that valid for dwarf ellipticals (see
Fig.~\ref{fig:logSN}).

\section{Nuclear Star Clusters}\label{sec:nuclear}

A detailed study of the photometric properties of nuclear star
clusters has been presented in \citet{wfpc2_I,wfpc2_II,nicmos_I}, here
we briefly discuss their inferred ages and dust content. A summary of
the photometry for the central sources is reported in
Table~\ref{tb:nuclearSC}, while the resulting mass, age and dust
content from our maximum likelihood fit are in
Table~\ref{tab:central_cluster_fit}. The nuclear star clusters in the
sample are typically the brightest clusters of their host galaxies,
with an inferred mass up to $\sim 10^7$ $\mathrm{M_{\sun}}$. The mass
of the central star cluster broadly correlates with the bulge
luminosity (and therefore mass), as observed in a wide sample of
galaxies, both photometrically (e.g. \citealt{weh06,cote06,rossa06})
and dynamically from spectroscopy
(e.g. \citealt{geha02,walcher06}). The correlation is consistent with
Fig.~1 in \citet{weh06}. Like in the \citet{weh06} work, we identify a
few clusters (NGC 406, NGC 1345 and NGC 2758) with a bright bulge
($\sim 10^{8.5}~\mathrm{L_{\sun}}$), but with a relatively small
nuclear star cluster mass ($\sim 2 \cdot 10^6~\msun$).

From the nuclear star cluster colors we infer a typical age $\gtrsim
1$ Gyr in most of the sample, even after accounting for a variable
dust extinction as described in section~\ref{sec:dust}. Two galaxies
(NGC 1483, NGC 2758) present a clear evidence that their nuclear star
cluster is composed of a young stellar population (age $5-50$ Myr)
associated with dust ($E(B-V)>0.2$). The nuclear star cluster of NGC
3259 has instead colors that are poorly fitted by both models with and
without dust extinction: no reliable constrain on the stellar age can
be obtained. The younger stellar ages observed in the nuclear clusters
of NGC 1483 and NGC 2758 might be due to a recent burst of star
formation at the center of these galaxies that has rejuvenated them.
This is not surprising as rejuvenation is observed in a large number
of nuclear star clusters \citep{rossa06,walcher06} and suggests a in
situ formation scenario for the clusters, as proposed by
\citet{milos04} and inferred from observations of local spirals by
\citet{seth06}. However the relative old ages ($\gtrsim 1$ Gyr) of
most of the sources are also consistent with a formation scenario
driven by mergers of star clusters that reach the center of the galaxy
by dynamical friction \citep{tremaine75,lotz01,andersen08}.

\section{Discussion and Conclusion}\label{sec:concl}

In this paper we studied the properties of (relatively) old star
clusters in a sample of 11 late-type spiral galaxies selected for the
presence of a pseudo-bulge \citep{wfpc2_I}. Star clusters were
classified based on their color and luminosity through comparison of
population synthesis models with HST photometry in three bands (ACS
F435W, F814W and WFPC2 F606W). By means of artificial source
detections we estimated a completeness of $\simeq 50\%$ down to
$V=25.5$. The clusters in our sample present a wide range of ages and
masses, from young blue clusters with ages of a few tens of Myrs to an
older, red population (age $> 100-250$ Myr). We focus on these older,
red clusters and identify them as young globular cluster candidates.

All these galaxies have massive nuclear star clusters with masses in
the $10^6-10^7~\mathrm{M_{\sun}}$ range. These sources are typically
the brightest star clusters in their host galaxy. They have a
relatively old stellar age ($\gtrsim$ 1 Gyr) except for two cases where
a younger and dusty stellar population is inferred. Overall, the
properties of our sample appear consistent with both proposed
formation scenarios for nuclear star clusters, namely merging of
stellar clusters driven to the galaxy center by dynamical friction
\citep{lotz01} or in situ formation \citep{milos04}.
  
The presence of a young star clusters allows us to set a
lower limit to the star formation rate in the galaxies, which turns
out to be $0.1-0.01$ $\mathrm{M_{\sun}/yr}$. This continuous, low rate
of star formation is consistent with the formation scenario for
pseudo-bulges, postulated to arise out of secular processes in the
disk \citep{kormendy_kennicutt2004}, compared to a violent burst of SF
needed to form a classical bulge.

Based on the \citet{kormendy_kennicutt2004} discussion of the
formation of pseudo-bulges, we would expect them to have a deficit of
massive star clusters compared to spirals with classical bulges, but
this is not what we find. On the contrary, the specific frequencies
(number of star clusters normalized to the galaxy luminosity ---
\citealt{spec_freq}) for the old population is consistent, within our
uncertainties, to published data for other spirals. This appears to be
a solid result, especially since we considered only central star
clusters (the WFCP2 high resolution detector limits our area of search
to the central $35\arcsec \times 35\arcsec$), normalizing the specific
frequency to the galaxy luminosity within the field of view: in
general star cluster systems are more spatially extended than their host
galaxy (e.g. see \citealt{djo94,jo09}), thus our specific frequencies
are probably lower limits to the global specific frequency.

When the specific frequency is computed with respect to the bulge
luminosity we get even higher $S_N$. Interestingly the specific
frequency vs. bulge magnitude ${S_N}_{bul}({M_{V_{JC}}}_{bul})$ trend is
similar to the one observed in dwarf ellipticals. Pseudo-bulges have
photometric and kinematic properties very similar to dwarf
ellipticals, thus it is suggestive that some dwarf ellipticals might
be the result of evolution of spiral galaxies with pseudo-bulges: the
galaxy disk might in fact be stripped during galaxy-galaxy
interactions (galaxy harassment --- \citealt{Moore96}). In this
scenario star clusters are also stripped away, but still a sizable
number might survive compared to the more spatially extended disk.  At
the same time the pseudo-bulge survives almost untouched by stripping,
protected as it sits at the center of the galaxy potential well. While
this scenario is overall appealing, detailed numerical simulations are
needed for a proper validation. These will be presented in a follow-up
paper.
  

\acknowledgments

We thank the referee for a thorough and constructive report. We
acknowledge the use of the NASA/IPAC Extragalactic Database (NED),
which is operated by the Jet Propulsion Laboratory, California
Institute of Technology, under contract with the National Aeronautics
and Space Administration. This publication was supported in part by
NASA grants No. HST-GO9395 and HST-AR10982, awarded by The Space
Telescope Science Institute, which is operated by the Association of
Universities for Research in Astronomy, Inc., under NASA contract
NAS5-26555. RFGW acknowledges a SUPA Distinguished Visitor award at
University of Edinburgh.






\clearpage

\clearpage

\clearpage

\clearpage

\clearpage

\clearpage

\clearpage

\clearpage

\clearpage

\clearpage

\begin{deluxetable}{lccccccccc}
\tabletypesize{\scriptsize}
\tablecaption{Basic parameters for the sample galaxies.\label{tb:sample}}
\tablewidth{0pt}
\tablehead{
  \colhead{Name}                      &   \colhead{$\alpha$(J2000)} &
  \colhead{$\delta$(J2000)}           &   \colhead{$B_{tot}$}        &
  \colhead{D\tablenotemark{a}}        &   \colhead{dm}              &
  \colhead{$M_{B_{tot}}$}                     &   \colhead{Type}            &
  \colhead{E(B-V)}                    &   \colhead{$R_{search}$} \\
                                      &   \colhead{(h m s)}         &
  \colhead{(deg \arcmin $ $ \arcsec)} &   \colhead{(mag)}           &
  \colhead{(Mpc)}                     &   \colhead{(mag)}           &
  \colhead{(mag)}                     &                             &
  \colhead{(mag)}                     &   \colhead{(kpc)} \\
}

\startdata ESO 498G5 & 09 24 41.10 & -25 05 33.0 & 13.96 & 37.7 & 32.88 &
-18.92 & SXS4P/SBbc & 0.107 & 3.254    \\
ESO 499G37 & 10 03 42.10 & -27 01 39.0 & 13.15 & 17.7 & 31.24 & -18.09 &
SXS7*/SBc & 0.075 & 1.529    \\
NGC  406 & 01 07 24.10 & -69 52 35.0 & 13.10 & 19.8 & 31.48 & -19.38 &
SAS5*/Sc & 0.024 & 1.708    \\
NGC 1345 & 03 29 31.70 & -17 46 42.0 & 14.02 & 19.1 & 31.41 & -17.39 &
SBS5P/SBa & 0.038 & 1.654 \\
NGC 1483 & 03 52 47.60 & -47 28 39.0 & 13.00 & 15.1 & 30.90 & -17.90 &
SBS4/Sb-Sc & 0.007 & 1.308  \\
NGC 2082 & 05 41 51.20 & -64 18 04.0 & 12.62 & 17.1 & 31.17 & -18.55 &
SBR3/SBb & 0.058 & 1.481    \\
NGC 2758 & 09 05 30.80 & -19 02 38.0 & 13.46 & 31.3 & 32.48 & -18.02 &
PSB.4P?/Sbc & 0.126 & 2.707   \\
NGC 3259 & 10 32 34.68 & +65 02 26.8 & 12.97 & 24.8 & 31.97 & -19.00 &
SXT4*/SBbc & 0.015 & 2.140    \\
NGC 3455 & 10 54 31.20 & +17 17 02.8 & 12.87 & 19.8 & 31.48 & -18.61 &
PSXT3/Sb & 0.033 & 1.708    \\
NGC 4980 & 13 09 10.20 & -28 38 28.0 & 13.19 & 23.9 & 31.89 & -18.70 &
SXT1P?/SBa & 0.072 & 2.063    \\
NGC 6384 & 17 32 24.42 & +07 03 36.8 & 11.14 & 22.4 & 31.75 & -20.61 &
SXR4/SBbc & 0.123 & 1.934    \\
\enddata
\tablecomments{Right Ascension ($\alpha$), declination ($\delta$) and
  total apparent blue magnitude ($B_{tot}$) are from RC3 catalog
  \citep{RC3}. Distance ($D$) and distance modulus ($dm$) are from the
  online NASA/IPAC Extragalactic Database (NED). The absolute
  magnitude $M_{B_{tot}}$ is obtained from $B_{tot}$ and $dm$. Morphological
  classifications are from the RC3 (\textit{left}) and from the UGC
  (\textit{right}) catalogs. The color excess E(B-V) is taken from
  \citet{red}. The last column lists the radius of the common
  ACS/WFC-WFPC2/PC field of view.}

\tablenotetext{a}{using $H_0=73$ $km$ $s^{-1}$ $Mpc^{-1}$.}
\end{deluxetable}



\begin{deluxetable}{lccc}
\tabletypesize{\scriptsize}
\tablecaption{Details of the photometric calibration.\label{tb:basic_phot}}
\tablewidth{0pt}
\tablehead{
  \colhead{Filter}         &   \colhead{Zeropoint}  &
  \colhead{Threshold}      &   \colhead{Reddening}    \\
                           &   \colhead{(mag)}      &
  \colhead{(mag)}          &   \colhead{(mag)}        \\
}
\startdata
F435W (ACS/WFC)  & 25.779\tablenotemark{a}  & 26.5  & 1.319  \\
F606W (WFPC2/PC) & 22.084\tablenotemark{b}  & 26    & 0.908  \\
F814W (ACS/WFC)  & 25.501\tablenotemark{a}  & 26    & 0.586  \\
\enddata
\tablecomments{Zeropoints used for calibrating the magnitudes,
  threshold magnitudes corresponding to a SNR $\sim 5$ and reddening
  coefficients expressed as $A_{HSTfilter}/A_{{V}_{JC}}$.}

\tablenotetext{a}{Derived from \citet{ACS}.}
\tablenotetext{b}{Derived from \citet{ZP_wfpc2}.}
\end{deluxetable}

\begin{deluxetable}{lccc}
\tabletypesize{\scriptsize}
\tablecaption{Number of observed Star Clusters.\label{tb:Ntot}}
\tablewidth{0pt}
\tablehead{
  \colhead{Name}           &   \colhead{All}  &
  \colhead{Age $ \leqslant$ 8 Myr} &   \colhead{Age $\geqslant$ 250 Myr}    \\
}
\startdata
ESO 498G5  &   477  &  225  &  62  \\
ESO 499G37 &   434  &  230  &  53  \\
NGC  406   &  1058  &  376  & 133  \\
NGC 1345   &   490  &  194  &  41  \\
NGC 1483   &  1208  &  551  &  59  \\
NGC 2082   &  1401  &  576  &  96  \\
NGC 2758   &   777  &  326  & 141  \\
NGC 3259   &   805  &  409  &  70  \\
NGC 3455   &  1317  &  623  &  67  \\
NGC 4980   &  1070  &  591  & 133  \\
NGC 6384   &   153  &   25  &  71  \\
\enddata
\tablecomments{For each galaxy (first column) the total number of
  sources identified as star clusters is given in the second column,
  while the number of young (age $\leqslant$ 8 Myr) and old (age
  $\geqslant$ 250 Myr) star clusters is in the third and fourth column
  respectively.}

\end{deluxetable}

\begin{deluxetable}{lccccc}
\tabletypesize{\scriptsize}
\tablecaption{Star Cluster counts with extinction. 
\label{tb:extinction}}
\tablewidth{0pt}
\tablehead{
  \colhead{Name}   &   \colhead{$(N\geqslant 100~\mathrm{ Myr})_{Gal-Ext}$}  &    
  \colhead{$(N\geqslant 100~ \mathrm{  Myr})_{LMC-Ext}$}  &    
  \colhead{$(N\geqslant 100~ \mathrm{  Myr})_{SMC-Ext}$}  &    
  \colhead{$(N\geqslant 250~ \mathrm{  Myr})_{NO-Dust}$}  \\
}
\startdata
 ESO 498G5 &    7       &    3       &    6      &    5 \\
 ESO 499G37 &    2       &    1       &    1      &    5 \\
 NGC 406 &   17       &   14       &   17      &   23 \\
 NGC 1345 &   10       &    8       &   12      &   18 \\
 NGC 1483 &    1       &    1       &    1      &   10 \\
 NGC 2082 &    8       &    3       &    7      &   18 \\
 NGC 2758 &   15       &   11       &   13      &   20 \\
 NGC 3259 &    7       &    5       &    6      &   14 \\
 NGC 3455 &    3       &    3       &    3      &    7 \\
 NGC 4980 &    7       &    5       &    8      &   19 \\
 NGC 6384 &    6       &    5       &    5      &    6 \\
\enddata

\tablecomments{Number of star clusters more massive than $10^5~
  \mathrm{M_{\sun}}$ and with age $\geqslant 100$ Myr identified in
  the central region of the galaxies within the NICMOS F160W field of
  view. The number of sources has been obtained using a least
  chi-squared fit of the four band photometry (F435W, F606W, F814W \&
  F160W) allowing for a variable amount of dust extinction with
  different extinction laws (Galactic: second column, LMC third column
  and SMC fourth column). The last column reports the number of
  sources within the same field of view that are older than $250$ Myr
  and more massive than $10^5 ~\mathrm{M_{\sun}}$ when the fit is
  forced to have no dust extinction.}
\end{deluxetable}

\begin{deluxetable}{lcccccc}
\tabletypesize{\scriptsize}
\tablecaption{Candidate Globular Cluster Luminosity Function fitted as a Gaussian. 
\label{tb:gaussian}}
\tablewidth{0pt}
\tablehead{
  \colhead{Name}       &   \colhead{$M_{V_{peak}}$}  &
  \colhead{FWHM}       &   \colhead{$N_{fit}$}   & 
  \colhead{$f_{cont}$} &   \colhead{$f_{ext-bias}$} & 
  \colhead{$N_{tot}$}\\
}
\startdata
\smallskip
ESO 498G5  & -9.32$\pm$0.09  &  1.46$\pm$0.14  &  67$\pm$10 & 0.1& $<$ 0.05 & $ 57^{+10}_{-10}$ \\
\smallskip
ESO 499G37 & -7.69$\pm$0.10  &  1.46$\pm$0.15  &  53$\pm$9  & 0.1& 0.70 & $ 14^{+10}_{-13}$ \\
\smallskip
NGC  406   & -7.86$\pm$0.09  &  2.15$\pm$0.15  & 147$\pm$15 & 0.1& 0.30 & $92^{+20}_{-31}$ \\
\smallskip
NGC 1345   & -8.56$\pm$0.14  &  1.80$\pm$0.26  &  33$\pm$6  & 0.1& 0.45 & $ 16^{+ 7}_{-10}$ \\
\smallskip 
NGC 1483   & -6.87$\pm$0.13  &  2.01$\pm$0.22  &  61$\pm$9 & 0.1& 0.90 & $  5^{+4}_{-4}$ \\
\smallskip 
NGC 2082   & -8.16$\pm$0.12  &  2.13$\pm$0.20  &  84$\pm$10 & 0.1& 0.65 & $ 26^{+13}_{-20}$ \\
\smallskip 
NGC 2758   & -8.75$\pm$0.08  &  2.08$\pm$0.13  & 171$\pm$17 & 0.1& 0.35 & $100^{+24}_{-39}$ \\
\smallskip 
NGC 3259   & -8.45$\pm$0.08  &  1.49$\pm$0.14  &  72$\pm$10 & 0.1& 0.55 & $ 29^{+13}_{-19}$ \\
\smallskip 
NGC 3455   & -7.91$\pm$0.12  &  2.00$\pm$0.19  &  69$\pm$10 & 0.1& 0.55 & $ 27^{+12}_{-18}$ \\
\smallskip 
NGC 4980   & -8.12$\pm$0.08  &  1.92$\pm$0.14  & 139$\pm$14 & 0.1& 0.65 & $ 43^{+20}_{-32}$ \\
\smallskip 
NGC 6384   & -8.34$\pm$0.11  &  1.93$\pm$0.19  &  76$\pm$11 & 0.1& 0.10 & $ 61^{+11}_{-12}$ \\
\enddata
\tablecomments{Completeness corrected GCLF gaussian fit for the
  galaxies in our sample (first column). Fitted peak $M_V$-magnitude and
  FWHM are in the second and third columns respectively. The fourth
  column gives the total number of candidate globular clusters as inferred from
  the fit ($N_{fit}$ ). The fifth column contains the estimated
  fraction of stellar contaminants in our sample (see section
  \ref{sec:contamination}). The sixth column gives the fractional bias
  in the number of old star clusters induced by dust extinction of
  their host galaxy (see section~\ref{sec:dust}). The last column
  reports our fiducial number $N_{tot}$ of candidate globular clusters
  obtained,  defined as $N_{tot} = N_{fit} \cdot
  (1-f_{cont})\cdot(1-f_{ext-bias})$. }

\end{deluxetable}

\begin{deluxetable}{lccccl}
\tabletypesize{\scriptsize}
\tablecaption{Specific frequencies of candidate GCs. \label{tb:spec_freq}}
\tablewidth{0pt}
\tablehead{
  \colhead{Name}             &   \colhead{$R_{search}$}   &
  \colhead{${M_{V_{JC}}}_{gal}$} &   \colhead{${S_N}_{gal}$}  &
  \colhead{${M_{V_{JC}}}_{bul}$} &   \colhead{${S_N}_{bul}$}  \\
                             &   \colhead{(kpc)}  &
  \colhead{(mag)}              &                  &
  \colhead{(mag)}              &    \\
}
\startdata
\smallskip 
ESO 498G5  & 3.254 & -19.04 & $ 1.4^{+0.2}_{-0.3}$  & -17.06 & $   8.6^{+  1.5}_{-   1.6}$ \\
\smallskip 
ESO 499G37 & 1.529 & -17.08 & $ 2.1^{+1.5}_{-1.9}$  & -12.79 & $ 110^{+ 75}_{- 101}$ \\
\smallskip 
NGC  406   & 1.708 & -18.01 & $ 5.3^{+1.2}_{-1.8}$  & -16.65 & $  20^{+  4}_{-   7}$ \\
\smallskip 
NGC 1345   & 1.654 & -17.68 & $ 1.4^{+0.6}_{-0.8}$  & -17.57 & $   1.5^{+  0.7}_{-   0.9}$ \\
\smallskip 
NGC 2082   & 1.481 & -18.20 & $ 1.4^{+0.7}_{-1.0}$  & -15.65 & $  14^{+  7}_{-  11}$ \\ 
\smallskip 
NGC 2758   & 2.707 & -18.87 & $ 2.8^{+0.7}_{-1.1}$  & -17.20 & $  13^{+  3}_{- 5}$ \\ 
\smallskip 
NGC 3259   & 2.140 & -18.55 & $ 1.1^{+0.5}_{-0.7}$  & -14.97 & $  30^{+ 13}_{-  20}$ \\
\smallskip 
NGC 3455   & 1.708 & -18.01 & $ 1.7^{+0.8}_{-1.1}$  & -13.66 & $ 96^{+ 43}_{-  63}$ \\ 
\smallskip 
NGC 4980   & 2.063 & -18.29 & $ 2.1^{+1.0}_{-1.5}$  & -17.39 & $  5.0^{+  2.2}_{-   3.5}$ \\ 
\smallskip 
NGC 6384   & 1.934 & -19.56 & $ 0.9^{+0.2}_{-0.2}$  & -19.44 & $   1.0^{+  0.2}_{-   0.2}$ \\
\enddata
\tablecomments{Candidate globular cluster specific frequencies with respect to
  the galaxy luminosity (fourth column) and to the bulge luminosity
  (sixth column). The relevant luminosities used here are given in the
  third and fifth columns and are expressed in the Johnson-Cousins
  system. The galaxy luminosity has been obtained by considering the
  light within a circular region of radius $R_{search}$ (second
  column), corresponding to the search area for the star clusters. The bulge
  luminosity is from \citet{acs_I}.}

\end{deluxetable}

\begin{deluxetable}{lccccccccl}
\tabletypesize{\scriptsize}
\tablecaption{Photometry of Nuclear Star Clusters. \label{tb:nuclearSC}}
\tablewidth{0pt}
\tablehead{
  \colhead{Name}                                 &   \colhead{$\alpha$(J2000)} &
  \colhead{$\delta$(J2000)}                  &   \colhead{$B$}   &
  \colhead{$V$}                                    &   \colhead{$I$}  &
  \colhead{$H$}                                    &   \colhead{$R_B$}  \\
                                                            &   \colhead{(h m s)}         &
  \colhead{(deg \arcmin $ $ \arcsec)}  &   \colhead{(mag)}           &
  \colhead{(mag)}                                 &   \colhead{(mag)}           &
  \colhead{(mag)}                                 &   \colhead{(arcsec)} \\
}
\startdata
ESO 498G5   & 09 24 40.68 & -25 05 31.7 & 20.93$\pm$0.04  & 19.72$\pm$0.02 & 18.93$\pm$0.17  & 17.23$\pm$0.33  & 0.04$\pm$0.02  \\
ESO 499G37 & 10 03 41.69 & -27 01 38.1 & 20.2$\pm$0.1  & 19.32$\pm$0.05 & 18.8$\pm$0.25  & 18.35$\pm$0.35  & 0.25$\pm$0.05  \\
NGC  406     & 01 07 24.48 & -69 52 30.2 & 22.29$\pm$0.06  & 21.22$\pm$0.04& 20.47$\pm$0.14  & 19.15$\pm$0.28  & 0.04$\pm$0.02  \\
NGC 1345    & 03 29 31.66 & -17 46 42.6 & 21.7$\pm$0.1  & 20.91$\pm$0.07 & 20.36$\pm$0.19  & 18.88$\pm$0.31  & 0.05$\pm$0.02  \\
NGC 1483    & 03 52 47.65 & -47 28 37.3 & 22.02$\pm$0.04  & 21.54$\pm$0.06 & 21.02$\pm$0.07  & 19.76$\pm$0.19  & 0.04$\pm$0.02  \\
NGC 2082    & 05 41 50.92 & -64 18 02.6 & 21.79$\pm$0.07  & 20.39$\pm$0.02 & 19.48$\pm$0.18  & 17.76$\pm$0.23  & 0.05$\pm$0.02  \\ 
NGC 2758    & 09 05 31.17 & -19 02 33.1 & 20.97$\pm$0.06  & 20.29$\pm$0.02 & 19.82$\pm$0.15  & 18.62$\pm$0.26  & 0.08$\pm$0.05  \\ 
NGC 3259    & 10 32 34.67 & +65 02 25.9 & 20.54$\pm$0.08  & 19.06$\pm$0.05 & 18.96$\pm$0.24  & 17.05$\pm$0.33  & 0.04$\pm$0.02  \\
NGC 3455    & 10 54 31.12 & +17 17 04.1 & 20.78$\pm$0.05  & 19.16$\pm$0.01 & 18.86$\pm$0.13  & 17.35$\pm$0.30  & 0.05$\pm$0.02  \\ 
NGC 4980    & 13 09 10.18 & -28 38 33.6 & 21.1$\pm$0.03  & 20.07$\pm$0.01 & 19.44$\pm$0.08  & 18.23$\pm$0.11  & 0.05$\pm$0.02  \\ 
NGC 6384    & 17 32 24.27 & +07 03 36.1 & 20.6$\pm$0.08  & 18.97$\pm$0.07 & 17.87$\pm$0.13  & 15.98$\pm$0.27  & 0.06$\pm$0.03  \\
\enddata
\tablecomments{Summary of the photometric properties of nuclear star
clusters for our sample of galaxies, listed in the first column. The
second and third columns report the coordinates of the central star
cluster, followed by apparent magnitudes ($B,~V,~I,~H$). The last
column is an estimate of the half-light radius in the $B$ band, after
deconvolution with the PSF.}

\end{deluxetable}

\begin{deluxetable}{lccc}
\tabletypesize{\scriptsize}
\tablecaption{Properties of Nuclear Star Clusters. \label{tab:central_cluster_fit}}
\tablewidth{0pt}
\tablehead{
  \colhead{Name}                                 &   \colhead{Mass} &
  \colhead{Age}                                    &   \colhead{$E(B-V)$}  \\
  }
\startdata
  ESO 498G5 & $5-12 \cdot 10^7 \msun$ & $5-13$ Gyr & 0.06-0.2 \\ 
  ESO 499G37 & $1.7-3.9 \cdot 10^6 \msun$ & $1-3$ Gyr & 0-0.15 \\ 
  NGC 406 &  $1.8-3.9 \cdot 10^6 \msun$ & $2-5$ Gyr & 0-0.2 \\  
  NGC 1345 &  $1.1-1.7 \cdot 10^6 \msun $ & $1-2$ Gyr & 0-0.25 \\ 
  NGC 1483 &  $0.2-1 \cdot 10^6 \msun$ & $8-50$ Myr & 0.2-0.25 \\
  NGC 2082 &  $8.6-15 \cdot 10^7 \msun$ & 5-12 Gyr & 0.1-0.17 \\
  NGC 2758 &  $6-31 \cdot 10^5 \msun$ & 5-50 Myr & 0.5-0.7 \\
  NGC 3259 &  $2.5-21 \cdot 10^6 \msun$ & 5-3000 Myr & 0-1.17 \\
  NGC 3455 &  $1.1-2.8 \cdot 10^7 \msun$ & 2-10 Gyr & 0-0.24 \\
  NGC 4980 &  $8.1-26 \cdot 10^6 \msun$ & 3-13 Gyr & 0-0.03 \\
  NGC 6384 &  $8-20 \cdot 10^7 \msun$ & $2-13$ Gyr & 0.2-0.8 \\
\enddata
\tablecomments{Properties of the nuclear star clusters derived from
the fit of photometric properties (see Tab.~\ref{tb:nuclearSC}) using
single-stellar population models. The range in Mass (second column),
Age (third column) and Extinction (fourth column) is based on the
range of acceptable $1~\sigma$ fits.}

\end{deluxetable}


\end{document}